\documentclass[numberedappendix]{emulateapj}
\usepackage{epsfig,amsmath,amssymb,graphics,lscape}
\usepackage[colorlinks=true,  citecolor=blue,  pagecolor=blue,
  linkcolor=blue,  menucolor=blue, urlcolor=blue,
  linkbordercolor={0 0 1}, frenchlinks=True, breaklinks]{hyperref}
\usepackage{multirow}

\begin{document}
\newcommand{\Ha}{H$\alpha$}
\newcommand{\Hb}{H$\beta$}
\newcommand{\Hg}{H$\gamma$}
\newcommand{\Hd}{H$\delta$}
\newcommand{\Hyd}{{\rm H}}
\newcommand{\Hae}{\Hyd\alpha}
\newcommand{\Hbe}{\Hyd\beta}
\newcommand{\Hge}{\Hyd\gamma}
\newcommand{\Hde}{\Hyd\delta}
\newcommand{\Lya}{Ly$\alpha$}
\newcommand{\NII}{[{\rm N}~\textsc{ii}]}
\newcommand{\OIII}{[{\rm O}~\textsc{iii}]}
\newcommand{\OII}{[{\rm O}~\textsc{ii}]}
\newcommand{\SII}{[S \textsc{ii}]}
\newcommand{\NeIII}{[{\rm Ne}~\textsc{iii}]}
\newcommand{\OIIIa}{\OIII\,$\lambda$4363}
\newcommand{\OIIIA}{[\textsc{O iii}]{\rm-A}}
\newcommand{\zphotf}{z_{\rm phot}}
\newcommand{\zspecf}{z_{\rm spec}}
\newcommand{\zphot}{$\zphotf$}
\newcommand{\zspec}{$\zspecf$}
\newcommand{\Rcf}{R_{\rm C}}
\newcommand{\Rc}{$\Rcf$}

\newcommand{\Pagel}{$R_{23}$}
\newcommand{\Oratio}{$O_{32}$}
\newcommand{\Te}{$T_e$}
\newcommand{\OH}{12\,+\,$\log({\rm O/H})$}
\newcommand{\OHm}{12\,+\,\log({\rm O/H})}
\newcommand{\zsun}{$Z_{\sun}$}
\newcommand{\MB}{$M_B$}
\newcommand{\fifth}{$Z$=0.004}
\newcommand{\solar}{$Z$=0.02}
\newcommand{\iz}{$i$\arcmin$z$\arcmin}
\newcommand{\Ri}{\Rc$i$\arcmin}

\newcommand{\rNB}{{\rm NB}}
\newcommand{\nOII}{$\sim$1,300}

\newcommand{\HaHbi}{2.86}

\newcommand{\EBV}{E(B-V)}
\newcommand{\EBVa}{$E$($B$--$V$)}
\newcommand{\cc}{cm$^{-3}$}
\newcommand{\mm}{$\mu$m}
\newcommand{\Msun}{$M_{\sun}$}
\newcommand{\Mstar}{$M_{\star}$}
\newcommand{\iyr}{yr$^{-1}$}

\newcommand{\MZ}{\Mstar--$Z$}
\newcommand{\MZR}{MZR}
\newcommand{\zmin}{0.065}

\newcommand{\za}{0.01}
\newcommand{\zb}{1.62}
\newcommand{\areaa}{870.4}
\newcommand{\areab}{788.7}

\newcommand{\Nem}{9264}
\newcommand{\Ndet}{20}
\newcommand{\Ndetf}{19}
\newcommand{\NMMTf}{14}
\newcommand{\NKeckf}{6}

\newcommand{\OHKecksix}{7.24$^{+0.45}_{-0.30}$}
\newcommand{\OHKecktwo}{7.54$^{+0.27}_{-0.22}$}
\newcommand{\OHMMT}{7.53$^{+0.52}_{-0.33}$}
\newcommand{\NXMPG}{4}

\newcommand{\SFRSDA}{0.5}
\newcommand{\SFRSDM}{0.04}
\newcommand{\SFRA}{8.1}
\newcommand{\SFRM}{2.3}
\newcommand{\sSFRA}{$10^{-8}$}
\newcommand{\sSFRt}{100 Myr}
\newcommand{\MassA}{$2\times10^8$}
\newcommand{\MassM}{$3\times10^8$}
\newcommand{\AgeA}{7.75 dex}
\newcommand{\Dproj}{100 kpc}

\newcommand{\GALEX}{{\it GALEX}}
\newcommand{\FUV}{{\it FUV}}
\newcommand{\NUV}{{\it NUV}}

\newcommand{\TA}{\tablenotemark{a}}
\newcommand{\TB}{\tablenotemark{b}}
\newcommand{\TC}{\tablenotemark{c}}
\newcommand{\TD}{\tablenotemark{d}}
\newcommand{\TE}{\tablenotemark{e}}
\newcommand{\TF}{\tablenotemark{f}}

\submitted{Received 2013 July 16; accepted 2013 November 15}
\title{``DIRECT'' GAS-PHASE METALLICITIES, STELLAR PROPERTIES, AND
  LOCAL ENVIRONMENTS OF EMISSION-LINE GALAXIES AT REDSHIFTS BELOW 0.90}
\author{Chun Ly,\altaffilmark{1,10,11}
  Matthew A. Malkan,\altaffilmark{2}
  Tohru Nagao,\altaffilmark{3,4,5}
  Nobunari Kashikawa,\altaffilmark{6,7}
  Kazuhiro Shimasaku,\altaffilmark{8,9} and
  Masao Hayashi\altaffilmark{6}}

\shorttitle{Extremely Metal-Poor Emission-line Galaxies}
\shortauthors{Ly et al.}
\email{chun.ly@nasa.gov}
\altaffiltext{1}{Space Telescope Science Institute, Baltimore, MD, USA}
\altaffiltext{2}{Department of Physics and Astronomy, UCLA, Los Angeles, CA, USA}
\altaffiltext{3}{Research Center for Space and Cosmic Evolution, Ehime University,
  Matsuyama, Japan}
\altaffiltext{4}{The Hakubi Project, Kyoto University, Kyoto, Japan}
\altaffiltext{5}{Department of Astronomy, Kyoto University, Kyoto, Japan}
\altaffiltext{6}{Optical and Infrared Astronomy Division, National Astronomical
  Observatory, Mitaka, Tokyo, Japan}
\altaffiltext{7}{Department of Astronomy, School of Science, Graduate University
  for Advanced Studies, Mitaka, Tokyo, Japan}
\altaffiltext{8}{Department of Astronomy, School of Science, University of Tokyo,
  Bunkyo, Tokyo, Japan}
\altaffiltext{9}{Research Center for the Early Universe, School of Science,
  University of Tokyo, Tokyo, Japan}
\altaffiltext{10}{Current Address: National Aeronautics and Space Administration,
  Goddard Space Flight Center, Greenbelt, MD, USA}
\altaffiltext{11}{Giacconi Fellow.}

\begin{abstract}
  Using deep narrow-band (NB) imaging and optical spectroscopy from the
  Keck telescope and MMT, we identify a sample of \Ndet\ emission-line
  galaxies (ELGs) at $z=\zmin$--0.90 where the weak auroral emission
  line, \OIIIa, is detected at $\geq$3$\sigma$.
  These detections allow us to determine the gas-phase metallicity
  using the ``direct'' method. With electron temperature measurements,
  and dust attenuation corrections from Balmer decrements, we find that
  \NXMPG\ of these low-mass galaxies are extremely metal-poor with
  \OH\ $\leq$ 7.65 or one-tenth solar. Our most metal-deficient galaxy has
  \OH = \OHKecksix\ (95\% confidence), similar to some of the lowest
  metallicity galaxies identified in the local universe.
  We find that our galaxies are all undergoing significant star
  formation with average specific star formation rate (SFR) of
  (\sSFRt)$^{-1}$, and that they have high central SFR surface
  densities (average of \SFRSDA\ \Msun\ \iyr\ kpc$^{-2}$).
  In addition, more than two-thirds of our galaxies have between one
  and four nearby companions within a projected radius of \Dproj,
  which we find is an excess among star-forming galaxies at
  $z=$0.4--0.85.
  We also find that the gas-phase metallicities for a given stellar
  mass and SFR lie systematically lower than the local \MZ--(SFR)
  relation by $\approx$0.2 dex (2$\sigma$ significance).
  These results are partly due to selection effects, since
  galaxies with strong star formation and low metallicity
  are more likely to yield \OIIIa\ detections.
  Finally, the observed higher ionization parameter and high electron density
  suggest that they are lower redshift analogs to typical $z\gtrsim1$ galaxies.
\end{abstract}


\defcitealias{ly07}{L07}
\defcitealias{calzetti00}{C00}
\defcitealias{chabrier03}{Chabrier}
\defcitealias{salpeter}{Salpeter}
\defcitealias{andrews13}{AM13}


\section{INTRODUCTION}\label{1}
The chemical enrichment of galaxies, driven by star formation and
regulated by gas outflows from supernovae and inflows from cosmic
accretion, is a key process in galaxy formation that remains to
be fully understood.  The greatest difficulty in measuring
chemical evolution across all galaxy populations is the need for
rest-frame optical spectroscopy.
Metallicity determinations can be obtained through (1) interstellar
absorption lines (e.g., \ion{Fe}{2}, \ion{Mg}{2}), and (2)
nebular emission lines (e.g., \OII, \OIII, and \NII).
While studies have used absorption lines to measure heavy-element
abundances \citep{savaglio04}, the need for deep spectroscopy and
complications with curve-of-growth analysis have made it difficult.
As such, the primary method used to measure the metal abundances
in galaxies has been nebular emission lines. This technique has
the advantage of being able to probe low-luminosity galaxies
since it does not require continuum detection.
In addition, these emission lines can be observed in the optical and
near-infrared (near-IR) at redshifts of $\sim$3 and below with current
ground \citep[see e.g.,][]{hayashi09,moustakas11,henry13,momcheva13}
and space-based capabilities \citep[see e.g.,][]{atek10,xia12}, and
the forthcoming IR capabilities of the {\it James Webb Space Telescope}
(JWST) will extend this further to $z\approx6$.

The most reliable metallicity determination is made possible by
measuring the flux ratio of the \OIIIa\ auroral line against
a lower excitation line, such as \OIII\,$\lambda$5007.
The technique is often called the \Te\ or ``direct'' method for its
ability to determine the electron temperature (\Te) of the ionized gas,
and hence the gas-phase metallicity \citep[see e.g.,][]{aller84}.
However, the detection of \OIIIa\ is difficult, as it is very
weak (and almost undetectable in metal-rich galaxies). 
For example, the first data release
(DR1) of the Sloan Digital Sky Survey \citep[SDSS;][]{york00} only
revealed 8 new extremely metal-poor galaxies (XMPGs;
\OH\ $\leq7.65$) among 250,000 galaxies \citep{kniazev03}.
Even with improved selection and a larger sample ($\approx$530,000),
\cite{izotov06a} only detected \OIIIa\ at $\geq$2$\sigma$ significance
in 310 galaxies (i.e., one in 1700).
While SDSS spectra can be stacked for average measurements of \OIIIa\
\citep[hereafter AM13]{andrews13}, this sacrifices knowledge of the
  intrinsic scatter in the mass-metallicity (\MZ) relation (\MZR), which
can also constrain galaxy evolution models \citep[see e.g.,][]{dave11}.

While it is unfortunate that the \Te\ method cannot be used for
the full dynamic range of metallicity, the detection in
a galaxy of \OIIIa\ alone is a strong indication that it
is extremely metal deficient.
These rare XMPGs are suspected to be primeval galaxies that are
undergoing rapid assembly at the observed redshift
\citep[and references therein]{kniazev03}.
Studying larger samples of them can provide a better
understanding of the early stages of galaxy assembly.

One possibility is that XMPGs have significant outflows
that are induced by supernova from massive star formation.
These outflows could drive large amounts of metal-rich gas
out of the galaxy, thus decreasing the metal abundances.
Studies have found that outflows are prevalent in metal-poor
galaxies through (1) detection of outflowing ionized gas from
integral field unit (IFU) spectroscopy of an XMPG \citep{izotov06b};
(2) blue-shifted absorption lines from slit spectroscopy of
star-forming galaxies, suggesting that outflows are ubiquitous
\citep[e.g.,][]{weiner09,martin12}; and (3) evidence that galaxies
with higher specific SFR (SFR per unit stellar mass;
sSFR $\equiv$ SFR/\Mstar) are more metal poor
\citep{ellison08,laralopez10,mannucci10}.

Another explanation is that metal-deficient gas is supplied
either from the circumgalactic medium (CGM) perhaps through
a ``cold-mode'' accretion phase \citep{keres05,dekel09} or
from the strong interaction with a nearby merging companion
\citep[see e.g.,][]{kewley06,rupke10}.

The two most metal-deficient galaxies known to date are I Zw 18
\citep{searle72} and SBS0335-052 \citep[hereafter, SBS0335;][]{izotov90}
with oxygen abundances of \OH\ = 7.14 \citep[\zsun/35;][]{izotov06a}
and 7.19--7.34 \citep[\zsun/(23--32);][]{izotov99}, respectively.
Efforts have been made to increase the galaxy sample with
direct metallicity determinations in the local universe
\citep{brown08,berg12,izotov12}, and at higher redshift
\citep{hoyos05,kakazu07,hu09,atek11}. These studies have either
targeted galaxies with low luminosity or high equivalent
width (EW) emission lines. The latter are found using NB
imaging, grism spectroscopy, or unusual broad-band colors to
select them.
For example, \cite{kakazu07} and \cite{hu09} utilized NB
imaging to select ELGs at $z\sim0.40$--0.85 and conducted
optical follow-up spectroscopy with Keck to detect \OIIIa\
and determine direct metallicity for 28 galaxies. They found
one galaxy where their measured metallicity is
\OH\ = $6.97\pm0.17$.
To date, only $\approx$70 galaxies are known to have
\OH\ $\leq 7.65$ with the majority (90\%) of them
at $z\lesssim0.1$.

In this paper, we focus on our spectroscopic detections of \OIIIa\
in \Ndet\ galaxies at $z=\zmin$--0.90 (average of $z=0.54\pm0.22$)
in the Subaru Deep Field \citep[SDF;][]{kashik04}.
These galaxies were initially selected for their excess flux in
NB and/or intermediate-band filters produced by emission lines.
In particular, we have rest-frame spectral coverage of
at least 3700--5010\AA, enabling metallicity determinations
using the \Te\ method.

The outline of the paper is as follows.
In Section~\ref{sec:SDF} we describe the imaging survey for the
SDF, the selection of over 9,000 ELGs, the follow-up optical
spectroscopy that we conducted, and our accurate 
flux calibration approach.
We then discuss in Section~\ref{sec:sample} our approach
for detecting and measuring nebular emission lines, which yields a
spectroscopic sample with \OIIIa\ detections that are significant
at $\geq$3$\sigma$. We also present arguments for why all
but one of our galaxies is ionized primarily by young stars.
Section~\ref{sec:Results} then describes how we determine: (1) the
dust attenuation properties in our galaxies; (2) the electron
temperature and the gas-phase oxygen metallicity; (3) the
dust-corrected SFRs; (4) the stellar properties from spectral
energy distribution (SED) fitting; (5) the nearby environment; and
(6) the SFR surface density.
In Section~\ref{sec:Disc}, we compare our results to other galaxies
that have direct metallicity determinations,  and discuss
  selection effects for our sample.
Finally, we estimate the space densities of compact, extreme star-forming,
metal-poor galaxies found in this survey, and consider their implications for
the broader context of galaxy evolution.

Throughout this paper, we adopt a flat cosmology with $\Omega_{\Lambda}=0.7$,
$\Omega_M=0.3$, and $H_0=70$ km s$^{-1}$ Mpc$^{-1}$ to determine
distance-dependent measurements, and magnitudes are reported on the
AB system \citep{oke74}. For reference, we adopt \OH$_{\sun}$ = 8.69
\citep{prieto01} for metallicity measurements quoted against the
solar value, \zsun.
Unless otherwise indicated, we report 95\% confidence measurement
uncertainties, and ``\OIII'' alone refers to the strong
  5007\AA\ emission line.


\section{THE SUBARU DEEP FIELD}\label{sec:SDF}
\begin{figure*}
  \plotone{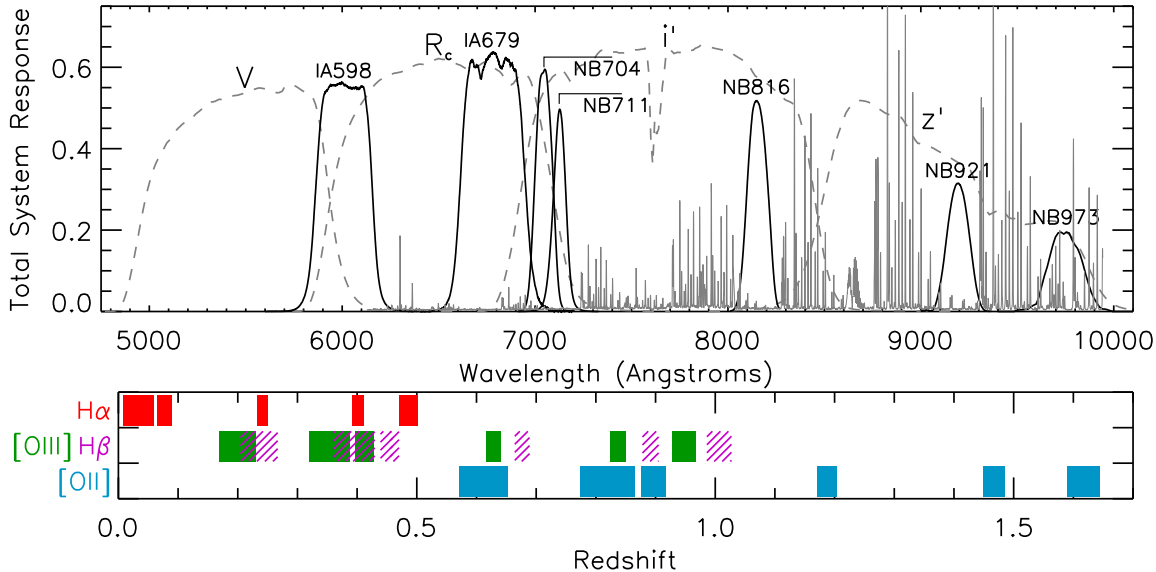}
  \caption{{\it Top:} Total system throughput for the
    $V\Rcf i\arcmin z\arcmin$ filters (gray dashed lines) and the
    IA598, IA679, NB704, NB711, NB816, NB921, and NB973 filters
    (black solid lines). {\it Bottom:} Redshift ranges probed
    when \Ha, \OIII, \Hb, and \OII\ are redshifted into these filters.
    The SDF probes 64\% of redshift space and 67\% of comoving volume
    at $z\leq1.03$.}
  \label{fig:response}
\end{figure*}

The SDF has the most sensitive optical imaging in several NB
and intermediate (IA) filters in the sky, and is further complemented
with ultra-deep multi-band imaging between 1500\AA\ and 4.5\mm. A summary
of the ancillary imaging is available in \cite{ly11a} and later
in Section~\ref{sec:SED}.
The results for this paper are based on data obtained in the NB704, NB711,
NB816, NB921, NB973, IA598, and IA679 filters.
A summary of their properties (i.e., central wavelength, sensitivity)
is reported in Table~\ref{tab:SDF_sample}, and Figure~\ref{fig:response}
shows the total system response through these filters and the surveyed
redshifts.


\newcommand{\pa}{\phm{1}}
\begin{deluxetable*}{cccccrccccc}
  \tabletypesize{\scriptsize}
  \tablewidth{0pc}
  \tablecaption{Summary of Filters, Emission-line Samples, and Redshift Windows}
  \tablehead{
    \colhead{Filter}&
    \colhead{$\lambda_c$}&
    \colhead{FWHM}&
    \colhead{$m_{\rm lim}(3\sigma)$}&
    \colhead{Area}&
    \colhead{$N_{\rm total}$}&
    \colhead{$N_{\rm target}$}&
    \colhead{$N_{\rm spec}$}&
    \colhead{$z$(\Ha)}&
    \colhead{$z$(\OIII)}&
    \colhead{$z$(\OII)}\\
    & \colhead{[\AA]} & \colhead{[\AA]} & \colhead{[mag]} & \colhead{[arcmin$^2$]}\\
    \colhead{(1)}&\colhead{(2)}&\colhead{(3)}& \colhead{(4)}&\colhead{(5)}&
    \colhead{(6)}&\colhead{(7)}&\colhead{(8)}&\colhead{(9)}&\colhead{(10)}&\colhead{(11)}}
  \startdata
  IA598 & 6007 &  303 & 26.79 & 870.4 &  641 &\pa31 &\pa21 &       \ldots & 0.170--0.230 & 0.571--0.652\\
  IA679 & 6780 &  340 & 27.39 & 870.4 &  790 &\pa76 &\pa54 & 0.007--0.059 & 0.320--0.388 & 0.774--0.865\\
  NB704 & 7046 &  100 & 26.71 & 870.4 & 1695 &  173 &  140 & 0.066--0.081 & 0.397--0.417 & 0.877--0.904\\
  NB711 & 7111 &\pa72 & 26.07 & 870.4 & 1480 &  111 &\pa92 & 0.078--0.089 & 0.413--0.427 & 0.898--0.918\\
  NB816 & 8150 &  120 & 26.90 & 870.4 & 1602 &  300 &  204 & 0.233--0.251 & 0.616--0.640 & 1.171--1.203\\
  NB921 & 9196 &  132 & 26.71 & 870.4 & 2361 &  251 &  185 & 0.391--0.411 & 0.824--0.850 & 1.450--1.485\\
  NB973 & 9755 &  200 & 25.69 & 788.7 & 1243 &\pa71 &\pa63 & 0.471--0.502 & 0.928--0.968 & 1.591--1.644\\\hline
  Total &\ldots&\ldots&\ldots &\ldots & \Nem &  870 &  713 & \ldots       & \ldots       & \ldots\\
  \vspace{-3mm}
  \enddata
  \label{tab:SDF_sample}
  \tablecomments{We report the numbers of excess emitters ($N_{\rm total}$) in Col. (6), 
    the numbers of galaxies with targeted MMT and/or Keck spectra ($N_{\rm target}$) in
    Col. (7), and the sub-sample with robust spectroscopic redshift ($N_{\rm spec}$) in Col. (8).}
\end{deluxetable*}

These SDF data were acquired with Suprime-Cam \citep{miyazaki02}, the
optical imager mounted at the prime focus of the Subaru telescope,
between 2001 March and 2007 May. The acquisition and reduction of
these data have been discussed extensively in \cite{kashik04},
\cite{kashik06}, \cite{ly07} (hereafter L07), and \cite{OIIpop} for
the NB data, and in \cite{nagao08} for the IA data.
In brief, data were obtained mostly in photometric conditions with
average seeing of 0\farcs9--1\farcs0 for all five NB and two IA filters.
These data were reduced following standard reduction procedures using
\textsc{sdfred} \citep{yagi02,ouchi04}, a software package designed
especially for Suprime-Cam data.

The most prominent emission lines entering these NB and IA filters are
\Ha, \OIII, \Hb, and \OII, at well-defined redshift windows between
$z=\za$ and $z=\zb$. This results in probing 64\% in redshift space and
67\% of the available comoving volume at $z\leq1.03$.
Compared to the previous NB survey for XMPGs at $z=0.24$--0.85 \citep{hu09},
our survey probes 4.7 (3.8) times more redshift (volume) space, and is
deeper by $\approx$1.5 mag in the NB filters.

\subsection{Selection of Emission-line Galaxies}\label{sec:NBem}
To select NB and IA excess emitters due to the presence of nebular
emission line(s), we use the standard color excess selection, where
photometric fluxes from these filters are compared against those for
adjacent broad-band filters that sample the continuum.
Since the technique has been extensively used, we briefly summarize
it below, referring readers to \cite{fujita03} and \cite{newha}.
We summarize our excess emitter sample in Table~\ref{tab:SDF_sample}.

The measured continuum adjacent to the line is determined by the two
broad-band filters closest to the narrow bandpass.
For NB921 and NB973, we start with the $z$\arcmin-band. Since the
central wavelengths of these filters are redder than what the $z$\arcmin\
filter measures, we correct for such differences using the
$i$\arcmin--$z$\arcmin\ color \citep{HaSFR,OIIpop}.
The remaining filters use a flux-weighted combination of either the
$V$- and \Rc-band (IA598), the \Rc- and $i$\arcmin-band
(NB704, NB711, and IA679), or the $i$\arcmin- and $z$\arcmin-band (NB816):
\begin{equation}
  f_{\rm cont} = \epsilon f_{\rm blue} + (1-\epsilon) f_{\rm red},
\end{equation}
where $f_{\rm blue}$ and $f_{\rm red}$ are the flux density in 
erg s$^{-1}$ cm$^{-2}$ Hz$^{-1}$ for the bluer and redder broad-band
filters, respectively (e.g., $i$\arcmin\ and $z$\arcmin\ for NB816),
and $\epsilon = $ 0.45 (IA598), 0.5 (NB704, NB711), 0.75 (IA679), and
0.6 (NB816).

Photometric measurements for broad-band data are obtained by running
SExtractor \citep{bertin96} in ``dual-image'' mode, where the
respective NB or IA image is used as the ``detection'' image.
This works well because all broad-band, IA, and NB mosaicked images
have very similar seeing, so that excess colors are determined
within the same physical scale of the galaxies. 
For the extraction of fluxes and selection of sources, we use a
2\arcsec\ diameter circular aperture.
For comparison, the 3$\sigma$ sensitivities for the $V$, \Rc,
$i$\arcmin, and $z$\arcmin\ data are between 26.27 and 27.53 mag,
which are generally deeper than the NB and IA imaging.

We also exclude sources which fall in regions affected by poor
coverage and contamination by bright foreground stars.
The unmasked regions cover \areaa\ arcmin$^2$ for all filters
with the exception of NB973, which covers \areab\ arcmin$^2$.
The latter is smaller due to higher systematic noise in one of
the ten CCDs, which we mask to avoid significant spurious
detections.
In total, 123123, 97632, 133273, 119541, 84786, 118097, and
139585 sources are detected in the unmasked regions of the
NB704, NB711, NB816, NB921, NB973, IA598, and IA679 mosaics,
respectively. Among these sources, 1695, 1480, 1602,
2361, 1243, 641, and 790 are identified as NB or IA
excess emitters.

In certain circumstances, our sources are selected by more than
one filter. This is due to some fortuitous redshift overlap of
our NB/IA filters such that different emission lines (e.g.,
\Ha\ and \OIII) are detectable at the same redshift
(see Figure~\ref{fig:response}).
Accounting for duplicate galaxies, the complete SDF ELG sample
consists of \Nem\ galaxies mostly at $z=\za$ to $z=\zb$ with
some at higher redshift due to Ly$\alpha$ emission.
We note that the sample presented in this paper supersedes our
earlier multi-NB studies. 
Previously, we had selected ELGs in four NB filters and examined
the evolution of the \Ha, \OIII, and \OII\ luminosity functions
\citepalias{ly07}. A fifth NB filter (NB973) was later included,
and that filter allowed us to study dust attenuation at
$z=0.4$--0.5 \citep{HaSFR} and stellar properties out to $z=1.5$--1.6
\citep{OIIpop}.

\subsection{Optical Spectroscopy}\label{sec:spectra}
The primary results of this paper are based on optical spectroscopy
taken over the past several years, with Keck's Deep Imaging Multi-Object
Spectrograph \citep[DEIMOS;][]{faber03} and MMT's Hectospec
\citep{fabricant05}.
In total, we obtained 945 optical spectra for 870 ELGs, and successfully
detected emission lines to determine redshift for 713 galaxies or 82\%
of the targeted sample. These spectra were initially obtained to confirm
that the NB technique efficiently identified ELGs. 
A summary of the spectroscopically confirmed excess emitters
is provided in Table~\ref{tab:SDF_sample}.

The majority (61\%) of our spectroscopic sample was obtained
from Hectospec. These spectra were obtained between 2008 Mar 13
and 2008 Apr 14, utilizing the 270 mm$^{-1}$ grating blazed at
5200\AA\ to yield spectral coverage of 3650\AA--9200\AA.
The combination of this grating with a fiber diameter of
1\farcs5 yields a spectral resolution of $\approx$6\AA.

The typical seeing for these observations was between 0\farcs7 and
1\farcs4, and the data were obtained at airmasses below 1.35,
with 95\% (77\%) of the data below 1.3 (1.2).
Four different fiber configurations were used, with on-source
integrations varying between 4 and 6 20-min exposures, which
is sufficient for cosmic ray rejection.
The MMT spectra were reduced following standard procedures with the
\textsc{iraf} Hectospec Reduction
Software\footnote{\url{http://tdc-www.harvard.edu/instruments/hectospec/}.}.
Since no order-blocking filter was used, the spectra at wavelengths
longward of 8200\AA\ have a significant amount of contaminating
second-order light (up to 30\% for the relatively blue flux standard stars).
The flux calibration in the far-red is therefore unreliable, and we do
not use it.

The Keck/DEIMOS observations were obtained in 2004, 2008, and 2009.
The 2004 spectroscopic observations have been discussed in
\cite{kashik06} and \citetalias{ly07}, and the more recent data have been
discussed in \cite{kashik11}. In brief, we constructed 13 slit-masks
with 1\arcsec\ slit widths. This typically corresponds to a spectral
resolution of $R\sim3600$ at 8500\AA, and spectral dispersion of
0.47\AA\ pixel$^{-1}$.
The 830 line mm$^{-1}$ grating and GG495 order-cut filter were used in
all DEIMOS observations.
This set-up resulted in a spectral coverage of $\approx$5000\AA--1\mm;
however the coverage varied along the dispersion axis of the
slit mask.
The typical seeing for these observations was 0\farcs5--1\farcs0
in 2004 and 0\farcs7-1\farcs1 for 2008--2009 
with integration times of 2--3 hr.
Almost all (86\%) of the DEIMOS spectra were obtained at low
airmasses ($\leq$1.3), with the remaining data taken at an airmass
of 1.35.

\subsubsection{Flux Calibration}
The metallicities that we will determine require measurements
between 3700\AA\ and 5010\AA\ in the rest-frame, thus accurate flux
calibration is critical. We follow a rigorous approach: we
(1) observe spectro-photometric standards to account for the
wavelength-dependent sensitivity of each spectrograph; (2) correct
for slit losses by comparing spectra against broad-band photometric
data; (3) compare emission-line fluxes against NB photometry
to assess the accuracy of flux calibration; and (4) compare spectra
obtained on different nights for a few dozen multiply-observed galaxies.
A more detailed description of the flux calibration for our MMT/Hectospec
and Keck/DEIMOS spectra is deferred to Appendix~\ref{sec:f_calib}. In
brief, our various independent tests and analyses yielded consistent
results, and demonstrated that the absolute flux calibration is reliable
at the 0.15--0.17 dex (0.12--0.17 dex) level for Hectospec (DEIMOS).


\section{THE \OIIIa\ SAMPLE}\label{sec:sample}

\begin{figure*}
  \epsscale{1.1}
  \plotone{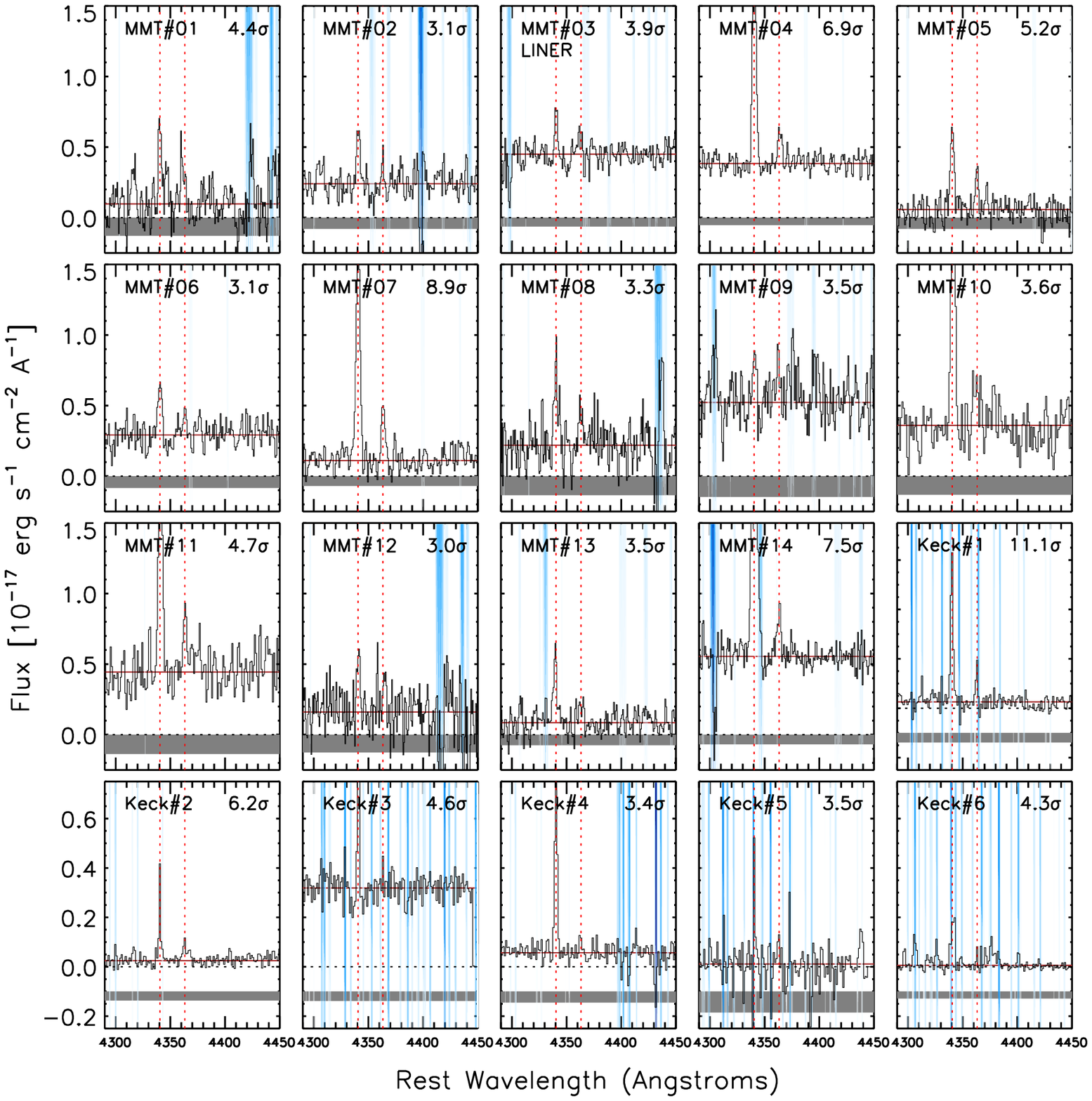}
  \vspace{-4mm}
  \caption{Detections of \OIIIa\ in \Ndet\ galaxies. The rest-frame spectra
    are shown in black, with vertical red dashed lines indicating the
    locations of \Hg\,$\lambda$4340 and \OIIIa. OH sky-lines are indicated
    by blue vertical bands, where the strength of the sky-lines is denoted
    by their shade of blue (darker is stronger).
    The grey-shaded regions indicate the locally measured 1$\sigma$ noise,
    while the horizontal brown lines correspond to the median of the continua.}
  \label{fig:zoom_in}
\end{figure*}

To extract fluxes for strong and weak emission lines in these spectra,
we fit each line with a Gaussian profile using the IDL routine
\textsc{mpfit} \citep{markwardt09}. The expected
location of emission lines was based on  {\it a priori} 
redshift determined by either the
\OIII\ or \Ha\ (for lower redshift). A local median,
$\left<f\right>$, is computed within a 200\AA-wide region,
excluding regions affected by OH skylines
and nebular emissions. In addition, the standard deviation
$\sigma(f)$ is measured locally.
Examples of the computed medians and standard deviations are
shown in Figure~\ref{fig:zoom_in}.
To determine the significance of emission lines, we integrate the
spectrum between $l_C-2.5\sigma_G$ and $l_C+2.5\sigma_G$, where
$\sigma_G$ is the Gaussian width:
\begin{equation}
  {\rm Flux} \equiv \sum_{-2.5\sigma_G}^{+2.5\sigma_G} \left[f(\lambda-l_C)-\left<f\right>\right] \times l\arcmin.
\end{equation}
Here, $l$\arcmin\ is the spectral dispersion (1.21\AA\ pixel$^{-1}$
for MMT and $\approx$0.47\AA\ pixel$^{-1}$ for Keck).
We then compute the signal-to-noise (S/N) of the line by dividing
the integrated flux by:
\begin{equation}
  {\rm Noise} \equiv \sigma(f) \times l\arcmin \times \sqrt{N_{\rm pixel}},
\end{equation}
where $N_{\rm pixel}=5\sigma_G/l$\arcmin.

Adopting a minimum significance threshold of 3$\sigma$, we identify
20 and 14 \OIIIa\ detections with MMT and Keck, respectively.
We visually inspected each \OIIIa\ detection. For MMT, we found that
OH sky-lines contaminated \OIIIa\ in three cases, and \Hb\ in three
other galaxies. This results in a final sample of \NMMTf\ \OIIIa\
detections.
For Keck, OH sky-lines contaminated \OIIIa\ in three cases and
\OII\ measurements in two other cases, while three sources lack
full spectral coverage (missing \OII, \Hb, and/or
\OIII\,$\lambda\lambda$4959,5007).\footnote{There are two galaxies
  (Keck\#2, \#4) that we include in our sample for various reasons
  discussed in Section~\ref{sec:dust} and Table~\ref{tab:em_lines}.}
This reduced the \OIIIa\ Keck sample to \NKeckf\ galaxies.

\begin{figure*}
  \plotone{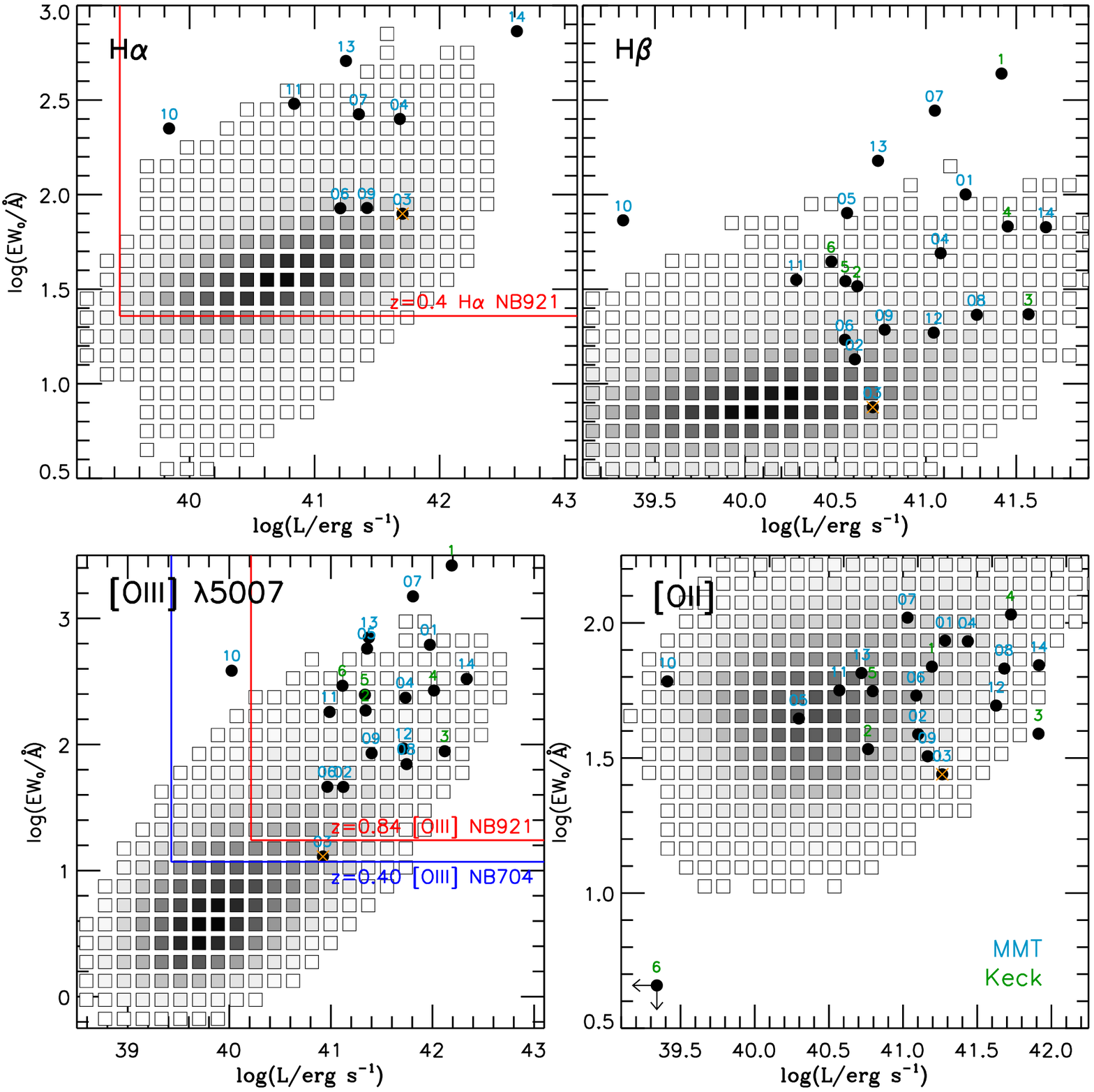}
  \caption{Emission-line luminosities and rest-frame EWs for our \OIIIA\ sample.
    All fluxes are observed, before correction for dust attenuation.
    Gray squares illustrate the SDSS DR7 emission-line sample with at
    least 5$\sigma$ detections in all four lines. The density of sources is
    shown with a linear scale from white (low) to black (highest).
    Our only LINER (MMT\#03) is denoted by the orange crosses.
    For comparative purposes, we show the EW and luminosity limits of our
    NB921 imaging (solid red lines) for $z=0.4$ \Ha\ (top left) and $z=0.84$
    \OIII\ (bottom left) and NB704 imaging (solid blue lines) for $z=0.4$
    \OIII. The MMT sample is labelled in blue while the Keck sample
    is indicated in green.}
  \label{fig:EW_Lum}
\end{figure*}

Our final sample of \OIIIa\ detections, hereafter the ``\OIIIA'' sample,
consists of \Ndet\ galaxies.
The MMT and Keck \OIIIa\ detections are shown in Figure~\ref{fig:zoom_in}
with the full MMT (Keck) spectra provided in
Figures~\ref{fig:MMT_spec1}--\ref{fig:MMT_spec2}
(Figure~\ref{fig:Keck_spec}).
We also summarize our \OIIIa\ sample in Table~\ref{tab:source_summary}.
For convenience, our galaxies are identified as ``MMT'' and
``Keck'' followed by a sequential number.
In Table~\ref{tab:em_lines}, we provide the \OII, \Hb, and \OIII\ emission-line
fluxes along with the \Oratio\ and \Pagel\ \citep{pagel79} flux ratios:
\begin{eqnarray}
  O_{32} & \equiv & \frac{\OIII\,\lambda\lambda4959,5007}{\OII\,\lambda\lambda3726,3729},{\rm~and}\\
  R_{23} & \equiv & \frac{\OII\,\lambda\lambda3726,3729 + \OIII\,\lambda\lambda4959,5007}{\Hbe}.
\end{eqnarray}
Also, emission-line luminosities and rest-frame EWs (EW$_0$) are illustrated
in Figure~\ref{fig:EW_Lum}. The latter are determined by measuring the
continuum from the SEDs with corrections for emission-line
contamination (see Section~\ref{sec:SED}).
The former is $L = 4\pi d_L^2 F_{\rm Line}$, where $d_L$ is the
luminosity distance, and $F_{\rm Line}$ is the emission-line flux.


We note that for two galaxies (MMT\#04 and MMT\#07), spectra were obtained with
two different MMT/Hectospec fiber configurations. In both cases, \OIIIa\
was detected in each spectrum, confirming that these detections are robust.
For these two galaxies, we combine the spectra for a higher S/N spectrum.
In addition, more recent MMT/Hectospec observations for MMT\#01,
\#03--\#05, \#07, \#10, and \#13 were taken in less ideal observing
conditions. These observations also detected \OIIIa\ in MMT\#04,
\#07, and possibly \#13.
Since these spectra are less sensitive than those obtained in 2008, we
do not combine them for the purpose of our analyses.
\begin{deluxetable*}{llcccclc}
  \tabletypesize{\scriptsize}
  \tablewidth{0pc}
  \tablecaption{Summary of Spectroscopic Sample}
  \tablehead{
    \colhead{ID}&
    \colhead{Name}&
    \colhead{Line Sel.}&
    \colhead{R.A.}&
    \colhead{Dec.}&
    \colhead{\zspec}&
    \colhead{Obs. Dates (UT)}&
    \colhead{$t_{\rm int}$}\\
    & & & \colhead{[hr]} & \colhead{[deg]} & & & \colhead{[min.]}\\
    \colhead{(1)}&\colhead{(2)}&\colhead{(3)}&\colhead{(4)}&\colhead{(5)}&\colhead{(6)}&\colhead{(7)}&\colhead{(8)}}
  \startdata
  MMT\#01 &                               NB816--140623 & \OIII         & 13:25:16.87 & 27:39:06.92 & 0.6380 & 2008 Apr 14         &  120\\
  MMT\#02 &                               NB711--064628 & \OIII         & 13:23:39.13 & 27:32:52.71 & 0.4327 & 2008 Mar 13         &  120\\
  MMT\#03\TA &                            NB973--104154 & \Ha           & 13:23:39.17 & 27:31:47.34 & 0.4809 & 2008 Mar 13         &  120\\
  MMT\#04 & NB704--088982\_NB921--126525\_IA679--112491 & \OIII,\Ha     & 13:24:46.63 & 27:34:56.98 & 0.3933 & 2008 Mar 13, Apr 11 &  240\\
  MMT\#05 &                               IA679--031637 & \OIII         & 13:25:03.37 & 27:17:23.77 & 0.3846 & 2008 Mar 13         &  120\\
  MMT\#06 &                NB704--036405\_NB921--063205 & \OIII,\Ha     & 13:23:54.77 & 27:20:12.53 & 0.3995 & 2008 Mar 13         &  120\\
  MMT\#07 & NB704--049936\_NB921--079428\_IA679--062450 & \OIII,\Ha     & 13:24:06.94 & 27:24:01.76 & 0.3896 & 2008 Mar 13, Apr 11 &  240\\
  MMT\#08 &                               NB816--081644 & \OIII         & 13:23:42.97 & 27:26:35.59 & 0.6335 & 2008 Apr 10         &\pa80\\
  MMT\#09 &                               NB973--094500 & \Ha           & 13:23:49.80 & 27:28:35.32 & 0.4788 & 2008 Apr 10         &\pa80\\
  MMT\#10 &                               NB704--009999 & \Ha           & 13:23:56.39 & 27:13:32.98 & 0.0683 & 2008 Apr 10         &\pa80\\
  MMT\#11 &                               IA598--079010 & \OIII         & 13:24:13.64 & 27:25:09.27 & 0.1752 & 2008 Apr 10         &\pa80\\
  MMT\#12 &                               NB816--112403 & \OIII         & 13:25:21.78 & 27:33:15.69 & 0.6405 & 2008 Apr 10         &\pa80\\
  MMT\#13 &                NB711--102472\_NB973--156739 & \Hb,\Ha       & 13:24:28.88 & 27:45:51.88 & 0.4696 & 2008 Apr 11         &  120\\
  MMT\#14 &                NB711--077774\_NB973--125003 & \Hb,\Ha       & 13:25:22.94 & 27:37:40.33 & 0.4644 & 2008 Apr 11         &  120\\
  Keck\#1 & NB711--049857\_IA679--079866\_NB921--995851 & \OIII         & 13:25:11.94 & 27:27:31.20 & 0.8390 & 2004 Apr 23         &  120\\
  Keck\#2 &                               NB816--070113 & \OIII         & 13:24:34.91 & 27:24:10.20 & 0.6230 & 2004 Apr 23         &  118\\
  Keck\#3 &                               IA679--077341 & \OII          & 13:23:53.54 & 27:27:13.01 & 0.7906 & 2004 Apr 23         &  118\\
  Keck\#4 &                               NB704--087569 & \OII          & 13:23:43.51 & 27:34:20.98 & 0.8829 & 2008 May 1          &  130\\
  Keck\#5 &                               NB921--078003 & \OIII         & 13:24:43.66 & 27:23:34.86 & 0.8353 & 2009 Apr 27         &  180\\
  Keck\#6 &                               NB704--060432 & \OIII\TB      & 13:24:58.62 & 27:26:40.47 & 0.8237 & 2009 Apr 27         &  180\\
  \vspace{-3mm}
  \enddata
  \label{tab:source_summary}
  \tablenotetext{1}{This source is likely a LINER (see Section \ref{sec:AGN}).}
  \tablenotetext{2}{This galaxy was previously targeted for its NB921 excess flux due to \OIII.
      It was not identified as a NB921 emitter in the re-selection because of its faintness.
      The NB704 excess is due to \NeIII\,$\lambda$3869.}
\end{deluxetable*}

\renewcommand{\pa}{\phm{$^a$}}
\begin{deluxetable*}{lrrrrcccr}
  \tabletypesize{\scriptsize}
  \tablewidth{0pc}
  \tablecaption{Emission-line Measurements and Flux Ratios}
  \tablehead{
    \colhead{ID}&
    \colhead{\OII}&
    \colhead{\Hb}&
    \colhead{\OIII\,$\lambda$4959}&
    \colhead{\OIII\,$\lambda$5007}&
    \colhead{\OIII\,$\lambda$4363}&
    \colhead{S/N(4363)}&
    \colhead{\Pagel}&
    \colhead{\Oratio}\\
    \colhead{(1)}&\colhead{(2)}&\colhead{(3)}& \colhead{(4)}&
    \colhead{(5)}&\colhead{(6)}&\colhead{(7)}&\colhead{(8)}&\colhead{(9)}}
  \startdata
  MMT\#01    &  11.05$\pm$0.59 &   9.55$\pm$0.78\pa &  19.12$\pm$1.04 &  54.30$\pm$0.70\pa &   2.20$\pm$0.50 &  4.44 & 8.85$^{+1.18}_{-1.12}$ & 6.64$^{+0.44}_{-0.41}$\\[1mm]
  MMT\#02    &  18.55$\pm$0.35 &   5.93$\pm$0.27\pa &   6.73$\pm$0.27 &  19.41$\pm$0.28\pa &   0.64$\pm$0.21 &  3.06 & 7.54$^{+0.53}_{-0.41}$ & 1.41$^{+0.05}_{-0.05}$\\[1mm]
  MMT\#03\TA &  20.94$\pm$0.46 &   5.79$\pm$0.27\pa &   3.13$\pm$0.39 &   9.51$\pm$0.31\pa &   0.84$\pm$0.21 &  3.95 & 5.80$^{+0.26}_{-0.34}$ & 0.60$^{+0.03}_{-0.03}$\\[1mm]
  MMT\#04    &  50.23$\pm$0.28 &  22.17$\pm$0.17\pa &  32.13$\pm$0.26 &  99.65$\pm$0.24\pa &   1.48$\pm$0.22 &  6.86 & 8.21$^{+0.05}_{-0.06}$ & 2.62$^{+0.02}_{-0.02}$\\[1mm]
  MMT\#05    &   3.84$\pm$0.36 &   7.09$\pm$0.25\pa &  15.15$\pm$0.24 &  43.97$\pm$0.26\pa &   1.30$\pm$0.25 &  5.22 & 8.88$^{+0.40}_{-0.32}$ & 15.38$^{+1.39}_{-1.68}$\\[1mm]
  MMT\#06    &  21.76$\pm$0.34 &   6.31$\pm$0.34\pa &   5.66$\pm$0.29 &  16.34$\pm$0.26\pa &   0.84$\pm$0.27 &  3.10 & 6.94$^{+0.41}_{-0.41}$ & 1.01$^{+0.03}_{-0.02}$\\[1mm]
  MMT\#07    &  20.15$\pm$0.28 &  21.06$\pm$0.20\pa &  35.97$\pm$0.24 & 120.77$\pm$0.24\pa &   2.28$\pm$0.26 &  8.91 & 8.40$^{+0.10}_{-0.10}$ & 7.78$^{+0.11}_{-0.08}$\\[1mm]
  MMT\#08    &  28.23$\pm$0.71 &  11.21$\pm$1.27\pa &  10.47$\pm$1.04 &  32.57$\pm$0.89\pa &   1.85$\pm$0.56 &  3.32 & 6.36$^{+0.58}_{-0.64}$ & 1.52$^{+0.08}_{-0.09}$\\[1mm]
  MMT\#09    &  16.88$\pm$0.82 &   6.81$\pm$0.62\pa &  10.37$\pm$1.12 &  28.96$\pm$1.17\pa &   1.45$\pm$0.41 &  3.53 & 8.26$^{+0.31}_{-0.71}$ & 2.33$^{+0.05}_{-0.10}$\\[1mm]
  MMT\#10    &  22.70$\pm$1.06 &  18.55$\pm$0.42\pa &  30.93$\pm$0.45 &  92.80$\pm$0.52\pa &   2.22$\pm$0.62 &  3.58 & 7.89$^{+0.14}_{-0.21}$ & 5.45$^{+0.21}_{-0.25}$\\[1mm]
  MMT\#11    &  43.34$\pm$0.75 &  22.30$\pm$0.57\pa &  35.26$\pm$0.60 & 112.99$\pm$0.63\pa &   2.08$\pm$0.45 &  4.67 & 8.59$^{+0.20}_{-0.24}$ & 3.42$^{+0.06}_{-0.07}$\\[1mm]
  MMT\#12    &  24.18$\pm$0.68 &   6.30$\pm$0.93\pa &  11.29$\pm$0.77 &  29.59$\pm$0.73\pa &   1.46$\pm$0.48 &  3.04 & 10.32$^{+1.84}_{-1.82}$ & 1.69$^{+0.07}_{-0.07}$\\[1mm]
  MMT\#13    &   6.33$\pm$0.54 &   6.55$\pm$0.26\pa &  10.94$\pm$0.39 &  28.50$\pm$0.45\pa &   1.07$\pm$0.31 &  3.46 & 6.99$^{+0.33}_{-0.34}$ & 6.23$^{+0.66}_{-0.63}$\\[1mm]
  MMT\#14    & 102.13$\pm$0.54 &  57.20$\pm$0.34\pa &  87.06$\pm$0.61 & 269.12$\pm$0.67\pa &   2.14$\pm$0.28 &  7.55 & 8.01$^{+0.07}_{-0.06}$ & 3.49$^{+0.02}_{-0.02}$\\[1mm]
  Keck\#1    &   4.61$\pm$0.14 &   7.72$\pm$0.09\pa &  15.35$\pm$0.09 &  45.52$\pm$0.08\pa &   0.74$\pm$0.07 & 11.13 & 8.49$^{+0.10}_{-0.09}$ & 13.21$^{+0.46}_{-0.45}$\\[1mm]
  Keck\#2    &   3.55$\pm$0.15 &   2.54$\pm$0.05\TB &   4.31$\pm$0.05 &  13.40$\pm$0.05\pa &   0.64$\pm$0.10 &  6.22 & 8.36$^{+0.21}_{-0.21}$ & 5.00$^{+0.30}_{-0.29}$\\[1mm]
  Keck\#3    &  27.89$\pm$0.16 &  12.60$\pm$0.11\pa &  14.78$\pm$0.07 &  44.94$\pm$0.08\pa &   0.30$\pm$0.06 &  4.64 & 6.95$^{+0.05}_{-0.06}$ & 2.14$^{+0.02}_{-0.02}$\\[1mm]
  Keck\#4    &  13.86$\pm$0.14 &   7.37$\pm$0.11\pa &   8.66$\pm$0.14 &  26.85$\pm$0.14\TC &   0.34$\pm$0.10 &  3.44 & 6.70$^{+0.09}_{-0.13}$ & 2.56$^{+0.04}_{-0.04}$\\[1mm]
  Keck\#5    &   1.86$\pm$0.07 &   1.07$\pm$0.04\pa &   1.63$\pm$0.04 &   6.43$\pm$0.05\pa &   0.55$\pm$0.16 &  3.46 & 9.30$^{+0.48}_{-0.40}$ & 4.34$^{+0.18}_{-0.18}$\\[1mm]
  Keck\#6    &$<$0.07          &   0.92$\pm$0.03\pa &   1.44$\pm$0.03 &   4.00$\pm$0.03\pa &   0.21$\pm$0.05 &  4.27 & 5.96$^{+0.20}_{-0.16}$ & $>$80.70\\[1mm]
  \enddata
  \vspace{-3mm}
  \enddata
  \label{tab:em_lines}
  \tablecomments{All emission-line fluxes are reported in units of 10$^{-17}$ erg s$^{-1}$ cm$^{-2}$ with 68\%
    confidence uncertainties. No dust attenuation corrections have been applied to these fluxes or flux ratios.}
  \tablenotetext{1}{This source is likely a LINER (see Section \ref{sec:AGN}).}
  \tablenotetext{2}{The \Hb\ line fell inside a CCD gap. The reported flux here is
    derived from \Hg\ (see Section~\ref{sec:dust}).}
  \tablenotetext{3}{\OIII\,$\lambda$5007 was outside the spectral coverage; however,
    \OIII\,$\lambda$4959 was detected at $\sim$60$\sigma$. We therefore adopt a flux
    that is 3.1 times the 4959\AA\ flux.}
\end{deluxetable*}

\subsection{Contamination from LINERs and AGN}\label{sec:AGN}
Because \OIIIa\ is more likely to be detected in higher temperature gas,
a common concern is whether these galaxies harbor low-ionization nuclear
emitting regions \citep[LINERs;][]{heckman80}, where the gas may be
shock-heated.
To determine if any of our galaxies are LINERs, the preferred method
is to use the [\ion{O}{1}]\,$\lambda$6300 emission line. Unfortunately,
this is redshifted out of our spectral coverage for the majority of
our galaxies.
Instead, we use a variety of emission-line flux ratios, including
\OII/\OIII, \OIII/\Hb, and \OII/\NeIII\,$\lambda$3869,
to determine if any of our galaxies could be a LINER. Four galaxies
(MMT\#02, \#03, \#06, and \#12) have \OII/\OIII\ ratios that are
similar to or above unity (0.82--2.2), which is a cautionary
LINER flag.
Upon comparing our emission-line fluxes to SDSS DR7 LINERs, we find
that only MMT\#03 is arguably a LINER. The three remaining galaxies
have too low (high) of an \OII/\OIII\ (\OIII/\Hb) ratio by at least
0.4 dex.
Further independent evidence supporting the idea that MMT\#03 is a
LINER is its stellar mass. As we will later show (Section~\ref{sec:SED}),
it is our most massive galaxy. In general, LINERs are primarily found
in more massive galaxies (e.g., 94\% of SDSS DR7 LINERs are above a
stellar mass of $10^{10}$ \Msun).

\begin{figure}
  \epsscale{1.1}
  \plotone{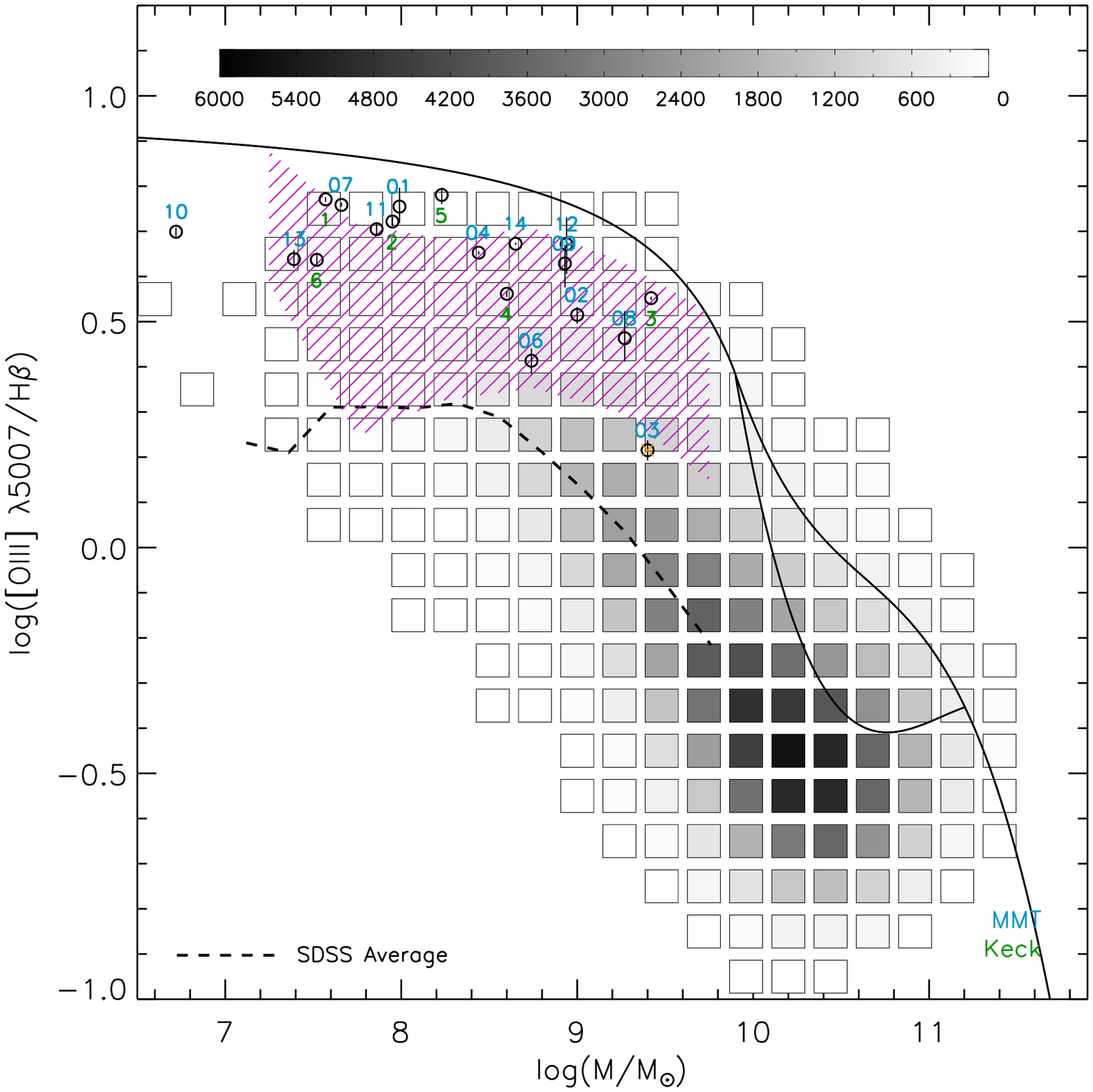}
  \vspace{-2mm}
  \caption{The MEx diagram for our \OIIIA\ sample (circles)
    plotted against the SDSS DR7 emission-line sample (grey
    boxes). Each of our galaxies is labelled by a
    two-digit (MMT; blue) or one-digit (Keck; green) number.
    The stellar masses are determined from SED fitting,
    discussed in Section~\ref{sec:SED}. The solid black lines
    defined by \cite{juneau11} separate star-forming (left),
    AGN (right), and composite (middle) galaxies. All of our
    galaxies are classified as star-forming based on the MEx
    classification.
    Our ELGs have \OIII/\Hb\ fluxes that are 0.2--0.5 dex
    higher than the average for SDSS galaxies with the
    same stellar mass. We also show the SDF spectroscopic
    sample of $\sim$200 $z=0.4$--1 galaxies in purple
    with the shaded region representing $\pm$1$\sigma$. This
    comparison indicates that ELGs at these redshifts
    have higher ionization in general, across 2.5 dex in
    stellar mass.
    Our only LINER (MMT\#03) is denoted by the orange cross.}
 \label{fig:MEx}
\end{figure}

Another possibility to consider is that some of our galaxies might
harbor an AGN. Supporting evidence would include very high ionization
lines (e.g., [\ion{Ne}{5}]\,$\lambda$3425, \ion{He}{2}\,$\lambda$4686),
although these lines have also been seen in some local blue compact
dwarf galaxies \citep{izotov12b}.
A search for [\ion{Ne}{5}] and \ion{He}{2} yielded non-detections,
arguing that our sample is free of Seyfert galaxies.
In addition, our two galaxies (MMT\#10 and \#11) with coverage of
\NII\,$\lambda$6583 have \NII/\Ha\ ratios of 0.02 and 0.04, which
places them in the star-forming region of the ``BPT'' diagram
\citep{baldwin81}. While \NII\ is not available for the remaining
18 galaxies, we are able to use an {\it analogous} diagnostic tool
called the ``Mass-Extinction'' diagram \citep[``MEx'';][]{juneau11}.
We illustrate in Figure~\ref{fig:MEx} our sample against star-forming
galaxies selected using the BPT diagram\footnote{We require at
  least 3$\sigma$ detections of \Ha, \NII, \OIII, and \Hb, which
  yielded 274,613 galaxies. The sample was further limited to
  203,630 galaxies by excluding AGN and LINERs with the
  \cite{kauffmann03} selection.} from the SDSS MPA-JHU DR7
sample\footnote{\url{http://www.mpa-garching.mpg.de/SDSS/DR7/}.}.
We find that all of our galaxies lie in the star-forming domain.
Compared to local galaxies of similar stellar masses, our ELGs all
show a higher \OIII/\Hb\ flux ratio by $\approx0.2$--0.5 dex.
We find that for more typical ELGs (i.e., those that lack
\OIIIa\ detections) in our spectroscopic sample, the
\OIII/\Hb\ ratios are similarly higher than in SDSS. This holds
for more massive galaxies with \Mstar\ $\sim10^{10}$ \Msun. We
overlay the range of the full sample as a purple shaded region in
Figure~\ref{fig:MEx}.
We later discuss this result for higher ionization in
Section~\ref{sec:compare}, and discuss selection effects
associated with our sample in Section~\ref{sec:bias}.


\section{RESULTS}\label{sec:Results}
In this section, we utilize our spectroscopic and photometric data
to estimate dust attenuation (\S\ref{sec:dust}), electron
temperature and gas-phase metallicity (\S\ref{sec:metals}),
de-reddened SFRs (\S\ref{sec:SFR}), stellar properties from
SED modeling (\S\ref{sec:SED}), the local environment 
(\S\ref{sec:companions}), and the SFR surface density
(\S\ref{sec:SFR_density}).

\subsection{Dust Attenuation Correction from Balmer Decrements}\label{sec:dust}
To correct the emission-line fluxes for dust attenuation, we use
Balmer decrement measurements obtained from a combination of our
spectroscopy and NB imaging. Since our ELGs possess high EWs, 12,
18, and 20 galaxies have \Hd, \Hg, and \Hb\ detected at
$\gtrsim$5$\sigma$, respectively.
In addition, \Ha\ measurements are available for 5 galaxies.
Two of them (MMT\#10 and MMT\#11) are at lower redshifts,
allowing us to use their spectroscopic \Ha\ measurements.
For the other 3 galaxies (MMT\#04, \#06, \#11), the NB \Ha\ flux
is determined using the following equation:
\begin{equation}
  F_{\Hae} = \Delta{\rm NB} \frac{f_{\lambda,{\rm NB}} - f_{\lambda,{\rm BB}}}{1-(\Delta{\rm NB}/\Delta{\rm BB})} f_{\rm corr},
  \label{eqn:Ha_flux}
\end{equation}
where $f_{\lambda,{\rm NB}}$ and $f_{\lambda,{\rm BB}}$ are the flux density
in erg s$^{-1}$ cm$^{-2}$ \AA$^{-1}$ for the narrow-band and broad-band,
$\Delta$'s are the respective FWHM of the filters
($\Delta z$\arcmin\ = 956\AA), and $f_{\rm corr}$ is the correction for
the non-tophat shape of the NB filter when the redshifted \Ha\ emission
is in the filter's wing. Using filter throughputs and spectroscopic
redshifts, we are able to determine $f_{\rm corr}$.
We note that for four galaxies (MMT\#03, \#07, \#13, and \#14),
the NB fluxes are not reliable for Balmer decrement determinations.
In particular, MMT\#03 suffers from significant \NII\ contamination
in the NB921 filter as a LINER candidate. For the other three sources,
the \Ha\ line falls in the wing of the NB921 or NB973 filter where
precise filter response corrections cannot be made.

A significant problem encountered with using Balmer decrements
to determine dust attenuation is the underlying stellar absorption. 
In three galaxies (MMT\#03, MMT\#04, and Keck\#3), the spectra are
of high S/N to sufficiently detect the continuum and fit and remove
the stellar absorption or use \textsc{iraf}'s splot command to
re-measure the continuum from the absorption trough.
For MMT\#03 (Keck\#3), we determine corrections of
EW$_{\rm abs}$(\Hd) = 2.7\AA\ (1.6\AA), EW$_{\rm abs}$(\Hg) = 1.9\AA\ (2.5\AA),
and EW$_{\rm abs}$(\Hb) = 1.6\AA\ (0.0\AA).
While for MMT\#04, we measure EW$_{\rm abs}$(\Hd) = 3.9\AA\ and
EW$_{\rm abs}$(\Hg) = 2.5\AA.

For the remaining galaxies, the S/N of our spectra are not sufficient
to model the stellar absorption. To correct these galaxies, we adopt
EW$_{\rm abs}$(\Hd) = 2\AA\ and EW$_{\rm abs}$(\Hg) = 1\AA.
We assume no stellar absorption for \Hb\ and \Ha, which is reasonable
since the measured rest-frame emission-line EWs are significantly large
(\Hb: median of 60\AA, average of 95\AA).
%


\begin{figure*}
  \epsscale{0.5}
  \plotone{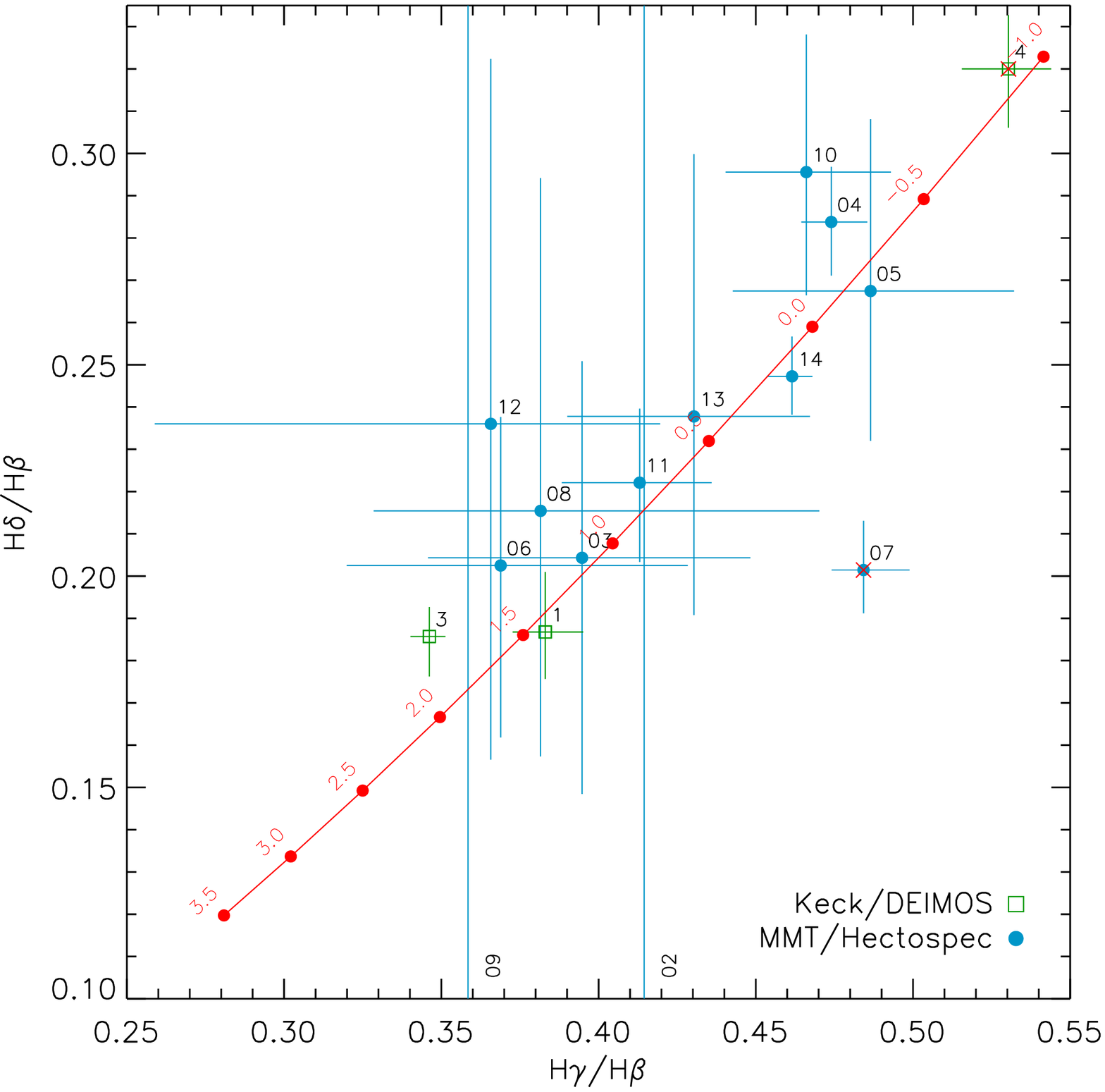}
  \plotone{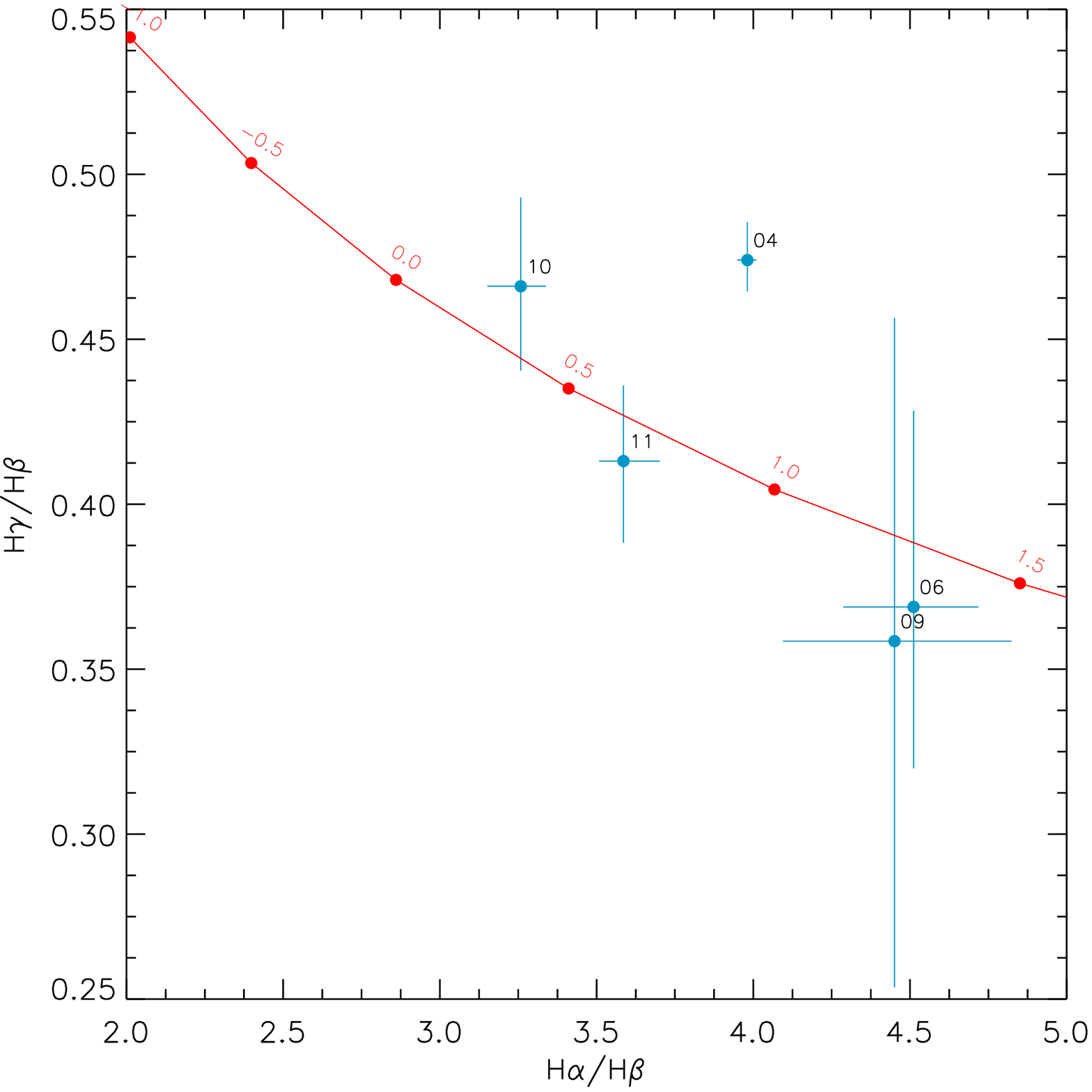}
  \caption{Balmer decrements for our \OIIIa\ galaxies. The MMT
    and Keck samples are shown as blue circles and green squares,
    respectively. Red crosses indicate that the \Hd\ fluxes are
    unreliable since it fell on an OH sky-line for MMT\#07 and
    Keck\#4.
    Vertical blue lines indicate non-detections in \Hd\ for MMT\#02
    and MMT\#09.
    Finally, red circles and the solid red curves show the effects
    on the Balmer decrements with increasing dust reddening following
    \citetalias{calzetti00}. The values reported in red are $A$(\Ha).}
  \label{fig:balmer}
\end{figure*}
With these corrections for stellar absorption, we illustrate the Balmer
decrements in Figure~\ref{fig:balmer}. In addition, we report in
Table~\ref{tab:balmer} the absorption-corrected Balmer decrements, and
the determined or assumed EW for stellar absorption.

Under the assumption that the hydrogen nebular emission originates
from an optically thick ionization-bounded \textsc{H ii} region
obeying Case B recombination, the intrinsic Balmer flux ratios
are: (\Ha/\Hb)$_0$ = \HaHbi, (\Hg/\Hb)$_0$ = 0.468, and
(\Hd/\Hb)$_0$ = 0.259 for \Te\ = 10$^4$ K and electron density of
$n_e$ = 100 cm$^{-3}$ (see Section~\ref{sec:metals} for further discussion).
We note that these values differ by less than 5\% for \Te\ = $2\times10^4$ K.
Dust absorption alters these observed ratios:
\begin{equation}
  \frac{(\Hyd n/\Hbe)_{\rm obs}}{(\Hyd n/\Hbe)_0} = 10^{-0.4\EBV[k(\Hyd n)-k(\Hbe)]},
  \label{eqn:balmer}
\end{equation}
where \EBVa\ is the {\it nebular} color excess, and 
$k(\lambda)\equiv A(\lambda)$/\EBVa\ is the reddening curve
at $\lambda$.
The latter is dependent on the dust reddening ``law.'' We illustrate
in Figure~\ref{fig:balmer} the observed Balmer decrements under
the \cite{calzetti00} (hereafter C00) dust reddening formalism with
$k(\Hae)=3.33$, $k(\Hbe)=4.60$, $k(\Hge)=5.12$, and $k(\Hde)=5.39$.
We find that our Balmer decrements are consistent with the
\citetalias{calzetti00} dust reddening formula. For the remainder
of our paper, all dust-corrected measurements adopt
\citetalias{calzetti00} reddening.

Our color excesses, which are tabulated in Table~\ref{tab:balmer},
are determined using either \Ha/\Hb\ (\#06, \#09--\#11)
or \Hg/\Hb\ (MMT\#02--\#05, \#08, \#12--\#14, and Keck\#1,
\#3, and \#4). We do not use \Hd/\Hb\ since it is not well
measured for the majority of our galaxies.
However, the \EBVa\ estimates from \Hd/\Hb\ are consistent with \Hg/\Hb,
as indicated in Figure~\ref{fig:balmer}.
In three galaxies (MMT\#04, MMT\#05, and Keck\#4), the \Hg/\Hb\ ratios
suggests negative reddening, so we adopt \EBVa\ = 0.0.
For four of our galaxies (MMT\#01, \#07, \#5--\#6), the dust reddening
could not be determined from Balmer decrements. For these galaxies, we
assume \EBVa\ = $0.24\pm0.21$ mag ($A(\Hae) \approx 0.8\pm0.7$
mag), which is the average of our sample\footnote{The median \EBVa\ is 0.25 mag.},
and is similar to typical reddening found for local galaxies
\citep[$A(\Hae)=0.8$--1.1;][]{kennicutt98a}.
This also agrees with what we found for ELGs at $z\approx0.4$ \citep{HaSFR}.

For Keck\#2, the \Hb\ line unfortunately fell at the edge of a CCD gap, so
the \Hb\ flux is not fully measured.
However, the \Hg\ and \Hd\ lines are robustly detected (19 and 9$\sigma$).
Assuming Case B recombination and no reddening, the \Hg/\Hb\ and \Hd/\Hb\
Balmer decrements yielded \Hb\ fluxes of $2.54\times10^{-17}$ and
$2.87\times10^{-17}$ erg s$^{-1}$ cm$^{-2}$, respectively. This excellent
agreement in predicted \Hb\ fluxes suggests that very little reddening
is present in this galaxy. We assume \EBVa\ = 0.0.

\newcommand{\pc}{\phm{$^c$}}

\begin{deluxetable*}{lccccccc}
  \tabletypesize{\scriptsize}
  \tablewidth{0pc}
  \tablecaption{Balmer Decrement Measurements and Derived Reddening}
  \tablehead{
    \colhead{ID}&
    \colhead{\Ha/\Hb}&
    \colhead{\Hg/\Hb}&
    \colhead{EW(\Hd)}&
    \colhead{EW(\Hg)}&
    \colhead{EW(\Hb)}&
    \colhead{Source}&
    \colhead{\EBVa}\\
    \colhead{(1)}&\colhead{(2)}&\colhead{(3)}&\colhead{(4)}&
    \colhead{(5)}&\colhead{(6)}&\colhead{(7)}&\colhead{(8)}}
  \startdata
  MMT\#01 &                 \ldots    &              \ldots\TA & 2.0 & 1.0 & 0.0 &  \ldots & 0.24$^{+0.21}_{-0.21}$\pc\\[1.0mm]
  MMT\#02 &                 \ldots    & 0.41$^{+0.03}_{-0.04}$ & 2.0 & 1.0 & 0.0 & \Hg/\Hb & 0.25$^{+0.20}_{-0.22}$\pc\\[1.0mm]
  MMT\#03 &              \ldots\TB    & 0.39$^{+0.04}_{-0.04}$ & 2.7 & 1.9 & 1.6 & \Hg/\Hb & 0.35$^{+0.26}_{-0.24}$\pc\\[1.0mm]
  MMT\#04 & 3.98$^{+0.04}_{-0.03}$\TC & 0.47$^{+0.01}_{-0.01}$ & 3.9 & 2.5 & 0.0 & \Hg/\Hb & 0.00$^{+0.05}_{-0.04}$\TD\\[1.0mm]
  MMT\#05 &                 \ldots    & 0.49$^{+0.05}_{-0.04}$ & 2.0 & 1.0 & 0.0 & \Hg/\Hb & 0.00$^{+0.21}_{-0.21}$\TD\\[1.0mm]
  MMT\#06 & 4.51$^{+0.16}_{-0.18}$    & 0.37$^{+0.06}_{-0.05}$ & 2.0 & 1.0 & 0.0 & \Ha/\Hb & 0.39$^{+0.04}_{-0.03}$\pc\\[1.0mm]
  MMT\#07 &                 \ldots\TC & 0.48$^{+0.02}_{-0.02}$ & 2.0 & 1.0 & 0.0 &  \ldots & 0.24$^{+0.21}_{-0.21}$\pc\\[1.0mm]
  MMT\#08 &                 \ldots    & 0.38$^{+0.05}_{-0.05}$ & 2.0 & 1.0 & 0.0 & \Hg/\Hb & 0.42$^{+0.43}_{-0.45}$\pc\\[1.0mm]
  MMT\#09 & 4.45$^{+0.45}_{-0.44}$    & 0.36$^{+0.10}_{-0.10}$ & 2.0 & 1.0 & 4.4 & \Ha/\Hb & 0.38$^{+0.09}_{-0.10}$\pc\\[1.0mm]
  MMT\#10 & 3.26$^{+0.08}_{-0.11}$    & 0.47$^{+0.04}_{-0.04}$ & 2.0 & 1.0 & 0.0 & \Ha/\Hb & 0.11$^{+0.02}_{-0.02}$\pc\\[1.0mm]
  MMT\#11 & 3.59$^{+0.12}_{-0.12}$    & 0.41$^{+0.03}_{-0.02}$ & 2.0 & 1.0 & 0.0 & \Ha/\Hb & 0.19$^{+0.03}_{-0.03}$\pc\\[1.0mm]
  MMT\#12 &                 \ldots    & 0.37$^{+0.07}_{-0.11}$ & 2.0 & 1.0 & 0.0 & \Hg/\Hb & 0.51$^{+0.72}_{-0.46}$\pc\\[1.0mm]
  MMT\#13 &                 \ldots\TC & 0.43$^{+0.05}_{-0.05}$ & 2.0 & 1.0 & 0.0 & \Hg/\Hb & 0.17$^{+0.26}_{-0.24}$\pc\\[1.0mm]
  MMT\#14 &              \ldots\TD    & 0.46$^{+0.01}_{-0.01}$ & 2.0 & 1.0 & 0.0 & \Hg/\Hb & 0.03$^{+0.03}_{-0.03}$\pc\\[1.0mm]
  Keck\#1 &                 \ldots    & 0.38$^{+0.01}_{-0.01}$ & 2.0 & 1.0 & 0.0 & \Hg/\Hb & 0.41$^{+0.07}_{-0.07}$\pc\\[1.0mm]
  Keck\#2 &                 \ldots    &              \ldots\TA & 2.0 & 1.0 & 0.0 &  \ldots & 0.00\TE\\[1.0mm]
  Keck\#3 &                 \ldots    & 0.35$^{+0.01}_{-0.01}$ & 1.6 & 2.6 & 0.0 & \Hg/\Hb & 0.62$^{+0.04}_{-0.04}$\pc\\[1.0mm]
  Keck\#4 &                 \ldots    & 0.53$^{+0.01}_{-0.01}$ & 2.0 & 1.0 & 0.0 & \Hg/\Hb & 0.00$^{+0.07}_{-0.07}$\TD\\[1.0mm]
  Keck\#5 &                 \ldots    &              \ldots\TA & 2.0 & 1.0 & 0.0 &  \ldots & 0.24$^{+0.21}_{-0.21}$\pc\\[1.0mm]
  Keck\#6 &                 \ldots    &              \ldots\TF & 2.0 & 1.0 & 0.0 &  \ldots & 0.24$^{+0.21}_{-0.21}$\pc\\[1.0mm]
  \vspace{-3mm}
  \enddata
  \label{tab:balmer}
  \tablecomments{Uncertainties are reported at the 68\% CL. Stellar absorption line corrections are reported in Cols. (4)--(6)}.
  \tablenotetext{1}{\Hb\ is affected by a weak OH sky-line.}
  \tablenotetext{2}{The NB921 flux is contaminated by significant \NII\ emission because this source is a LINER.}
  \tablenotetext{3}{The \Ha\ emission falls at the very edge of NB921 or NB973 filter, and therefore extinction
    determinations are unreliable using \Ha/\Hb.}
  \tablenotetext{4}{Negative reddening was seen. We adopt \EBVa\ = 0.0.}
  \tablenotetext{5}{The \Hg\ and \Hd\ fluxes suggests no reddening}
  \tablenotetext{6}{\Hg\ is affected by an OH sky-line.}
\end{deluxetable*}

\subsection{\Te\ and Metallicity Determinations}\label{sec:metals}
To determine the gas-phase metallicity for our galaxies, we use the
empirical relations of \cite{izotov06a}.
This follows the approach of most direct metallicity studies.
The first equation estimates \Te(\OIII) using the nebular-to-auroral
\OIII\ flux ratio:
\begin{equation}
  \log{\left(\frac{\OIII\,\lambda\lambda4959,5007}{\OIII\,\lambda4363}\right)}
    = \frac{1.432}{t_3}+\log{C_T},
\end{equation}
where $t_3$ = \Te(\OIII)/$10^4$ K,
\begin{equation}
  C_T = (8.44-1.09t_3+0.5t_3^2-0.08t_3^3)\frac{1+0.0004x}{1+0.044x},
\end{equation}
and $x = 10^{-4}n_e t_3^{-0.5}$. For the majority of our galaxies, the
\SII\,$\lambda\lambda$6717,6732 doublet, an estimate for $n_e$, is
redshifted out of our spectral coverage.
For one of our galaxies, MMT\#10, both \SII\ lines are weakly detected.
The $\lambda$6717/$\lambda$6732 flux ratio of $\approx$1.1 corresponds to
$n_e\approx400$ cm$^{-3}$ (assuming \Te\ = $10^4$ K). In addition, DEIMOS
has the spectral resolution to separate \OII\,$\lambda\lambda$3726,3729.
The $\lambda$3729/$\lambda$3726 flux ratios vary between 0.87 and 1.35,
which correspond to $n_e=70$--600 cm$^{-3}$.
To determine \Te, we assume $n_e$ = 100 cm$^{-3}$.
We note that $C_T$ is only affected by $n_e$ in the high density regime
($n_e\gtrsim10^{4}$ cm$^{-3}$), and therefore assuming $n_e$ = 10, 100,
or 10$^3$ cm$^{-3}$ yields nearly identical \Te.

We correct the nebular-to-auroral \OIII\ flux ratio for dust
attenuation using our dust attenuation prescriptions
(see Section~\ref{sec:dust}). Correcting for dust attenuation
increases estimates of \Te, since dust extincts shorter
wavelengths (e.g., 4363\AA) more than longer
wavelengths (e.g., 5007\AA).

In addition, it has been known that \Te\ determinations using
(1) various direct metallicity prescriptions, including \cite{izotov06a};
and (2) strong-line metallicity diagnostics do not yield consistent
results \citep{kewley08}.
These discrepancies have been recently explored, and it is believed
that outdated effective collision strengths for the various O$^{++}$
excitation states, and a non-equilibrium distribution of the electron
energies result in overestimation of the electron temperature
\citep{nicholls12,nicholls13}. Those authors provided applicable
corrections for each method. We adopt their corrections,
which correspond roughly to a 5\% reduction for \Te\ estimates
using the \cite{izotov06a} approach.

For the very strong ELGs we are studying, our \OIII\ measurements
have a very large dynamic range. The strongest \OIIIa\ line is
as much as 0.089 of the \OIII\ flux (MMT\#03),
while the weakest is 0.007 (Keck\#3). We find that the average
and median $\lambda$4363/$\lambda$5007 flux ratio for our
sample are both 0.037.
The derived electron temperatures for our galaxy sample, also
have a wide range, from 10$^4$ K to $4.5\times10^4$ K.
Among our \Ndetf\ galaxies, 11 have \Te\ estimates of
10$^4$--$2.1\times10^4$ K, which are similar to those measured
in local galaxies\footnote{90\% of \cite{nagao06}'s sample
  has \Te\ = 9,400--20,000 K.}. The large tail toward high \Te\
yields an average $\log(T_e/{\rm K})$ that is higher than the
average of local galaxies by 0.15 dex.

Other higher redshift studies have measured high \Te\ in
star-forming galaxies. For example, \cite{yuan09} detected
\OIIIa\ in a lensed $z=1.7$ galaxy and measured \Te\
$\approx2.35\times10^4$ K, and \cite{hu09} have also measured
large \OIII\,$\lambda$4363/$\lambda$5007 flux ratios in a few ELGs.
The higher values of \Te\ are not well-determined, and have a
large tail (0.2--0.3 dex) toward lower temperatures.
This is likely the case for other published measurements,
as \cite{yuan09} yielded a 3$\sigma$ detection of \OIIIa\ and
the measurement uncertainties of \cite{hu09} result
in a significant tail toward low \Te.

MMT\#03 has the most extreme \Te\ value, but as we mentioned in
Section~\ref{sec:AGN}, it is a LINER with shock-heated, rather
than photo-ionized gas producing most of its emission-line
spectrum.
Other galaxies with high \Te\ ($\gtrsim2.5\times10^4$ K)
are MMT\#06, \#08, \#09, \#12, Keck\#5, and \#6. To
examine if these temperatures are reasonable, we generate
low-metallicity models with CLOUDY \citep{ferland98}. In
our modeling, we input Starburst99 \citep{leitherer99} SEDs
that adopt Geneva stellar evolution models with five stellar
metallicities between $Z$/\zsun\ = 0.05 and 2.0.
Coupling the stellar and gas metallicity, we find that the
highest plausible \Te(\OIII) temperature is $\approx$25,000 K
for $Z$/\zsun\ = 0.02--0.03.\footnote{An extrapolation is made from
  $Z$/\zsun\ = 0.05.}
Given this upper limit, we fix the dust-corrected \Te(\OIII) to
not exceed this temperature for these six galaxies. We note
that without this \Te\ offset, the resulting oxygen metallicity
will be lower by 0.1--0.4 dex for these six galaxies.

\begin{figure*}
  \vspace{-7.5mm}
  \plotone{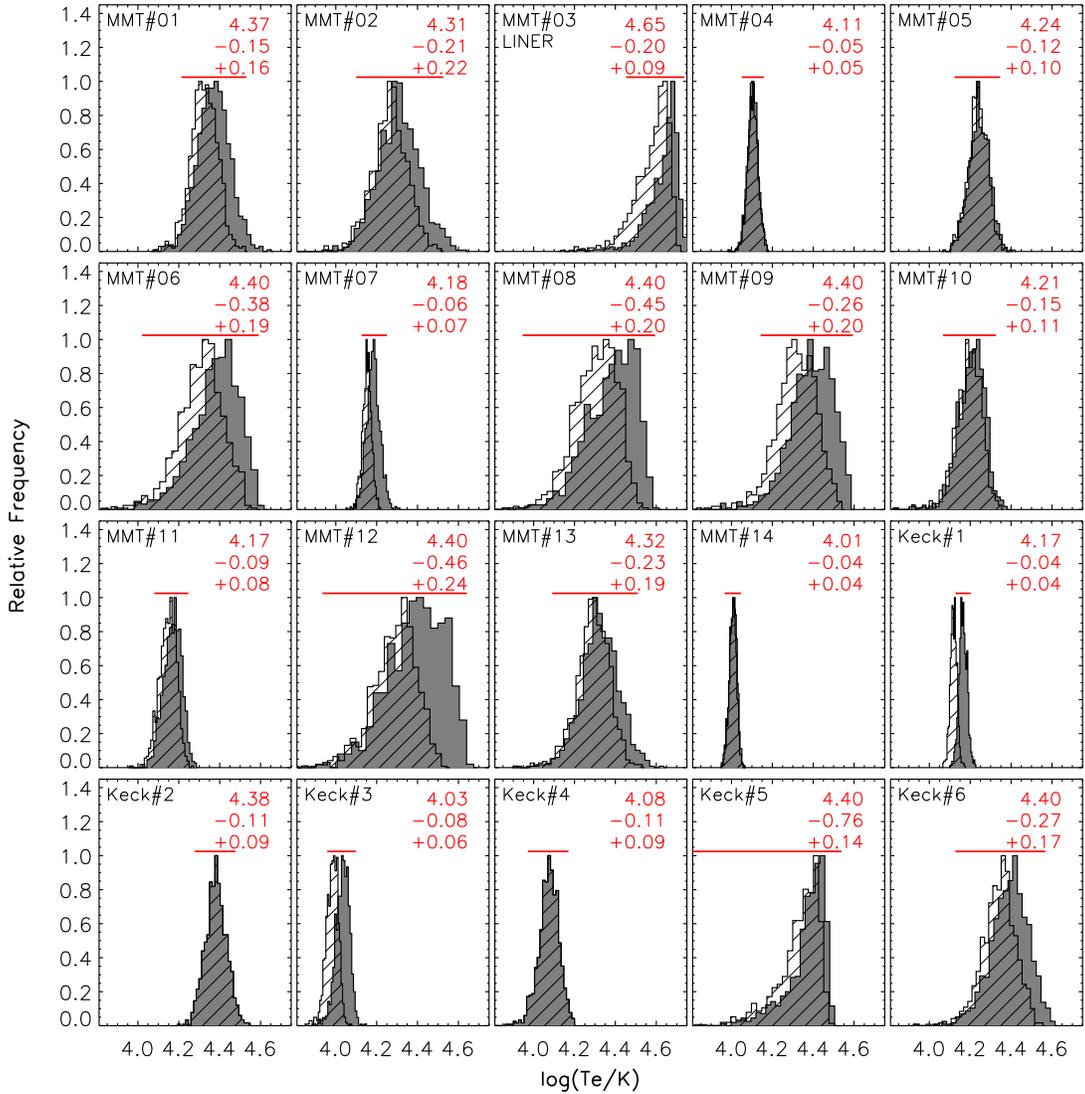}
  \vspace{-3mm}
  \caption{Probability distribution functions for the electron
    temperature determined from the \OIIIa\ line.
    The peak values and $\pm$95\% confidences are reported in
    each panel. \Te's are derived from the observed flux ratios
    (hashed histograms) and the dust-corrected flux ratios
    (grey histograms).}
  \label{fig:Te}
\end{figure*}

The \Te\ measurements for all \Ndet\ objects are shown in
Figure~\ref{fig:Te}. Throughout this paper, we generate
realizations of the emission-line fluxes to construct
probability distribution functions (PDFs) for all observed and
derived measurements. This is critical, as the distributions
are non-Gaussian in the domain of low S/N ($\lesssim$5), and
it allows us to propagate our measurement uncertainties,
including \EBVa.


To determine the ionic abundances of oxygen, we use two
emission-line flux ratios, \OII\,$\lambda\lambda$3726,3729/\Hb\
and \OIII\,$\lambda\lambda$4959,5007/\Hb:
\begin{eqnarray}
  12+\log{\left(\frac{{\rm O}^+}{{\rm H}^+}\right)} = \log{\left(\frac{\OII}{{\rm H}\beta}\right)}
  + 5.961 + \frac{1.676}{t_2}\\
  \nonumber
  - 0.4\log{t_2} - 0.034t_2+\log{(1+1.35x)}\\
  12+\log{\left(\frac{{\rm O}^{++}}{{\rm H}^+}\right)} = \log{\left(\frac{\OIII}{{\rm H}\beta}\right)}
  + 6.200+ \frac{1.251}{t_3}\\
  - 0.55\log{t_3} - 0.014t_3.
  \nonumber
\end{eqnarray}
Here, $t_2$ refers to the singly-ionized oxygen electron
temperature, \Te(\OII). For our metallicity estimation, we
adopt a standard two-zone temperature model with $t_2 \equiv $
\Te(\OII)/10$^4$ K = $-0.577 + t_3(2.065-0.498t_3)$ 
\citep{izotov06a}.
In computing O$^+$/H$^+$, we also correct the \OII/\Hb\ ratio
for dust attenuation. We do not correct O$^{++}$/H$^+$
since the effects are negligible (e.g., the \OIII/\Hb\ ratio
changes by less than 0.02 dex for $A$(\Ha) = 1 mag).

Since the most abundant ions of oxygen in \textsc{H ii} regions
are O$^+$ and O$^{++}$, we can determine the oxygen abundances as:
\begin{equation}
  \frac{\rm O}{\rm H} = \left(\frac{\rm O^+}{\rm H^+}\right) + \left(\frac{\rm O^{++}}{\rm H^+}\right).
\end{equation}

\begin{figure*}
  \vspace{-7.5mm}
  \plotone{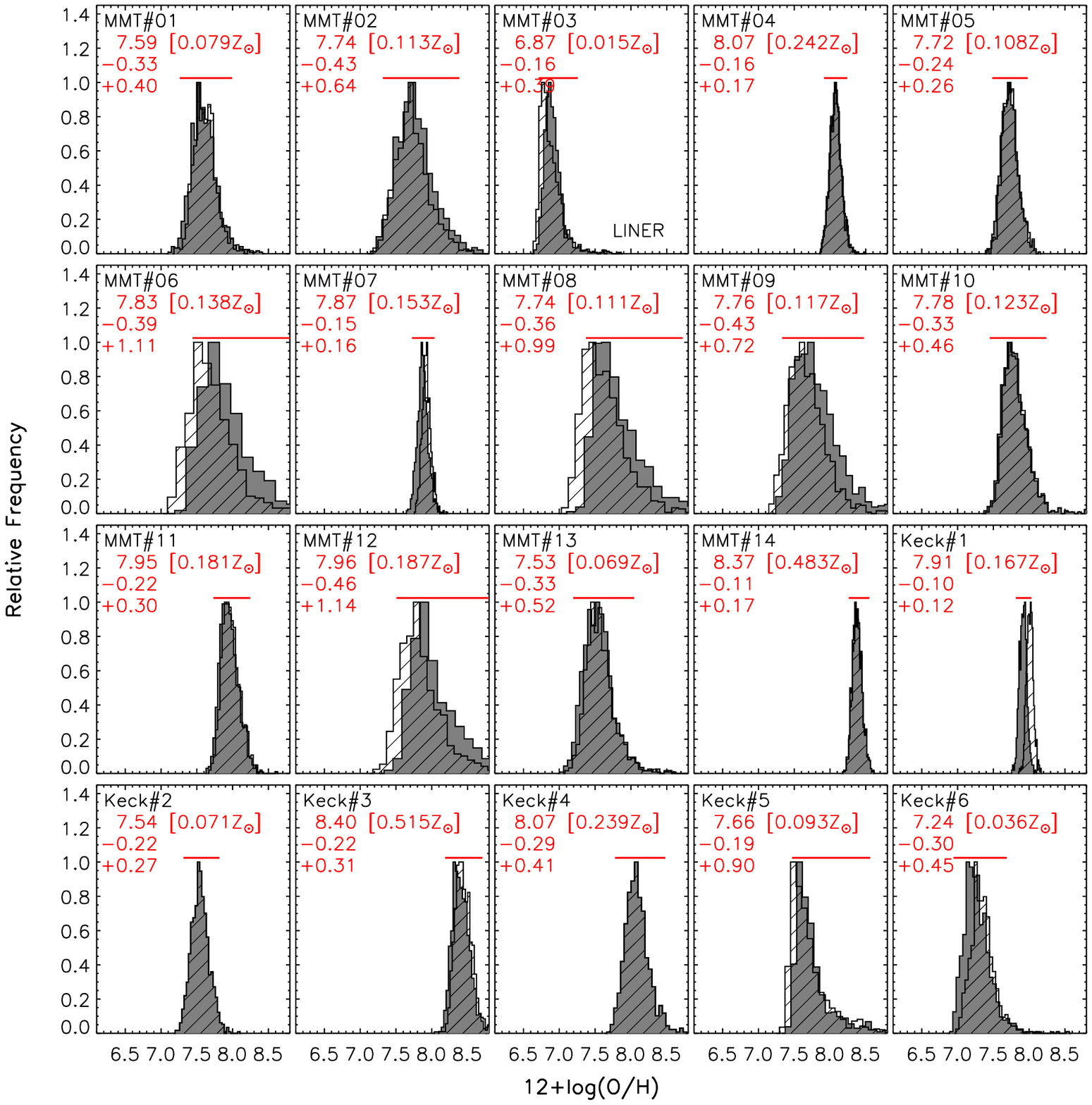}
  \vspace{-3mm}
  \caption{Probability distribution functions for the gas-phase metallicity,
    \OH. The peak values and $\pm$95\% confidences are reported in each
    panel. We illustrate the observed and de-reddened determinations by
    the hashed and grey filled histograms, respectively. The upper
    limit of the \OII\ flux is adopted for calculating the metallicity
    for Keck\#6.}
  \label{fig:logOH}
\end{figure*}

In Table~\ref{tab:metals}, we provide observed and de-reddened flux ratios,
estimates of \Te(\OIII), $\log({\rm O^+/H^+})$, $\log({\rm O^{++}/H^+})$,
and \OH\ for our sample, and 95\% confidence uncertainties (i.e., two 
standard deviations). In addition,
we illustrate our PDFs of \OH\ in Figure~\ref{fig:logOH}.
Our three most metal-poor galaxies are MMT\#13, Keck\#2 and
\#6 with \OH\ = \OHMMT, \OHKecktwo, and \OHKecksix, respectively.
For our sample of \Ndetf\ star-forming galaxies, \NXMPG\ of them
can be classified as an XMPG as their metallicity is below
\OH\ = 7.65.


\subsection{Dust-Corrected Star Formation Rates}\label{sec:SFR}

In addition to gas-phase metallicity determinations, our
extensive data allow us to determine dust-corrected SFRs
using the Hydrogen recombination lines, which are sensitive
to the shortest timescale of star formation, $\lesssim$10 Myr
(i.e., an instantaneous SFR estimate).

Assuming a \cite{chabrier03} (hereafter Chabrier) initial mass
function (IMF) with minimum and maximum masses of 0.1 and
100 \Msun, and solar metallicity
\citep{kennicutt98a}\footnote{We assume a factor of 1.8 between
the integrated masses for the \citetalias{chabrier03} and
\cite{salpeter} (hereafter Salpeter) IMFs.}, the SFR
can be determined from the dust-corrected \Ha\ luminosity as:
\begin{equation}
  {\rm SFR}(M_{\sun}~{\rm yr}^{-1}) = 4.4\times10^{-42} L({\rm erg~s}^{-1}).
\end{equation}

\noindent Similarly, SFRs can be determined from \Hb\ by first
correcting for dust reddening ($A$(\Hb) = 4.60 \EBVa), and then
adopting the intrinsic Case B flux ratio, (\Ha/\Hb)$_0$ = \HaHbi.

\begin{figure*}
  \vspace{-3mm}
  \plotone{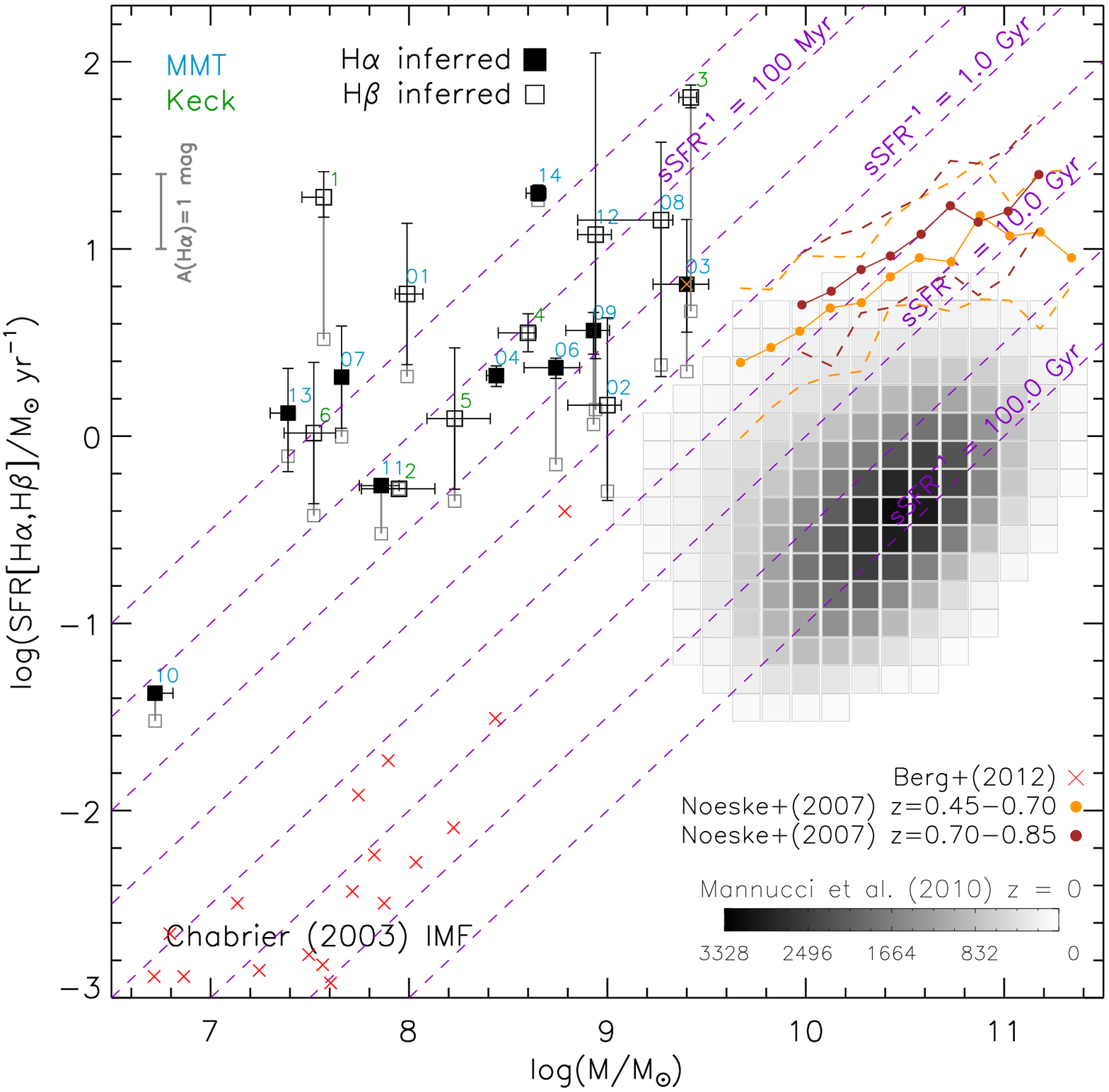}
  \vspace{-3mm}
  \caption{The SFR versus stellar mass for our \OIIIA\ sample. The stellar
    masses are obtained from SED fitting (Section~\ref{sec:SED}) where
    we adopt a \citetalias{chabrier03} IMF.
    The SFRs are determined from either \Ha\ (filled squares)
    or \Hb\ (open squares), which are sensitive to a timescale
    of $\lesssim$10 Myr. The observed SFRs are shown as grey open
    squares with the degree of dust attenuation corrections
    indicated by the solid grey vertical lines.  These dust
    corrections are discussed further in Section~\ref{sec:dust}.
    The average (median) sSFR of our galaxies is \sSFRA\ yr$^{-1}$
    ($8\times10^{-9}$ yr$^{-1}$).
    Each of our galaxies are labelled by a two-digit (MMT; blue)
    or one-digit (Keck; green) number.
    Our only LINER (MMT\#03) is denoted by the orange cross.
    Overlaid in red crosses are the low-luminosity local sample
    of \cite{berg12}, and the greyscale demonstrates the SDSS
    sample used by \cite{mannucci10}. The ``main sequence''
    relation of \cite{noeske07} at $z=0.45$--0.7 and $z=0.7$--0.85
    are illustrated in orange and brown, respectively.
    For direct comparisons, we have corrected \cite{noeske07} and
    \cite{berg12} measurements to a \citetalias{chabrier03} IMF.}
  \label{fig:SFR_Mass}
\end{figure*}

Our SFR estimates are summarized in Table~\ref{tab:SFR_Mass} and
illustrated in Figure~\ref{fig:SFR_Mass}. We find that our galaxies
have dust-corrected SFRs of 0.04--64 \Msun\ \iyr\ with an average
(median) of \SFRA\ (\SFRM) \Msun\ \iyr.


\subsection{Spectral Energy Distributions and Estimated Stellar
  Population Properties}\label{sec:SED}


\begin{figure*}
  \plotone{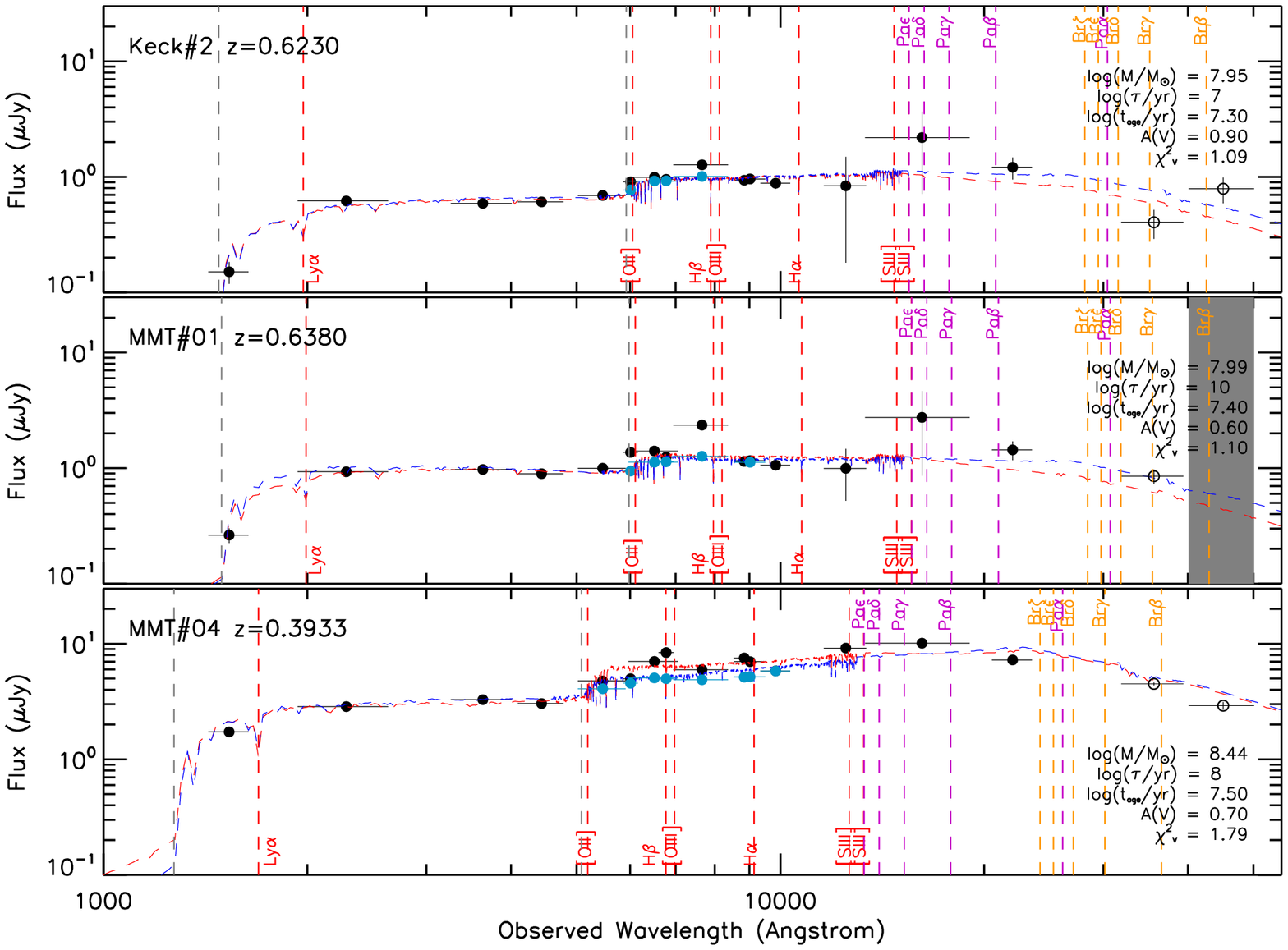}
  \caption{The SED and the best-fitting stellar population results
    from modeling the SED in three galaxies: Keck\#2, MMT\#01, and
    \#04. The observed fluxes are shown by the black circles while the
    blue circles illustrate the corrections for nebular emission lines.
    Errorbars along the $x$-axis demonstrate the FWHM of the filters
    while those along the $y$-axis are 1$\sigma$.
    The best $\chi^2$ fits to the original and emission-line corrected
    SEDs are shown in red and blue, respectively.
    In fitting the SED, we exclude the IRAC photometry,
    but include them in this figure for comparison.
    The emission-line corrected fitting results are reported on the right.
    Grey vertical lines indicate the locations of the Balmer and Lyman
    continuum breaks, while red, purple, and orange vertical lines show
    the location of various redshifted emission lines. When correcting for
    emission lines, the $\chi^2$ values are reduced by a factor of 3.5--10.3
    in these three cases. In MMT\#04, we find that the stellar mass and
    age are overestimated by 0.3 dex and 0.7 dex, respectively.}
  \label{fig:SED}
\end{figure*}

One significant advantage of studying low-mass galaxies in the SDF
is the ultra-deep imaging in twenty-four bands from the UV to the
IR, which allows us to characterize their stellar
properties.
The SDF is imaged with: (1) \GALEX\ \citep{martin05} in both the
\FUV\ and \NUV\ bands\footnote{Details on the \GALEX\ imaging
are available in \cite{ly09} and \cite{ly11a}.}; (2) KPNO's Mayall using
MOSAIC in $U$; (3) Subaru with Suprime-Cam in 14 bands
($BV\Rcf i\arcmin z\arcmin z_b z_r$, and the five NB and two
IA filters as mentioned previously);
(4) KPNO's Mayall using NEWFIRM \citep{probst08} in $J$ and $H$,
(5) UKIRT using WFCAM in $K$, and (6) {\it Spitzer} in the four
IRAC bands (3.6\mm, 4.5\mm, 5.8\mm, and 8.0\mm).
As discussed in \cite{ly11a}, a more complete photometric catalog
from SExtractor was constructed from an ultra-deep mosaic that
consisted of stacked optical and near-IR images with a normalization
that takes into account the rms (i.e., $\chi^2$ stacking).

Upon cross-referencing our \Ndet\ galaxies with the complete
photometric sample, all but one of them have direct matches.
The missing source (MMT\#05) was confused with a nearby bright extended
galaxy. We exclude MMT\#05 in our stellar population analyses.
For the remaining 19 galaxies, aperture photometry measurements
in 15 bands from the \FUV\ to $K$ are used to construct their complete
broad-band SEDs.
There are two modifications that we make to these measurements.

First, since all our galaxies are virtually point sources for \GALEX,
we obtain more accurate photometry in the \FUV\ and \NUV\ bands by
PSF-fitting with \textsc{daophot} \citep{stetson87}.
Our examination of the residuals for each source suggests that the
PSF-fitting is extremely successful, and that these galaxies are not
affected by contamination of nearby sources. Among the Keck sample,
we note that only Keck\#2 is detected in the \FUV.
The lack of \FUV\ detections is un-surprising, since the Lyman continuum
break occurs redward of the filter for the remaining five Keck sources.
Also, Keck\#6 is not detected in the \NUV\ because of its intrinsic
faintness.
Because the MMT \OIIIa\ galaxies are brighter with $z\lesssim0.65$, they are
all detected in the \NUV\ and \FUV\ bands, allowing for robust UV SFR
determinations.

Second, since our galaxies all have very high emission-line EWs, we correct
the broad-band photometry for the contribution from nebular emission lines
using emission-line measurements from our spectroscopy and NB imaging.
Here we generate a spectrum for each galaxy with zero continuum and emission
lines located at the redshifted wavelengths for \OII, \NeIII, \Hb, \OIII,
\Ha, and higher order Balmer lines.
These spectra are then convolved with the filter bandpasses to determine
excess fluxes, and then are removed from the broad-band photometry.
The correction for emission lines to SEDs has been recently explored
in other higher redshift studies where high-sSFR galaxies are
more prominent at $z\gtrsim1$ \citep{atek11,pirzkal13}.
For two galaxies (MMT\#10 and MMT\#11), the redshifts are low enough
to have spectroscopic detections of \Ha. For a subset of MMT galaxies
(\#03, \#04, \#06, \#07, \#09, \#13, and \#14), we use the NB921 or
NB973 excess fluxes for \Ha\ flux estimates.
However, for galaxies at $z\geq0.5$ and $0.43\lesssim z\lesssim0.47$, 
\Ha\ is redshifted into the near-IR or not available from optical
spectroscopy.
To correct \Ha\ in these galaxies (Keck\#1--\#6, MMT\#1, \#2, \#8, and
\#12), we use spectroscopic information for \Hb, but with the assumption
that \Ha\ is three times stronger.
Of course, higher dust reddening will yield stronger \Ha\ corrections,
so we are adopting a minimum correction of $A(\Hae)$ = 0.14 mag.

Both our observed SEDs and those corrected for nebular emission lines are
tabulated in Table~\ref{tab:SED}, and representative examples are shown
in Figure~\ref{fig:SED}.
The black circles show the original broad-band photometry, while the blue
circles show the corrected continuum fluxes after removal of emission lines.
These three examples illustrate how the SEDs change when
the emission-line corrections are small, average, and among the largest.

Both the original and corrected SEDs were fitted with stellar population
synthesis models \citep{bc03} with the Fitting and Assessment of Synthetic
Templates \citep[FAST;][]{kriek09} code.  We use exponentially-declining
star formation histories (i.e., $\tau$ models) similar to
previous fitting by us and many other groups \citep{ly11a}.
We have chosen this star-formation history (SFH) for its simplicity
since broad-band data are generally unable to distinguish against more
complicated SFHs (e.g., a constant SFR with a recent burst, which may
be more representable for our galaxies). As we later find, these fits
adopting an exponentially declining SFH are consistent with the data
with nearly unity $\chi^2_{\nu}$ values. Also, the primary purpose of
our SED modeling is to obtain stellar masses of our galaxies. This is
best traced from the rest-frame optical light from the older stellar
population. Thus the inclusion of a recent burst does not significantly
alter stellar mass estimates.

In these models, we continue to adopt \citetalias{calzetti00} reddening,
and inter-galactic medium attenuation following \cite{madau95}.
The only differences are that we adopt a \citetalias{chabrier03} IMF,
allow for an extremely short burst model, $\tau=10^7$ yr, and use
stellar atmospheres with
one-fifth (\fifth) solar abundances. The latter is motivated by
the fact that our galaxies are extremely metal-poor.
In addition we considered solar abundance (\solar) for completeness.
We find that for the majority of our galaxies, the \fifth\ models
yield similar results to \solar\ models. For consistency with our
metallicity determinations, we adopt the SED fitting results
with \fifth\ models.

We find that with the emission-line corrections, the SED fits are
improved with lower $\chi^2$ values by a factor 1.1--10.4 in the
\fifth\ models with an average (median) improvement of 3.3 (2.1).
In general, the results of fitting the SEDs with emission-line
corrections yielded lower stellar masses by 0.1 dex and
younger ages by 0.1 dex on average.
However, these differences vary from one source to another.
In particular, our analyses also reveal a moderate correlation
between \OIII\ EW and the difference in stellar mass
determinations, which others have reported \citep{atek11}.

Our SED-fitting results are summarized in Table~\ref{tab:SFR_Mass}.
We find that these galaxies typically are low-mass systems (median of
\MassM\ \Msun, average of \MassA\ \Msun).
However, there is dispersion in the mass distribution (extending from
$5\times10^6$ to $3\times10^9$ \Msun).
The estimated light-weighted stellar ages of
$\log{(t_{\rm age}/{\rm yr})}=7.1$--8.7 dex (average: \AgeA) suggest
that these galaxies have undergone most of their star formation in the
recent past.

We compare the de-reddened SFRs estimated from SED fitting
against the Balmer-derived de-reddened SFRs and find good
agreement, albeit with large uncertainties.
We also compare the dust reddening determinations from SED
fitting and Balmer decrements and find a weak correlation. With 
large uncertainties in determining reddening with both methods,
it is not possible for us to investigate any difference between
nebular and stellar reddening. The weak correlation is consistent
with \EBVa$_{\rm gas}\approx1.5$\EBVa$_{\rm star}$.
This result is similar to what we found for a larger sample
of ELGs in \cite{HaSFR}.

Finally, we combine our dust-corrected SFRs (Section~\ref{sec:SFR})
and stellar mass determinations to illustrate our galaxies on the
SFR--\Mstar\ plane in Figure~\ref{fig:SFR_Mass}. Relative to their
stellar mass, we find that our ELGs are all undergoing strong star
formation.
In particular, the sSFRs are between $10^{-9}$ yr$^{-1}$ and
$7\times10^{-7}$ yr$^{-1}$ with an average of \sSFRA\ yr$^{-1}$.
This is 2.0--3.0 dex higher than the so called ``main sequence'' of
star formation for $z\sim0$ SDSS galaxies \citep{mannucci10} and
low-luminosity local galaxies \citep{berg12}.
Extrapolating the main sequence of \cite{noeske07} toward lower
stellar mass, we find that the sSFR of our ELGs are
$\approx$0.5--1.0 dex higher than ``typical'' galaxies at
$z\sim0.45$--0.85, but they are consistent with \cite{noeske07}'s
sample at the 1$\sigma$ level.
We emphasize that the inverse of the sSFRs, is consistent with
the stellar ages derived from SED fitting.

With the strong nebular emission lines seen in our galaxies,
the nebular continuum from free-free, free-bound, and two-photon
emissions can be important \citep[see e.g.,][]{izotov11}. To
estimate the contribution toward the total light, at rest-frame
optical wavelength, we generate spectral synthesis models from
Starburst99. In our models we assume a constant SFH, a \cite{kroupa01}
IMF\footnote{This is similar to a \citetalias{chabrier03} IMF.}, Geneva
stellar evolutionary models, and $Z$/\zsun\ = 0.05. For EW$_0$(\Hb) =
45\AA\ (the median of our sample), we estimate that the nebular
continuum is responsible for $\lesssim$10\% of the total
optical/IR light, and thus the stellar masses (sSFRs)
that we have reported are $\lesssim$0.05 dex overestimated
(underestimated). We note that this correction can be as large
as 20--50\% (0.1--0.3 dex) in galaxies (Keck\#1, MMT\#07) with
EW$_0$(\Hb) of $\approx$250--400\AA. However, these corrections
for only two galaxies do not significantly alter our \MZR.

\subsection{Nearby Emission-line Galaxy Companions}
\label{sec:companions}
Utilizing our deep NB/IA images, we constructed postage stamps
that contain only the emission-line excess flux by differencing
the NB/IA images with adjacent broad-band images (as previously
discussed in Section~\ref{sec:NBem}).
For proper removal of the continuum, we normalize the images
by their respective zeropoints.
We find that the majority (14 of 20) of our sample has between
1 and 4 NB/IA excess emitters within a projected radius of
$\approx$\Dproj. The closest separation is 12 kpc with an
average projected distance of 48 kpc.

To ensure that these nearby NB/IA excess sources are real, we
conducted two checks. First, we examine if any of these nearby
galaxies are also classified as an NB/IA excess emitter in our
source catalogs. We find that more than half of them (16 of 34)
are NB/IA excess emitters. Upon further inspection, we find
that those that are missed did not satisfy our NB/IA excess
selections because they have low emission-line EW or
are faint, but many could also lie at the same redshift as
our \OIIIa\ galaxies.
Second, some of our galaxies are identified as an excess emitter
in two or more filters (e.g., NB704 and NB921).
This is due to multiple emission lines falling in these filters.
We find that in these four galaxies, six of the nearby excess
emitters are also detected in two or three filters.
This result strongly suggests that these nearby NB/IA excess
emitters surrounding our \OIIIa-selected galaxies are physically real, and
at a similar redshift.

While our sample size is small, the presence of nearby
($\lesssim$\Dproj) companions in 70\% of our galaxies
suggests that this is unlikely to be a coincidence.
To confirm this, we examine how frequent nearby emission-line
companions are using our sample of 401 $z=0.4$ \Ha\ emitters,
which has been previously discussed in \cite{HaSFR}.
We determine the projected distance to the nearest \Ha\ emitter, and
find that the average of this distance is $\approx$350 kpc
(median is $\approx$280 kpc).
Furthermore we find only 49 of 401 (12\%) \Ha\ emitters have a
companion that is within \Dproj.
This low percentage suggests that \Ha\ emitters, while clustered,
are infrequently found with a nearby satellite.
An identical analysis with 715 $z=0.84$ \OIII\ emitters yielded
similar numbers.

Three plausible explanations for the significant excess of
nearby companions are that galaxy-galaxy interactions can:
(1) stimulate a ``starburst'', (2) tidally strip out metal-rich
gas into the CGM, or (3) cause accretion of metal-poor gas from
the outskirts of the galaxy.

In the first case, the intense star formation can trigger
strong winds that can entrain metal-rich gas into the CGM.
If so, this would explain why these galaxies are identified
to be extremely metal-deficient.
Our measurements of SFR and masses (see Figure~\ref{fig:SFR_Mass})
suggest that the current SFR could form most of the stellar population in
$\sim$\sSFRt\ (average), supporting this hypothesis.
Since our \OIIIa\ sample is inhomogenous
(i.e., selection in different NB/IA filters), a more detailed
study is warranted from the complete NB/IA sample.
Forthcoming work will examine further these nearby companions
and any correlation seen with the properties of the central galaxies.

In the case of ``tidal disruption,'' one would expect to
see evidence of gas stripping, in the form of tidal tails
and asymmetric morphologies. An examination of our NB/IA-excess
images reveal no tidal tails or any ``bridge'' between our
targeted sources and a nearby companion. We do notice that the
light distribution is asymmetric in half of our galaxies;
however, the seeing-limited resolution of our data makes it
difficult to interpret such results.
Higher resolution imaging with {\it Hubble Space Telescope} (HST)
is needed to quantitatively measure the morphological properties
of these galaxies and detect any tidal tail features.

One of our galaxies (Keck\#05) fortuitously has HST/WFC3 broad-band
near-IR imaging \citep{jiang13}. Keck\#05 is in fact
two galaxies that are separated along the NE--SW direction
by 0.9\arcsec\ (projected distance of 6.9 kpc at $z=0.84$) with the
nebular emission coming from the SW component. Our examination
of the HST imaging reveal that the NE galaxy is diffuse along
the P.A. toward the SW component. More careful analysis is needed
to understand Keck\#05 and its nearby companion, and as we
emphasized, HST imaging for a larger fraction of our \OIIIA\ sample
is required to examine the possibility of tidal disruption for
the case of low metallicities in our galaxies.
We note however that this scenario is unlikely to result in low
metallicity in the centers of these galaxies, as these dynamical
effects are expected to have a stronger impact on the gas that
is external to the galaxy, which is relatively more metal-poor,
than the metal-rich gas that is located closer to the center.

Another contrasting possibility is that the interaction with
the nearby galaxy induces gravitational torques that drive
metal-deficient gas from the outskirts of galaxies into the
centers \citep[e.g.,][]{mihos96}. Support for this idea has
been seen in local interacting galaxies where the radial
metallicity gradient is unusually flat
\citep{kewley06,rupke10}.
This would explain (1) the deficiency of metals, (2) the excess of
nearby galaxies, and (3) the high SFRs and SFR surface densities
(see Section~\ref{sec:SFR_density}) in our galaxies.
  
Our data do not decisively distinguish between these
models without dynamical information. In particular, IFU
spectroscopy of the ionized gas would reveal the presence of
inflowing or outflowing gas.
\subsection{Star Formation Surface Densities}\label{sec:SFR_density}

In addition to subtracting of the continuum to identify nearby
companions, the NB/IA excess flux images allow us to determine
their SFR surface density, $\Sigma_{\rm SFR}$.  These images are
first converted to an emission-line luminosity, then integrated
across the galaxy to determine a total \Ha\ or \Hb\ observed
luminosity, and finally corrected for dust attenuation
(Section~\ref{sec:dust}).
Six of our galaxies (MMT\#04, \#06, \#07, \#09--\#11) have NB
imaging or spectroscopy that observes \Ha. In addition, three
other galaxies (MMT\#13, \#14, and Keck\#4) have \Hb\ measurements.
The remaining galaxies (MMT\#01, \#02, \#08, \#12, Keck\#1,
\#2, and \#5) are \OIII-selected. To obtain an estimate of the
\Hb\ luminosity for the latter galaxies, we use the integrated
\OIII/\Hb\ ratio from our spectroscopy (see Table~\ref{tab:em_lines}).
Finally to obtain the \Ha\ luminosities, we scale the
de-reddened \Hb\ measurements by a factor of \HaHbi.
We integrate the source flux over a region that is
$\pi({\rm FWHM}/2)^2$, which is converted to physical
distance using the angular diameter distance
(1\arcsec\ = 5.4 kpc at $z=0.4$ and 7.8 kpc at $z=0.85$).
This measured area is larger due to the degradation from
seeing. To correct for it, we determine the effective FWHM
in quadrature:
FWHM$_{\rm eff}^2$ = FWHM$^2$ - FWHM$_{\rm seeing}^2$.
Due to the slightly extended nature of our galaxies, this
correction to the area is typically a factor of 2 or less.
We find a wide dispersion in the $\Sigma_{\rm SFR}$ between
0.005 and 5 \Msun\ yr$^{-1}$ kpc$^{-2}$ with an average
(median) of \SFRSDA\ (\SFRSDM) \Msun\ yr$^{-1}$ kpc$^{-2}$.
Compared to local galaxies \citep{kennicutt98b},
our $\Sigma_{\rm SFR}$'s are on average an order of magnitude
higher than normal spirals, and at the low end
of IR-selected circumnuclear starbursts.

\section{DISCUSSION}\label{sec:Disc}

\subsection{Our \OIIIa\ Sample}
By construction, the \Ndet\ galaxies presented in this paper were
selected to be extreme. First, our SDF follow-up spectroscopy
concentrated on ELGs identified from our NB/IA imaging.
Then in this paper we have selected only a few percent of these
spectra, where the lines were strong enough to include detectable
\OIIIa.
Half of our galaxies were solely targeted because of the
strength of their \OIII\,$\lambda\lambda$4959,5007 doublet
(see Table~\ref{tab:SDF_sample}).
This favors emission from more highly ionized gas. Although the
\OIIIa\ line is occasionally seen in LINERs, we only found
one of them based on various emission-line ratios
(see Section~\ref{sec:AGN}).
The remaining \Ndetf\ are all extreme starburst galaxies, which tend
to be overlooked by traditional broad-band photometric selections.
Aside from deep narrow-band imaging, the only other windows on
galaxy evolution which provides access to many extreme SFR galaxies
are extensive spectroscopy with as little pre-selection as possible,
preferably none \citep[such as slitless spectroscopy;][]{atek10}
or selecting galaxies with unusual colors \citep{cardamone09,vanderWel11}.

Thus these \Ndet\ galaxies lie far above the established correlation
of star formation rate with stellar mass. In other words, their
sSFR is between two and three orders of magnitude larger than the
average observed in the local universe, or even than what is
``normal'' for galaxies at $z=0.4$--0.9. These \OIIIa-detected
galaxies are extremely rare in the local universe, making up only
0.05\% of those with SDSS spectra.

Since sSFR is the inverse of the star production time-scale, continuing
at their high observed SFRs, our extreme galaxies would produce their
entire stellar contents in only 1\%--10\% of the time it takes in normal
galaxies.
This strongly suggests that we are observing them in a highly atypical
evolutionary phase, which could last \sSFRt\ or even less. 
If this is a phase that most galaxies pass through occasionally, then
the `duty cycle' of this extreme burst of star formation would be only
$\sim$1\%--10\%.  
\subsection{Selection Effects}
\label{sec:bias}
The NB technique has two observable limitations: (1) a minimum EW
excess at the bright end ($\lesssim$24 mag), and (2) an
emission-line flux limit at the faint end. In the former case,
our selections are limited to galaxies with rest-frame EWs of
EW$_0$(\OIII) = 11\AA--17\AA\ ($z=0.4$--0.85) and
EW$_0$(\Ha) $\approx$ 20\AA\ ($z=0.24$--0.40). With a 2.5$\sigma$
emission-line flux limit of approximately $4.5\times10^{-18}$
erg s$^{-1}$ cm$^{-2}$, the luminosity limit corresponds to
$3\times10^{39}$ and $2\times10^{40}$ erg s$^{-1}$ at $z=0.40$
and $z=0.84$, respectively. These EW and luminosity limits are
illustrated in Figure~\ref{fig:EW_Lum}, and are compared to
the ELGs in the SDSS.
It can be seen that our NB imaging cannot detect the \OIII\ emission
from many SDSS galaxies, simply because of the lower equivalent widths.
For example, only 16\% of the full SDSS galaxy sample would be
identified as NB704 excess emitters at $z=0.40$. However, this
fraction increases to 90\% for ELGs with moderate line strengths,
EW$_0$(\OIII) $\gtrsim11$\AA. For \Ha, the majority (62\%) of the
SDSS would be detected at $z=0.4$ with our NB921 imaging, and
almost all (96\%) of galaxies with EW$_0$(\Ha) $\gtrsim22$\AA.
Thus, our NB survey is biased against weak emission lines. We note
that the lower success of detecting \OIII\ in SDSS ELGs at
$z=0.4$ compared to \Ha\ is due to the higher metallicities in
local galaxies. In metal-rich systems (i.e., lower \Te), O$^+$
is a more effective coolant for the ISM than O$^{++}$, resulting
in flux ratios of \OII/\Ha\ $\approx0.5$ and \OIII/\Ha\ $\approx0.1$.
Thus, our \OIII\ NB selection is less sensitive to metal-rich
galaxies.

In addition to our general NB selection, the primary restriction
for our sample is an \OIIIa\ detection. First, this limits the
sample toward high-EW galaxies, which generally have high sSFRs.
Second, the detection alone will further bias our sample against
moderately metal-poor galaxies (\OH\ $\approx8.0$). For example,
in our spectroscopic survey, we have identified an additional
$\sim$20 galaxies that possess strong \OIII\ emission
(EW$_0 \gtrsim50$\AA). These galaxies, however, have very
weak detections of \OIIIa\ (1--3$\sigma$).

While an \OIIIa\ detection restricts our ELG sample, we find that
our \Ndet\ galaxies have many similarities to our more general
ELG sample at $z\approx0.4$--1. For example, the
median (average) stellar mass of our spectroscopic sample of
over 200 galaxies is $\log(M_{\star}/M_{\sun})$ = 8.9 (8.7). These
values are moderately higher (0.4 dex) than our \OIIIA\ sample.
Also, as illustrated in Figure~\ref{fig:MEx}, our \OIIIA\ sample
shows similar \OIII/\Hb\ ratios, which is a measure of the
ionization. This demonstrates that ELGs selected from NB surveys
at $z=0.4$--1 all have higher ionization by $\approx0.2$--0.5 dex
compared to local SDSS galaxies with similar stellar masses.

\subsection{Similarities and Differences to Green Peas and
  Luminous Compact Galaxies}
\label{sec:LCGs}

The SDSS has found rare populations of strongly star-forming
galaxies called ``Green Peas'' \citep{cardamone09} and
``Luminous Compact Galaxies'' (LCGs) \citep{izotov11}. The former
population was identified by unusual optical colors due to significant
contribution of very strong nebular emission lines (\OIII\ and \Hb)
in the $r$-band (hence they appear ``green'').
Ancillary data indicated that these luminous ($r\lesssim20.5$)
galaxies are at $z\approx0.1$--0.35, have stellar masses of
$\sim$10$^{8.5}$--10$^{10}$ \Msun\ and SFRs of $\sim$10\ \Msun\
yr$^{-1}$, are relatively metal-rich \OH\ $\sim$
8.7\footnote{\cite{izotov11} find that these strong-line metallicities
  are overestimated by $\approx$0.5 dex.}, and are very rare (2 deg$^{-2}$).
Because of the unusual color selection, selection effects are
significant for this sample.

Analogously, LCGs were identified from the SDSS DR7 spectroscopic
sample to have EW(\Hb) $\gtrsim50$\AA, at least 2$\sigma$ detections
of \OIIIa, and appear compact or unresolved in the SDSS images.
These galaxies have redshifts of $z=0.02$--0.63 (most of them are
at $z\lesssim0.35$), median stellar masses of $\sim$10$^9$ \Msun,
median SFRs of $\sim$4 \Msun\ yr$^{-1}$, and oxygen abundance of
\OH\ $\approx$ 8.2. Since it is believed that green peas are a
subset of LCGs \citep{izotov11}, we will only compare our sample
against the LCG population.

LCGs show many similarities and differences to our galaxies
with \OIIIa\ detections. First, both samples occupy a distinct
region in BPT and MEx diagnostics diagrams with large
\OIII/\Hb\ and low \NII/\Ha\ (or stellar mass).
While LCGs have low stellar masses and high sSFRs, they are on
average $\sim$0.5 dex more massive than our sample and with lower
sSFR by $\sim$0.5 dex. These LCGs can be found with stellar masses
above $10^{10}$ \Msun, while our most massive galaxy has
\Mstar\ $\approx3\times10^9$ \Msun. In addition, these galaxies are
relatively metal-rich by $\sim$0.4--0.5 dex compared to our sample.
Also, 90\% of our \OIIIA\ sample is found above $z=0.4$, while the
majority of LCGs are identified at $z\lesssim0.35$. Therefore, while
LCGs are low-mass, metal-poor galaxies, our \OIIIA\ sample appears to be
an extension of LCGs toward lower masses, lower metallicity, and
higher redshift.
These differences might suggest that metal-poor ELGs identified
with deep NB imaging may evolve into LCGs. Finally, in terms
of surface density, it is estimated that there are $\approx$20 LCGs
per deg$^{2}$. The SDF has already detected \OIIIa\ in \Ndet\
galaxies over 0.25 deg$^{2}$, and more of them are expected to
be found (see Section~\ref{sec:numbers}). The higher surface
density is because our survey extends toward lower mass galaxies,
which are more common, and because higher sSFR are seen for
galaxies at higher redshift (see Section~\ref{sec:numbers}).

\subsection{Comparison with Other \OIIIa\ Studies}\label{sec:compare}
While these metal-poor galaxies are rare, there have been numerous
efforts to identify them. Here, we compare our sample to previous
\OIIIa\ studies.
Two of the first studies that probed a large number of low-mass galaxies
in the local universe were \cite{lee04} and \cite{lee06}. First,
\cite{lee04} identified and detected \OIIIa\ for 24 local galaxies
from the KPNO International Spectroscopic Survey. These galaxies,
spanning \MB\ = --15 and --19, were pre-selected to have gas-phase
metallicities (determined from strong-line methods) to be below
\OH\ = 8.2. The sample provided the first measurement of the
\MZR\ using the ``direct'' method.
Targeting 25 dwarf irregular galaxies, \cite{lee06} extended the
\MZR\ toward lower stellar masses.
Since then, more extensive spectroscopy has been conducted targeting
low-luminosity galaxies within 11 Mpc \citep{berg12}, and mining
the SDSS \citep{brown08,izotov06a,izotov12}.

In addition, a few other studies have succeeded at extending
the search toward higher redshift where the redshift evolution
of the \MZR\ implies that more metal-poor galaxies should be
found in the early universe.
\cite{hoyos05} was the first study to detect \OIIIa\ in 17 
$\sim$$L_{\star}$ galaxies at $z=0.51$--0.86 with gas metallicity
ranging from \OH\ = 7.8--8.3.
These galaxies were selected from the DEEP2 Survey \citep{davis03} and
TKRS \citep{wirth04}, which are luminosity-limited to $I=24.5$ and
$R=24.3$ mag, respectively.
Then \cite{kakazu07} conducted follow-up spectroscopy of ultra-strong
ELGs selected from NB imaging. While some level of detection for
\OIIIa\ was available in 17 galaxies, only six had $\geq$3$\sigma$
\OIIIa\ detection\footnote{One of these six does not have
  metallicity determination, since \OII\ was not observed.}.
The rest were too weak, with a median detection of $\approx2\sigma$.
Among those with robust detections ($\geq$3$\sigma$), their most
metal-poor galaxy has \OH\ = $7.36^{+0.22}_{-0.18}$ (1$\sigma$).
They have since extended their \OIIIa\ sample to a total of 31
galaxies at $z=0.40$--0.85 \citep{hu09}. About three-fourths
(23/31) of their sample are detected above 3$\sigma$ with 9
galaxies above 5$\sigma$. Their most metal-poor galaxies have
\OH\ = 6.97$\pm$0.17 and \OH\ = 7.25$\pm$0.03 (1$\sigma$).

In addition, \cite{atek11} identified a sample of high-EW ELGs
at $z=0.35$--2.3 from space-based grism spectroscopy. With higher
spectral resolution follow-up observations, they detected \OIIIa\ to
determine a gas-phase metallicity of \OH\ = $7.5\pm0.1$ for a galaxy
at $z=0.7$.

For a complete comparison, we compile available information for these
studies. For the \cite{berg12} sample, dust-corrected \Ha\ SFRs are
determined from the \Ha\ luminosities reported in \cite{kennicutt08},
adopting the extinction determinations of \cite{lee09}, and assuming
a \citetalias{chabrier03} IMF. We also correct their reported
stellar masses from a \citetalias{salpeter} to \citetalias{chabrier03} IMF.
For the \cite{hoyos05} sample, we limit their sample to the seven
galaxies where the \OII\ line is observed.
Stellar masses for their sample were obtained from \cite{bundy06},
where a simple cross-matching of sources yielded only one match.
SFRs reported by \cite{brown08} and \cite{lee04} are also
corrected from a \citetalias{salpeter} to \citetalias{chabrier03}
IMF.
For the samples of \cite{kakazu07} and \cite{hu09}, we are only able
to report the redshifts and gas-phase metallicity for the former.
In their final sample, only metal abundances and luminosities
are immediately available.

Since the rest-frame $B$-band absolute magnitude, \MB, is commonly
reported, we compute it for our galaxies using our emission-line
corrected SED. The \MB\ magnitudes are determined as:
\begin{equation}
  M_B = B + 2.5\log{(1+z)} + 5.0\log{(d_L/{\rm 10pc})},
\end{equation}
where $B$ is the apparent magnitude at 4450\AA\ determined from
interpolating between adjacent broad-band filters.
These magnitudes are reported in Table~\ref{tab:SFR_Mass}.

\begin{figure*}
  \vspace{-0.5cm}
  \epsscale{1.1}
  \plotone{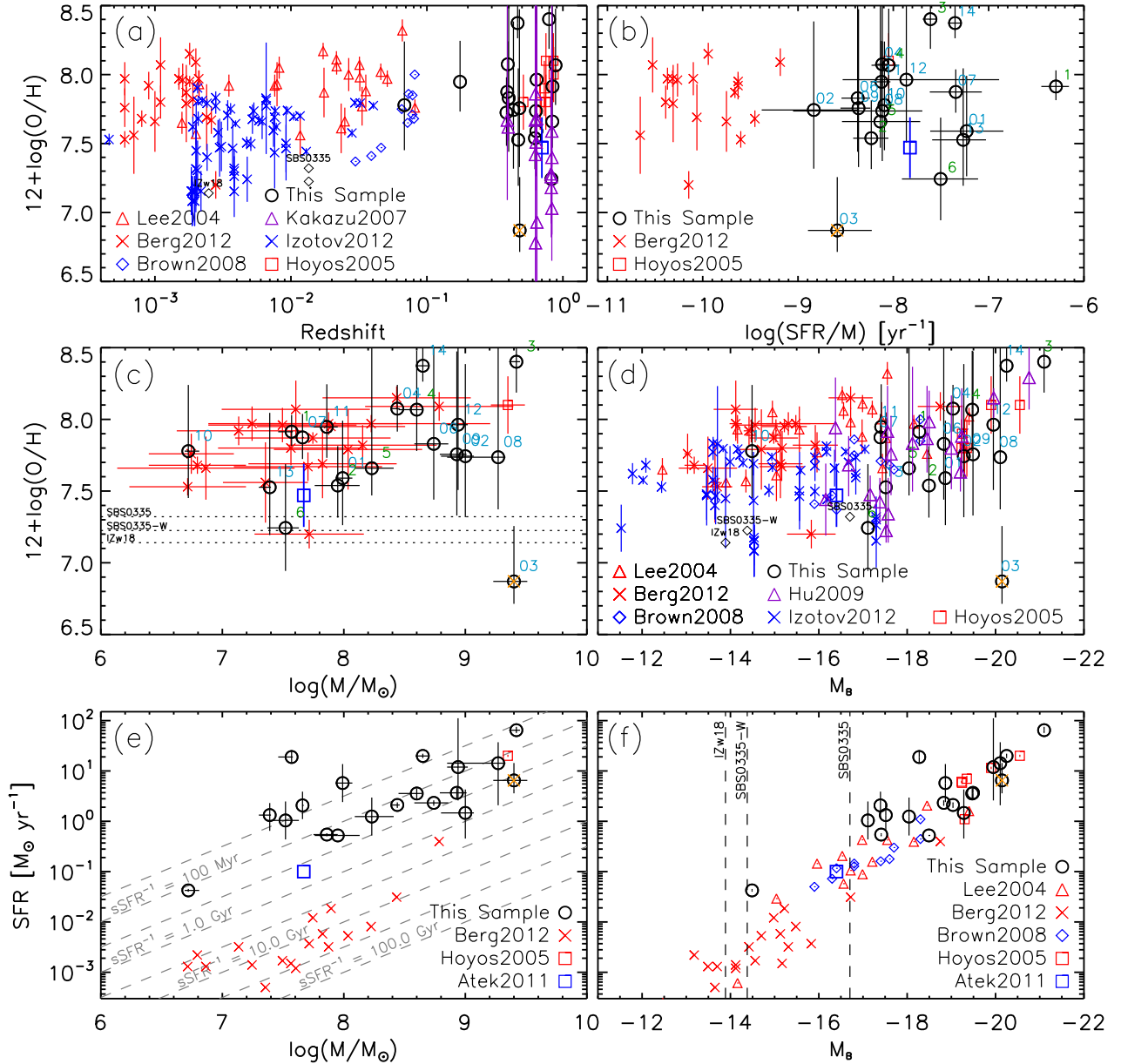}
  \vspace{-0.3cm}
  \caption{A comparison of the properties of our \OIIIA\ sample
    (circles) to published direct metallicity
    studies. The top and middle rows show the gas-phase
    metallicity vs. (a) redshift; (b) specific SFR; (c) stellar
    mass; and (d) rest-frame $B$-band absolute magnitude, \MB.
    The bottom row shows the SFR vs. (e) stellar mass, and (f) \MB.
    Our sample is shown as black circles while the samples of
    \cite{lee04}, \cite{hoyos05}, \cite{brown08}, \cite{kakazu07}
    and \cite{hu09}, \cite{atek11}, \cite{berg12} and \cite{izotov12}
    are shown as red triangles, red squares, blue diamonds, purple
    triangles, blue squares, red crosses, and blue crosses,
    respectively.
    Our only LINER (MMT\#03) is denoted by orange crosses.
    Measurements for I Zw18 and SBS0335 are also overlaid for
    comparison.
    We provide 95\% C.L. for our metallicity determinations, as well as for
    published results.
    Uncertainties on stellar masses, SFRs and other
    properties are 68\% C.L. Where applicable, we apply
    corrections to SFRs and stellar masses to a consistent
    \citetalias{chabrier03} IMF.}
  \label{fig:EMPG}
\end{figure*}

In Figure~\ref{fig:EMPG}, we illustrate the redshifts, SFRs, stellar
masses, specific SFRs, and \MB's for \OIIIa-detected galaxies.
Our sample surveys a redshift domain that is similar to the
studies of \cite{hoyos05} and \cite{hu09}. In addition,
our galaxies span \MB\ between --17 and --21 (excluding MMT\#10,
our lowest redshift galaxy), and have SFRs that are similar to
local dwarf irregulars and $z\sim1$ $L_{\star}$ galaxies.

Comparing our metallicities against their masses or luminosities,
we find that half of our sample follows the \MB--$Z$ or \MZ\ relation
of local and higher redshift galaxies.
The remaining galaxies are located systematically below these
relations by 0.2--0.3 dex. These galaxies have lower significance
detections (3.1--3.6$\sigma$) of \OIIIa.
So while they deviate from these relations, the are consistent
within the 95\% measurement uncertainties.
Deeper follow-up spectroscopy is needed to determine if these
galaxies are extremely metal-poor or just moderately metal-poor.

Our three most metal-deficient galaxies are MMT\#13 ($z=0.47$),
Keck\#2 ($z=0.62$) and Keck\#6 ($z=0.82$) with \OH\ = \OHMMT,
\OHKecktwo, and \OHKecksix, respectively.
While these galaxies have similar metallicities (within the errors)
to the most metal-deficient galaxies in the local universe, I Zw 18 and
SBS0335 (see Figure~\ref{fig:EMPG}), there are some notable differences
seen in their emission-line fluxes.

First, I Zw 18 and SBS0335 have \OIII/\Hb\ (\OIII/\OII) ratios of 2.2
and 3.2 (9.8 and 15.5), respectively.
Keck\#6, which has a more robust detection (4.3$\sigma$) of \OIIIa, has
a higher \OIII/\Hb\ ratio by a factor of 1.3--2, but similar to what
is seen for green peas and LCGs (see Section~\ref{sec:LCGs}).
Second, the \OII\ line is {\it not} detected, yielding an observed
\OIII/\OII\ ratio of $\gtrsim$80.
This non-detection is due to the high ionization and low
metallicity of the ISM.
In these circumstances, the excitation energies shift nearly
completely toward O$^{++}$, resulting in extremely low \OII\
emission.
For Keck\#6, estimates using the strong-line \Pagel\ and 
\Oratio\ flux ratios also suggest that the metallicity of this
galaxy is very low: \OH\ = 7.65 using the \cite{KK04}
calibration\footnote{This calibration yields higher metal
  abundances by more 0.2--0.3 dex against other strong-line
  calibrations \citep{kewley08}.} with a high ionization
parameter, $\log(q/{\rm cm~s}^{-1})\gtrsim8.8$ ($\log(U)\gtrsim-1.7$).

These galaxies are rare, but other studies have found them.
For example, \cite{kakazu07} identified a few galaxies where
\OII\ is undetectable but very high S/N detections of \OIII\
exist. There are also cases where both lines are detected,
yielding \OIII/\OII\ flux ratios of $\approx$10--60.
In addition, rest-frame optical spectroscopy of \Lya\ emitters
at $z=2.2$ have measured high \OIII/\OII\ ratios \citep{nakajima13},
which yield high ionization parameter estimates.

Furthermore, several studies have shown that high-$z$
star-forming galaxies are offset from local star-forming
galaxies in the BPT diagram with systematically higher
\OIII/\Hb\ ratios \citep[e.g.,][]{hainline09,rigby11}.
A recent comparison between theoretical models and
emission-line measurements of $z=0.5$--2.6 galaxies
by \cite{kewley13a} suggests that the ISM conditions
at higher redshifts are far more extreme with high
ionization and density.
These conditions are more analogous to those seen in dense,
clumpy \ion{H}{2} regions of local starburst galaxies.
As illustrated in Figure~\ref{fig:MEx}, our \OIIIA\
sample is offset by 0.2--0.5 dex in \OIII/\Hb\ from the
average of local galaxies with the same stellar mass.
In addition we find high electron densities of
$n_e=70$--600 cm$^{-3}$. These observables suggest that
XMPGs at $z=0.4$--1 are similar to typical $z\gtrsim1$
galaxies, and can be used as a powerful tool for studying
the ISM conditions in the early stages of galaxy formation.
%


\subsection{Dependence on Metallicity with Galaxy Properties}
It has been well established that the metal abundance of galaxies
correlates with their stellar mass \citep[e.g.,][]{tremonti04}.
The presence of such a correlation and the lack of significant
scatter have provided a key constraint in modeling the evolution
of galaxies.
For example, the shape of the \MZR, which shows a turnover in the
metallicity at high masses and a steep decline at lower masses, can
be explained by a simple model where massive star formation enriches
the ISM through stellar feedback. However, such enrichment can drive
metals beyond the ISM in low-mass galaxies, and thus lowering their
effective metallicity \citep[e.g.,][]{dave11}.

\begin{figure*}
  \epsscale{1.0}
  \plottwo{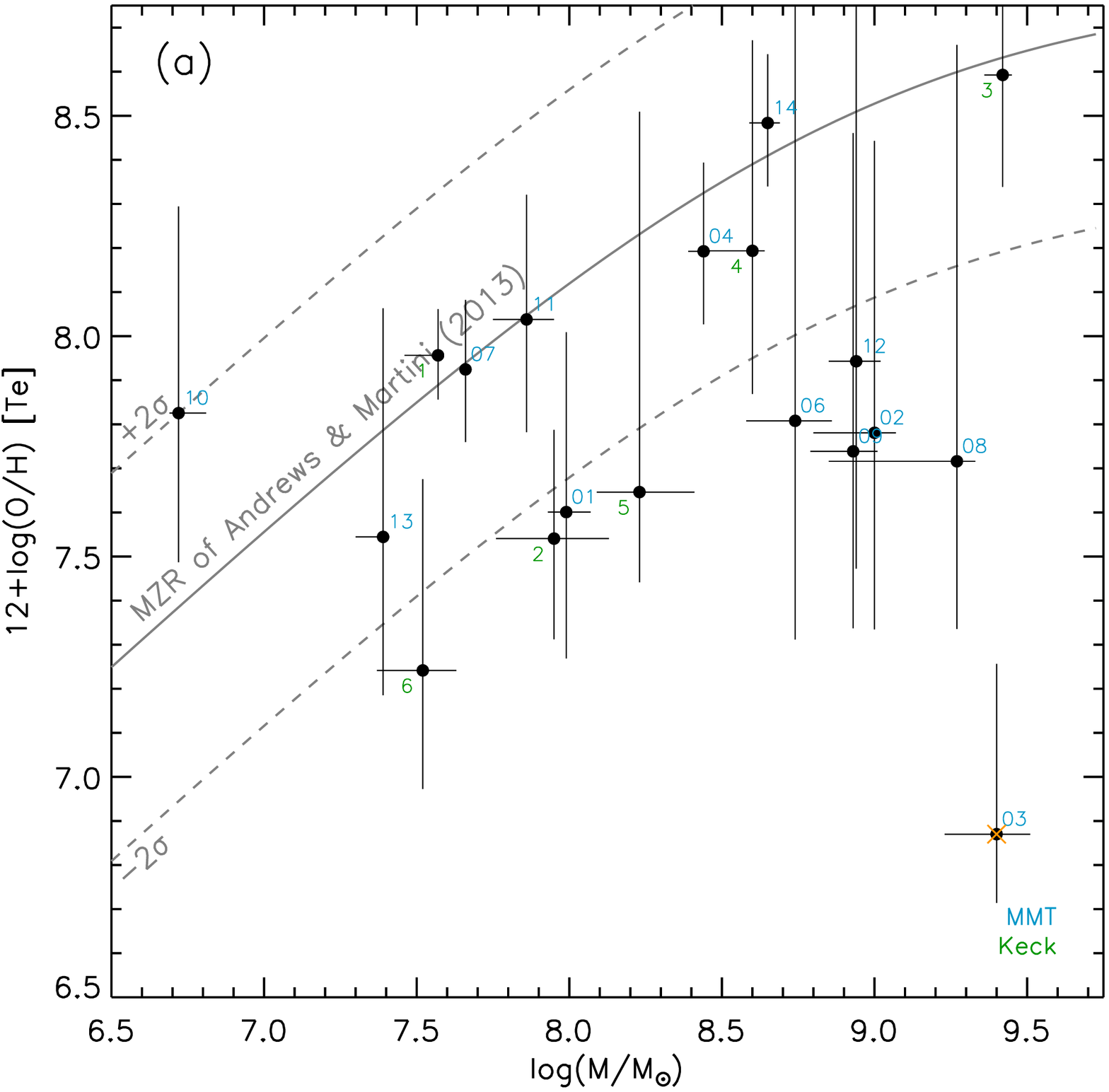}{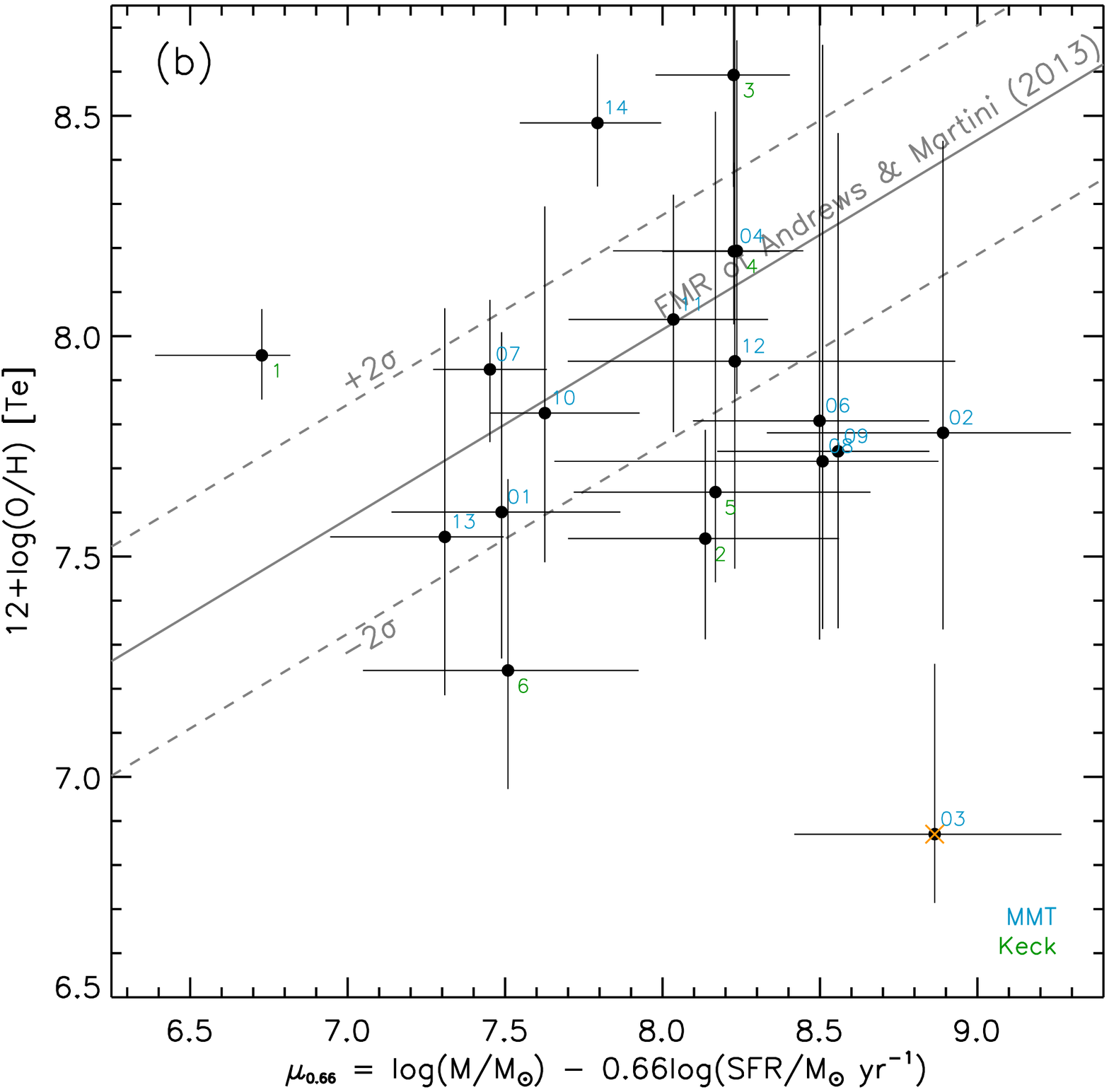}
  \vspace{-0.3cm}
  \caption{Comparison of \Te-based gas metallicity against (a) stellar
    mass, and (b) a projection of the ``FMR'' with $\mu=0.66$
    for our \OIIIA\ sample (black circles). The \MZR\ and FMR
    of the SDSS derived using the \Te\ method by \citetalias{andrews13}
    are shown in grey with dashed lines indicating $\pm2\sigma$.
    Here, the scatter of \citetalias{andrews13} is not the true
    intrinsic scatter from individual galaxies. Rather, it
    reflects the dispersion for stacked spectra in various
    \Mstar--SFR bins. This $\sigma$ is likely to be larger
    than the intrinsic scatter because it is weighted more
    toward high-SFR outliers (fewer galaxies are available
    in these bins; B. Andrews 2013, priv. comm.).
    Metallicity uncertainties are reported at 95\% C.L., while
    68\% C.L. are reported for mass and SFR uncertainties.
    For direct comparisons, we adopt the same $t_2$--$t_3$
    relation as \citetalias{andrews13}.}
  \label{fig:M_Z}
\end{figure*}

Studies in the past have been mostly limited toward more massive
galaxies ($\gtrsim$10$^9$ \Msun), and therefore whether the
\MZR\ extends toward lower masses or not is unclear
with existing limited samples.
In Figure~\ref{fig:M_Z}(a), we illustrate the \MZR\ of
\citetalias{andrews13}, which stacked SDSS spectra to detect
\OIIIa\ and determine metallicity using the same method
for easier comparison.
Here, the \MZR\ is fitted with a logarithmic functional
form devised by \cite{moustakas11}:
\begin{equation}
  \OHm = \OHm_{\rm asm}-\log{\left[1+\left(\frac{M_{\rm TO}}{M_{\star}}\right)^{\gamma}\right]},
\end{equation}
where \OH$_{\rm asm}$ is the asymptotic metallicity at high masses,
$M_{\rm TO}$ is the turnover mass, and $\gamma$ controls the
slope of the \MZR\ at low masses.

We note that while there are other determinations of the \MZR\
using different strong-line diagnostics that do not require
spectra stacking, there are nearly 1 dex
differences in metallicity determinations using the same
dataset \citep{kewley08}. For this reason, we chose to use
measurements obtained with the same method.
Since \citetalias{andrews13} used a different
\Te(\OII)--\Te(\OIII) relation to determine O$^+$/H$^+$, 
we adopt the same relation, $t_2 = 0.7t_3 + 0.1$,\footnote{For $t_3\geq2.0$, we set $t_2=1.5$.} for direct comparisons. We note that this
difference only raises metallicity on average by 0.05 dex,
with $\approx$0.2 dex as the largest correction.

We find that 9 of our \Ndetf\ galaxies follow the
local \Te-based \MZR\ to within $\pm$2$\sigma$. However, it
appears that our entire sample has systematically lower
metallicities at all masses. This may be the result of
selection effects, since more metal-rich galaxies are
less likely to have \OIIIa\ detections.

Recently, some have argued that the \MZR\ is in fact a
projection of a more ``fundamental'' plane
between stellar mass, metallicity, and SFR
\citep[i.e., \MZ--SFR relation;][]{laralopez10}. 
One such projection, called the ``Fundamental Metallicity Relation''
\citep[FMR;][]{mannucci10}, suggests that galaxies with higher
sSFR have lower metallicities compared to those at similar
stellar masses. Efforts to describe the FMR have defined a
plane in which a tighter correlation exists by considering a
combination of stellar mass and SFR:
\begin{equation}
  \mu_{\alpha} = \log{\left(\frac{M_{\star}}{M_{\sun}}\right)}
  - \alpha\log{\left(\frac{{\rm SFR}}{M_{\sun}~{\rm yr}^{-1}}\right)},
\end{equation}
where $\alpha$ is the coefficient that minimizes the
scatter against metallicity.
We illustrate the \citetalias{andrews13}'s determination
($\alpha = 0.66$) in Figure~\ref{fig:M_Z}(b).
A comparison against our sample reveals again that roughly
half of our galaxies follow the FMR; however, the
rest of our metallicities fall below it. 
Even more interesting, some of our galaxies (i.e., Keck\#1, \#3,
and MMT\#14), which followed the \MZR\ are no longer within +2$\sigma$
of the \Te-based FMR. In these galaxies, the \OIIIa\ detections are
robust (S/N = 11.1, 4.6, and 7.6, respectively), and their SFRs are high.
If anything, we see a positive or flat correlation between
metallicity and sSFR (see Figure~\ref{fig:EMPG}(b)),
contrary to what is expected from the FMR.
This might suggest that a \MZ--SFR relation as seen for local galaxies
may not hold in low-mass galaxies at $z=0.4$--0.9, and that there
is a larger scatter compared to $z\sim0$. 


\subsection{Number Densities of \OH\ $\lesssim7.65$ Galaxies}
\label{sec:numbers}

With a statistically meaningful sample of \OIIIa\ detections, we can
infer the space density of XMPGs from the largest emission-line sample
at intermediate redshifts. As illustrated in Figure~\ref{fig:EW_Lum},
we find that our \OIIIa\ sample has rest-frame EWs of
EW$_0({\rm H}\alpha)\gtrsim100$\AA\ and EW$_0(\OIII)\gtrsim50$\AA.
Limiting galaxies to these EWs, we find a number density at
$z\sim0.45$ of $4.5\times10^{-3}$ Mpc$^{-3}$ for \Ha\ and
$5.1\times10^{-3}$ Mpc$^{-3}$ for \OIII.
Our spectroscopic follow-up sampled a broadly representative range,
reaching down to EW$_0$ = 20\AA, and targeted about $\approx$25\% of
the high-EW galaxies. Among these spectra, 16\% and 11\% of the
\Ha-selected and \OIII-selected galaxies yielded \OIIIa\ detections,
respectively.
Extending these statistics to the full high-EW sample, we estimate
that the expected number density of galaxies with \OIIIa\ detections
at $z=0.3$--0.65 is $7\times10^{-4}$ Mpc$^{-3}$ for \Ha-selected
galaxies and $6\times10^{-4}$ Mpc$^{-3}$ for \OIII-selected galaxies.
This corresponds to a surface density of one per 13 arcmin$^2$
(\Ha; $z=0.3$--0.5) and one per 8 arcmin$^2$ (\OIII; $z=0.3$--0.65).
About one-fifth of these galaxies will have \OH\ $\leq$ 7.65.

The higher rate of XMPGs ($\approx$1\%) among star-forming
galaxies at high redshift compared to the local universe (recall
1 in 1700 SDSS galaxies are XMPGs) could be a reflection of galaxy
evolution. In particular, several lines of evidence have shown
that the SFRs and the sSFR's are an order magnitude higher at
$z\gtrsim1$ than the local universe
\citep[e.g.,][]{noeske07,newha,OIIpop,lee12}.
Future surveys at $z\gtrsim1$ with JWST and ``WFIRST'' should find
many more XMPGs simply because: (1) the \MZ\ relation evolves toward
lower metallicity at high redshift \citep[see e.g.,][]{erb06}; and
(2) the epoch of where the SFR density peaked
(\citealt{hopkins04}; \citetalias{ly07}), and where most of the
mass assembly occurred \citep{dickinson03,rudnick03},
is at $z\sim1$--3.

\cite{morales11} reported that the number density of XMPGs in the
local universe from the SDSS is $\approx1.3\times10^{-4}$ Mpc$^{-3}$,
which is similar to what we find in our study. This comparison would
suggest that no evolution is seen in the number density of XMPGs.
However, we caution against making such a comparison since the local
estimate is highly uncertain due to a statistically small sample
that suffers from incompleteness and complex selection functions.

\section{CONCLUSIONS}\label{sec:Conclusions}
We have conducted an extensive optical spectroscopic survey of
nearly 900 ELGs identified in the SDF from NB or IA imaging with
Keck/DEIMOS and MMT/Hectospec. 
Upon inspecting our spectra, we found that we have \Ndet\ galaxies
where the weak auroral \OIIIa\ line is detected at $\geq$3$\sigma$.
These detections allow us to determine the electron temperature
of the ionized gas, and directly determine gas-phase metallicity,
using the \Te\ method.
Excluding our one object which we believe is a LINER, we find that
the remaining \Ndetf\ galaxies are metal-poor with a median \OH\ of
7.75 (0.13 \zsun), and the most metal-rich galaxy having
\OH\ = 8.40 (0.5 \zsun). We find that \NXMPG\ of our galaxies
are XMPG (\OH\ $\leq$ 7.65).

Our most metal-deficient galaxy with a $\geq$4$\sigma$ detection
of \OIIIa\ has an oxygen abundance (\OH\ = \OHKecksix) that
is similar to I Zw 18 and SBS0335.
This galaxy lacks an \OII\ detection, which is caused by a
high ionization parameter and low metallicity that drive most
of the oxygen species to be in a doubly ionized state.
In addition, we find that all of our galaxies have high
ionization, as measured by \OIII/\Hb, and high electron
density ($n_e$ = 70--600 cm$^{-3}$).
While this is not typical for SDSS ELGs, there are rare
populations of highly star-forming low-$z$ galaxies found in the
SDSS that have high ionization. In addition, these properties
are seen for strongly star-forming $z\gtrsim1$ galaxies.
This suggests that XMPGs at $z\sim0.4$--1 are analogs to high-$z$
star-forming galaxies in terms of their ISM properties.

Our exceptionally deep and complete multi-wavelength dataset in the
SDF also allows us to determine other key properties of these
galaxies.
What we found is that these galaxies are exceptional in several ways.

First, with stellar masses between $5\times10^6$ and $3\times10^9$
\Msun\ (with a median and average of \MassM\ and \MassA\ \Msun,
respectively), we find that our galaxies have high specific SFRs,
which suggests that these galaxies have assembled their stellar
contents in only $\sim$\sSFRt.
Second, their vigorous star formation is highly concentrated into a
small central starburst of high surface density (an average of
$\Sigma_{\rm SFR}$ = \SFRSDA\ \Msun\ yr$^{-1}$ kpc$^{-2}$).
Third, they have an excess probability of having nearby
companions (within a projected \Dproj), which we have
determined to have strong emission lines at a similar redshift.
The presence of nearby companions in two-thirds of our galaxies
suggests that galaxy-galaxy interactions may be responsible
for the metal deficiency and the observed high sSFRs and high SFR
surface densities. In particular, outflowing or inflowing
gas can occur through stellar feedback that drives metal-rich
gas out of the ISM or gravitational torques that can cause
metal-poor gas to lose angular momentum and fall into the
center of the galaxies.

Some researchers have proposed that the \MZ\ relation
must be extended to include a third parameter, the SFR
\citep{laralopez10,mannucci10}.
Since these relations predict lower metallicities for galaxies with
unusually high sSFRs, about half of our galaxies are indeed
consistent with the FMR. 
However, there are two inconsistencies that we find. First,
our most star-forming galaxies (3 of them) have higher
metallicity than what is expected if they followed the FMR.
This suggests a flatter or positive correlation between
sSFR and $Z$, contrary to an inverse correlation found for the
FMR. Second, it appears that our sample has systematically lower
metallicity by $\approx$0.2 dex than the FMR. This result is only
robust at 2$\sigma$ confidence, and we suspect that selection
effects (i.e., required detection of \OIIIa) may be biasing our
sample toward lower metallicity.

The failure of any simple relation to predict O/H can be seen by the
large scatter of our metallicities versus stellar mass or sSFR in
Figure~\ref{fig:M_Z}. The large intrinsic scatter overwhelms any other
trends with metallicity which might be present.
For example, our sample contains \Ndetf\ galaxies with redshifts ranging
from $z$=0.38--0.88 (excluding the lowest redshift galaxy), a range of
look-back times from 4 to 7 Gyr.
This period of cosmic history is known to span a great deal of evolution
in many properties of the overall galaxy population as a whole.
Nonetheless our galaxy sample shows no trend of metallicity with redshift.
Across all of these redshifts, we find galaxies with half solar
abundances mixed in with very metal-poor galaxies.  We believe that
this large cosmic scatter of abundances is larger than our measurement
uncertainties, and is genuine.

\acknowledgements
The DEIMOS data presented herein were obtained at the W.M. Keck Observatory,
which is operated as a scientific partnership among the California Institute
of Technology, the University of California and the National Aeronautics
and Space Administration (NASA). The Observatory was made possible by the
generous financial support of the W.M. Keck Foundation.
The authors wish to recognize and acknowledge the very significant cultural
role and reverence that the summit of Mauna Kea has always had within the
indigenous Hawaiian community. We are most fortunate to have the opportunity
to conduct observations from this mountain.
Hectospec observations reported here were obtained at the MMT Observatory,
a joint facility of the Smithsonian Institution and the University of Arizona.
MMT telescope time was granted by NOAO, through the NSF-funded
Telescope System Instrumentation Program (TSIP).
We gratefully acknowledge NASA's support for construction, operation,
and science analysis for the GALEX mission.
This work is based in part on observations made with the Spitzer Space
Telescope, which is operated by the Jet Propulsion Laboratory, California
Institute of Technology under a contract with NASA.

C.L. acknowledges financial support through the Giacconi Fellowship program and
T.N. acknowledges financial support by JSPS (grant nos. 23654068 and 25707010).
This research was supported by the Japan Society for the Promotion of
Science through Grant-in-Aid for Scientific Research 23340050.

We thank the anonymous referee for their comments that improve the paper.
We thank B. Andrews, D. Berg, R. Dave, A. Henry, L. Kewley, J. Lotz,
K. Noeske, M. Pena-Guerrero, and R. Ryan for interesting discussions
that help improve this paper.
We also thank K. Bundy for providing proprietary stellar mass
estimates for DEEP2 galaxies.

{\it Facilities:} \facility{Subaru (Suprime-Cam)}, \facility{MMT (Hectospec)},
\facility{Keck:II (DEIMOS)}, \facility{GALEX},
\facility{Mayall (MOSAIC, NEWFIRM)}, \facility{UKIRT (WFCAM)},
\facility{Spitzer (IRAC)}, \facility{HST (WFC3)}

\clearpage
\newcommand{\phc}{\phantom{$<$}}
\begin{landscape}
\begin{deluxetable*}{lcccccccccccc}
  \tabletypesize{\scriptsize}
  \tablewidth{0pc}
  \tablecaption{Emission-line Ratios, Electron Temperatures, and Gas-Phase Metallicities}
  \tablehead{
    \colhead{ID}&
    \multicolumn{2}{c}{$\log{\left(\frac{\OIII}{\OIII\ \lambda4363}\right)}$}&
    \multicolumn{2}{c}{$\log{(T_e/{\rm K})}$}&
    \multicolumn{2}{c}{$\log{\left(\frac{\OII}{\Hbe}\right)}$}&
    \colhead{$\log{\left(\frac{\OIII}{\Hbe}\right)}$}&
    \multicolumn{2}{c}{$\log{\left(\frac{{\rm O}^+}{{\rm H}^+}\right)}$}&
    \colhead{$\log{\left(\frac{{\rm O}^{++}}{{\rm H}^+}\right)}$}&
    \multicolumn{2}{c}{$12+\log{\left(\frac{{\rm O}}{{\rm H}}\right)}$}\\
    \cline{2-3}
    \cline{4-5}
    \cline{6-7}
    \cline{9-10}
    \cline{12-13}
    \colhead{(1)}&\colhead{(2)}&\colhead{(3)}&\colhead{(4)}&
    \colhead{(5)}&\colhead{(6)}&\colhead{(7)}&\colhead{(8)}&
    \colhead{(9)}&\colhead{(10)}&\colhead{(11)}&\colhead{(12)}&\colhead{(13)}
  }
   \startdata
   MMT\#01    & 1.52$^{+0.24}_{-0.17}$ & 1.46$^{+0.23}_{-0.21}$ & 4.32$^{+0.12}_{-0.13}$ & 4.37$^{+0.16}_{-0.15}$ & +0.06$^{+0.09}_{-0.08}$ & +0.18$^{+0.20}_{-0.17}$ & 0.89$^{+0.08}_{-0.07}$ & -4.99$^{+0.50}_{-0.34}$ & -4.84$^{+0.54}_{-0.39}$ & -4.61$^{+0.29}_{-0.30}$ & 7.60$^{+0.37}_{-0.26}$ & 7.59$^{+0.40}_{-0.33}$\\[1.5mm]
   MMT\#02    & 1.61$^{+0.40}_{-0.24}$ & 1.55$^{+0.39}_{-0.28}$ & 4.27$^{+0.16}_{-0.18}$ & 4.31$^{+0.22}_{-0.21}$ & +0.50$^{+0.04}_{-0.04}$ & +0.62$^{+0.29}_{-0.17}$ & 0.64$^{+0.04}_{-0.04}$ & -4.54$^{+0.70}_{-0.44}$ & -4.43$^{+0.74}_{-0.49}$ & -4.74$^{+0.47}_{-0.38}$ & 7.71$^{+0.60}_{-0.42}$ & 7.74$^{+0.64}_{-0.43}$\\[1.5mm]
   MMT\#03 & 1.17$^{+0.24}_{-0.21}$ & 1.09$^{+0.29}_{-0.27}$ & 4.60$^{+0.10}_{-0.22}$ & 4.65$^{+0.09}_{-0.20}$ & +0.56$^{+0.05}_{-0.04}$ & +0.74$^{+0.21}_{-0.20}$ & 0.34$^{+0.05}_{-0.05}$ & \ldots\TA & \ldots\TA & \ldots\TA & \ldots\TA & \ldots\TA\\[1.5mm]
   MMT\#04    & 1.95$^{+0.14}_{-0.12}$ & 1.95$^{+0.14}_{-0.13}$ & 4.11$^{+0.05}_{-0.05}$ & 4.11$^{+0.05}_{-0.05}$ & +0.36$^{+0.01}_{-0.01}$ & +0.36$^{+0.03}_{-0.01}$ & 0.77$^{+0.01}_{-0.01}$ & -4.37$^{+0.20}_{-0.18}$ & -4.37$^{+0.21}_{-0.17}$ & -4.12$^{+0.15}_{-0.15}$ & 8.07$^{+0.17}_{-0.15}$ & 8.07$^{+0.17}_{-0.16}$\\[1.5mm]
   MMT\#05    & 1.66$^{+0.19}_{-0.17}$ & 1.66$^{+0.19}_{-0.16}$ & 4.24$^{+0.08}_{-0.12}$ & 4.24$^{+0.10}_{-0.12}$ & -0.27$^{+0.09}_{-0.10}$ & -0.27$^{+0.15}_{-0.10}$ & 0.92$^{+0.03}_{-0.03}$ & -5.27$^{+0.36}_{-0.27}$ & -5.27$^{+0.38}_{-0.25}$ & -4.32$^{+0.22}_{-0.24}$ & 7.72$^{+0.26}_{-0.20}$ & 7.72$^{+0.26}_{-0.24}$\\[1.5mm]
   MMT\#06    & 1.42$^{+0.41}_{-0.25}$ & 1.32$^{+0.41}_{-0.25}$ & 4.32$^{+0.22}_{-0.33}$ & 4.40$^{+0.19}_{-0.38}$ & +0.54$^{+0.06}_{-0.05}$ & +0.73$^{+0.07}_{-0.06}$ & 0.54$^{+0.05}_{-0.05}$ & -4.52$^{+0.89}_{-0.46}$ & -4.24$^{+1.16}_{-0.51}$ & -5.01$^{+0.72}_{-0.27}$ & 7.65$^{+0.81}_{-0.46}$ & 7.83$^{+1.11}_{-0.39}$\\[1.5mm]
   MMT\#07    & 1.84$^{+0.11}_{-0.08}$ & 1.78$^{+0.12}_{-0.13}$ & 4.15$^{+0.04}_{-0.05}$ & 4.18$^{+0.07}_{-0.06}$ & -0.02$^{+0.01}_{-0.02}$ & +0.10$^{+0.17}_{-0.13}$ & 0.87$^{+0.01}_{-0.01}$ & -4.87$^{+0.17}_{-0.13}$ & -4.81$^{+0.19}_{-0.15}$ & -4.23$^{+0.16}_{-0.18}$ & 7.92$^{+0.13}_{-0.12}$ & 7.87$^{+0.16}_{-0.15}$\\[1.5mm]
   MMT\#08    & 1.37$^{+0.31}_{-0.24}$ & 1.26$^{+0.41}_{-0.33}$ & 4.31$^{+0.22}_{-0.35}$ & 4.40$^{+0.20}_{-0.45}$ & +0.40$^{+0.09}_{-0.10}$ & +0.61$^{+0.36}_{-0.27}$ & 0.58$^{+0.10}_{-0.10}$ & -4.65$^{+0.79}_{-0.49}$ & -4.36$^{+1.08}_{-0.43}$ & -4.97$^{+0.61}_{-0.29}$ & 7.58$^{+0.70}_{-0.46}$ & 7.74$^{+0.99}_{-0.36}$\\[1.5mm]
   MMT\#09    & 1.43$^{+0.29}_{-0.22}$ & 1.34$^{+0.29}_{-0.26}$ & 4.32$^{+0.19}_{-0.28}$ & 4.40$^{+0.20}_{-0.26}$ & +0.39$^{+0.10}_{-0.09}$ & +0.58$^{+0.13}_{-0.12}$ & 0.76$^{+0.09}_{-0.09}$ & -4.66$^{+0.64}_{-0.51}$ & -4.39$^{+0.79}_{-0.47}$ & -4.79$^{+0.42}_{-0.33}$ & 7.65$^{+0.55}_{-0.42}$ & 7.76$^{+0.72}_{-0.43}$\\[1.5mm]
   MMT\#10    & 1.75$^{+0.33}_{-0.21}$ & 1.72$^{+0.33}_{-0.21}$ & 4.20$^{+0.10}_{-0.14}$ & 4.21$^{+0.11}_{-0.15}$ & +0.09$^{+0.04}_{-0.04}$ & +0.14$^{+0.04}_{-0.06}$ & 0.82$^{+0.02}_{-0.02}$ & -4.85$^{+0.51}_{-0.32}$ & -4.82$^{+0.47}_{-0.32}$ & -4.35$^{+0.38}_{-0.30}$ & 7.80$^{+0.41}_{-0.25}$ & 7.78$^{+0.46}_{-0.33}$\\[1.5mm]
   MMT\#11    & 1.85$^{+0.21}_{-0.15}$ & 1.80$^{+0.21}_{-0.16}$ & 4.15$^{+0.08}_{-0.08}$ & 4.17$^{+0.08}_{-0.09}$ & +0.29$^{+0.03}_{-0.03}$ & +0.39$^{+0.04}_{-0.03}$ & 0.82$^{+0.02}_{-0.02}$ & -4.55$^{+0.29}_{-0.27}$ & -4.50$^{+0.31}_{-0.27}$ & -4.24$^{+0.27}_{-0.20}$ & 7.97$^{+0.26}_{-0.21}$ & 7.95$^{+0.30}_{-0.22}$\\[1.5mm]
   MMT\#12    & 1.45$^{+0.42}_{-0.24}$ & 1.32$^{+0.54}_{-0.44}$ & 4.29$^{+0.21}_{-0.29}$ & 4.40$^{+0.24}_{-0.46}$ & +0.58$^{+0.14}_{-0.12}$ & +0.84$^{+0.60}_{-0.36}$ & 0.81$^{+0.13}_{-0.12}$ & -4.46$^{+0.90}_{-0.46}$ & -4.13$^{+1.24}_{-0.42}$ & -4.74$^{+0.80}_{-0.42}$ & 7.80$^{+0.78}_{-0.43}$ & 7.96$^{+1.14}_{-0.46}$\\[1.5mm]
   MMT\#13    & 1.57$^{+0.37}_{-0.20}$ & 1.52$^{+0.31}_{-0.27}$ & 4.30$^{+0.15}_{-0.20}$ & 4.32$^{+0.19}_{-0.23}$ & -0.02$^{+0.08}_{-0.08}$ & +0.07$^{+0.23}_{-0.16}$ & 0.78$^{+0.03}_{-0.04}$ & -5.06$^{+0.57}_{-0.46}$ & -4.98$^{+0.69}_{-0.45}$ & -4.64$^{+0.43}_{-0.34}$ & 7.54$^{+0.52}_{-0.30}$ & 7.53$^{+0.52}_{-0.33}$\\[1.5mm]
   MMT\#14    & 2.22$^{+0.12}_{-0.10}$ & 2.21$^{+0.15}_{-0.11}$ & 4.01$^{+0.03}_{-0.04}$ & 4.01$^{+0.04}_{-0.04}$ & +0.25$^{+0.01}_{-0.01}$ & +0.27$^{+0.03}_{-0.02}$ & 0.79$^{+0.01}_{-0.01}$ & -4.11$^{+0.20}_{-0.15}$ & -4.10$^{+0.17}_{-0.13}$ & -3.80$^{+0.16}_{-0.10}$ & 8.38$^{+0.16}_{-0.12}$ & 8.37$^{+0.17}_{-0.11}$\\[1.5mm]
   Keck\#1    & 1.92$^{+0.08}_{-0.08}$ & 1.81$^{+0.09}_{-0.09}$ & 4.12$^{+0.03}_{-0.03}$ & 4.17$^{+0.04}_{-0.04}$ & -0.22$^{+0.03}_{-0.03}$ & -0.02$^{+0.06}_{-0.08}$ & 0.90$^{+0.01}_{-0.01}$ & -4.99$^{+0.13}_{-0.12}$ & -4.89$^{+0.14}_{-0.13}$ & -4.16$^{+0.10}_{-0.09}$ & 8.01$^{+0.09}_{-0.08}$ & 7.91$^{+0.12}_{-0.10}$\\[1.5mm]
   Keck\#2    & 1.44$^{+0.15}_{-0.13}$ & 1.44$^{+0.15}_{-0.13}$ & 4.38$^{+0.09}_{-0.11}$ & 4.38$^{+0.09}_{-0.11}$ & +0.14$^{+0.04}_{-0.04}$ & +0.14$^{+0.04}_{-0.04}$ & 0.84$^{+0.02}_{-0.02}$ & -4.86$^{+0.36}_{-0.30}$ & -4.86$^{+0.36}_{-0.30}$ & -4.68$^{+0.20}_{-0.19}$ & 7.54$^{+0.27}_{-0.22}$ & 7.54$^{+0.27}_{-0.22}$\\[1.5mm]
   Keck\#3    & 2.30$^{+0.21}_{-0.18}$ & 2.14$^{+0.22}_{-0.16}$ & 3.98$^{+0.05}_{-0.06}$ & 4.03$^{+0.06}_{-0.08}$ & +0.35$^{+0.01}_{-0.01}$ & +0.66$^{+0.04}_{-0.04}$ & 0.68$^{+0.01}_{-0.01}$ & -3.90$^{+0.33}_{-0.23}$ & -3.82$^{+0.28}_{-0.23}$ & -4.00$^{+0.26}_{-0.18}$ & 8.44$^{+0.29}_{-0.22}$ & 8.40$^{+0.31}_{-0.22}$\\[1.5mm]
   Keck\#4    & 2.02$^{+0.29}_{-0.23}$ & 2.02$^{+0.29}_{-0.23}$ & 4.08$^{+0.09}_{-0.11}$ & 4.08$^{+0.09}_{-0.11}$ & +0.27$^{+0.02}_{-0.02}$ & +0.27$^{+0.02}_{-0.02}$ & 0.68$^{+0.01}_{-0.01}$ & -4.37$^{+0.46}_{-0.34}$ & -4.37$^{+0.46}_{-0.34}$ & -4.13$^{+0.36}_{-0.26}$ & 8.07$^{+0.41}_{-0.29}$ & 8.07$^{+0.41}_{-0.29}$\\[1.5mm]
   Keck\#5    & 1.16$^{+0.35}_{-0.24}$ & 1.10$^{+0.37}_{-0.25}$ & 4.36$^{+0.16}_{-0.77}$ & 4.40$^{+0.14}_{-0.76}$ & +0.24$^{+0.05}_{-0.05}$ & +0.36$^{+0.22}_{-0.16}$ & 0.88$^{+0.03}_{-0.04}$ & -4.79$^{+1.04}_{-0.28}$ & -4.61$^{+1.10}_{-0.33}$ & -4.67$^{+0.60}_{-0.15}$ & 7.61$^{+0.86}_{-0.26}$ & 7.66$^{+0.90}_{-0.19}$\\[1.5mm]
   Keck\#6    & 1.41$^{+0.24}_{-0.18}$ & 1.35$^{+0.32}_{-0.23}$ & 4.35$^{+0.16}_{-0.24}$ & 4.40$^{+0.17}_{-0.27}$ & -1.14$^{+0.51}_{-0.83}$\TB & -1.02$^{+0.59}_{-0.94}$\TB & 0.77$^{+0.03}_{-0.03}$ & -6.18$^{+0.99}_{-1.02}$\TB & -5.99$^{+1.06}_{-1.03}$\TB & -4.78$^{+0.39}_{-0.29}$ & 7.32$^{+0.47}_{-0.25}$\TB & 7.24$^{+0.45}_{-0.30}$\TB\\[1.5mm]
\vspace{-3mm}
\enddata
\label{tab:metals}
\tablenotetext{1}{This source is likely a LINER, so metallicity determinations are not trustworthy (see Section \ref{sec:AGN}).}
  \tablenotetext{2}{The 1$\sigma$ upper limit of the \OII\ flux is adopted.}
  \tablecomments{Columns (3), (5), (7), (10), and (13) include dust attenution corrections (see Section~\ref{sec:dust}).
    Errors are reported at the 95\% C.L.}
  
\end{deluxetable*}
\clearpage
\end{landscape}

\begin{center}
  \begin{deluxetable*}{lccccccccccccc}
  \tabletypesize{\scriptsize}
  \tablewidth{0pc}
  \tablecaption{Luminosities, Stellar Properties, and Star Formation Rates}
  \tablehead{
    \colhead{ID}&
    \colhead{\MB}&
    \multicolumn{2}{c}{$\log{\left[\frac{{\rm SFR}(\Hae)}{M_{\sun}/{\rm yr}}\right]}$}&
    \multicolumn{2}{c}{$\log{\left[\frac{{\rm SFR}(\Hbe)}{M_{\sun}/{\rm yr}}\right]}$}&
    \multicolumn{2}{c}{$\log{({\rm M}_{\star}/M_{\sun})}$\TA}&
    \multicolumn{2}{c}{$\log{(t_{\rm age}/{\rm yr})}$\TA}&
    \multicolumn{3}{c}{$A_V$\TA}\\
    \cline{3-4}
    \cline{5-6}
    \cline{7-8}
    \cline{9-10}
    \cline{11-12}
    & & \colhead{Obs.}&\colhead{De-red.} &\colhead{Obs.}&\colhead{De-red.}&
    \colhead{0.2$Z_{\sun}$} & \colhead{$Z_{\sun}$} & \colhead{0.2$Z_{\sun}$} &
    \colhead{$Z_{\sun}$} & \colhead{0.2$Z_{\sun}$} & \colhead{$Z_{\sun}$}\\\hline
    \colhead{(1)}&\colhead{(2)}&\colhead{(3)}&\colhead{(4)}&\colhead{(5)}&
    \colhead{(6)}&\colhead{(7)}&\colhead{(8)}&\colhead{(9)}&\colhead{(10)}&
    \colhead{(11)}&\colhead{(12)}}
  \startdata
MMT\#01    & --18.67 & \ldots    & \ldots                     &  +0.32 & +0.76$^{+0.38}_{-0.38}$ & 7.99$^{+0.08}_{-0.06}$ & 8.23$^{+0.08}_{-0.33}$ & 7.40$^{+0.17}_{-0.29}$ & 8.10$^{+0.24}_{-0.98}$ & 0.6$^{+0.1}_{-0.1}$ & 0.3$^{+0.5}_{-0.2}$\\[1mm]
MMT\#02    & --19.28 & \ldots    & \ldots                     & --0.29 & +0.17$^{+0.37}_{-0.41}$ & 9.00$^{+0.07}_{-0.20}$ & 9.23$^{+0.01}_{-0.09}$ & 8.60$^{+0.22}_{-0.93}$ & 9.00$^{+0.12}_{-0.29}$ & 0.9$^{+0.3}_{-0.1}$ & 0.7$^{+0.2}_{-0.1}$\\[1mm]
MMT\#03\TB & --20.15 &  +0.35    & +0.81$^{+0.35}_{-0.32}$    & \ldots & \ldots                  & 9.40$^{+0.11}_{-0.17}$ & 9.59$^{+0.09}_{-0.04}$ & 8.50$^{+0.30}_{-0.80}$ & 8.90$^{+0.24}_{-0.20}$ & 1.1$^{+0.2}_{-0.2}$ & 0.9$^{+0.1}_{-0.1}$\\[1mm]
MMT\#04    & --19.07 &  +0.32    & +0.32$^{+0.06}_{-0.06}$    & \ldots & \ldots                  & 8.44$^{+0.02}_{-0.05}$ & 8.60$^{+0.08}_{-0.11}$ & 7.50$^{+0.35}_{-0.10}$ & 8.20$^{+0.32}_{-0.50}$ & 0.7$^{+0.2}_{-0.3}$ & 0.6$^{+0.1}_{-0.3}$\\[1mm]
MMT\#05    & \ldots & \ldots    & \ldots                     & --0.34 &--0.34$^{+0.39}_{-0.38}$ & \ldots                 & \ldots & \ldots                 & \ldots & \ldots              & \ldots\\[1mm]
MMT\#06    & --18.83 & --0.15    & +0.37$^{+0.05}_{-0.05}$    & \ldots & \ldots                  & 8.74$^{+0.12}_{-0.16}$ & 9.01$^{+0.02}_{-0.09}$ & 8.40$^{+0.40}_{-0.71}$ & 9.00$^{+0.12}_{-0.24}$ & 0.9$^{+0.2}_{-0.2}$ & 0.6$^{+0.1}_{-0.1}$\\[1mm]
MMT\#07    & --17.61 & --0.00\TC & +0.32$^{+0.27}_{-0.27}$\TC & \ldots & \ldots                  & 7.66$^{+0.00}_{-0.00}$ & 7.62$^{+0.00}_{-0.00}$ & 7.40$^{+0.00}_{-0.00}$ & 7.40$^{+0.00}_{-0.00}$ & 0.5$^{+0.0}_{-0.0}$ & 0.4$^{+0.0}_{-0.0}$\\[1mm]
MMT\#08    & --20.13 & \ldots    & \ldots                     &  +0.38 & +1.15$^{+0.79}_{-0.84}$ & 9.27$^{+0.06}_{-0.42}$ & 9.35$^{+0.11}_{-0.00}$ & 8.20$^{+0.29}_{-1.20}$ & 8.30$^{+0.42}_{-0.14}$ & 1.2$^{+0.4}_{-0.4}$ & 0.8$^{+0.3}_{-0.0}$\\[1mm]
MMT\#09    & --19.50 &  +0.06    & +0.56$^{+0.12}_{-0.13}$    & \ldots & \ldots                  & 8.93$^{+0.08}_{-0.14}$ & 9.14$^{+0.09}_{-0.02}$ & 8.50$^{+0.34}_{-0.80}$ & 8.90$^{+0.29}_{-0.12}$ & 0.7$^{+0.1}_{-0.2}$ & 0.5$^{+0.1}_{-0.2}$\\[1mm]
MMT\#10    & --14.57 & --1.52    &--1.37$^{+0.02}_{-0.03}$    & \ldots & \ldots                  & 6.72$^{+0.09}_{-0.03}$ & 7.01$^{+0.33}_{-0.04}$ & 7.10$^{+0.31}_{-0.10}$ & 7.80$^{+1.28}_{-0.04}$ & 1.6$^{+0.1}_{-0.2}$ & 0.7$^{+0.2}_{-0.3}$\\[1mm]
MMT\#11    & --17.28 & --0.52    &--0.26$^{+0.04}_{-0.03}$    & \ldots & \ldots                  & 7.86$^{+0.09}_{-0.11}$ & 8.09$^{+0.18}_{-0.08}$ & 7.80$^{+0.39}_{-0.52}$ & 8.50$^{+0.54}_{-0.44}$ & 0.9$^{+0.3}_{-0.4}$ & 0.6$^{+0.2}_{-0.3}$\\[1mm]
MMT\#12    & --19.92 & \ldots    & \ldots                     &  +0.14 & +1.08$^{+1.33}_{-0.84}$ & 8.94$^{+0.08}_{-0.09}$ & 9.11$^{+0.09}_{-0.09}$ & 8.10$^{+0.42}_{-0.44}$ & 8.60$^{+0.22}_{-0.39}$ & 0.6$^{+0.3}_{-0.4}$ & 0.5$^{+0.1}_{-0.2}$\\[1mm]
MMT\#13    & --17.52 & --0.11\TC & +0.12$^{+0.35}_{-0.31}$\TC & \ldots & \ldots                  & 7.39$^{+0.01}_{-0.09}$ & 7.28$^{+0.32}_{-0.09}$ & 7.30$^{+0.10}_{-0.10}$ & 7.10$^{+0.96}_{-0.10}$ & 0.3$^{+0.1}_{-0.0}$ & 0.5$^{+0.1}_{-0.5}$\\[1mm]
MMT\#14    & --20.25 &  +1.26\TC & +1.30$^{+0.04}_{-0.04}$\TC & \ldots & \ldots                  & 8.65$^{+0.04}_{-0.06}$ & 8.49$^{+0.01}_{-0.03}$ & 7.30$^{+0.12}_{-0.08}$ & 7.10$^{+0.05}_{-0.10}$ & 0.6$^{+0.0}_{-0.1}$ & 0.7$^{+0.1}_{-0.1}$\\[1mm]
Keck\#1    & --18.08 & \ldots    & \ldots                     &  +0.52 & +1.28$^{+0.14}_{-0.13}$ & 7.57$^{+0.00}_{-0.11}$ & 7.39$^{+0.00}_{-0.07}$ & 7.30$^{+0.07}_{-0.08}$ & 7.00$^{+0.04}_{-0.00}$ & 0.0$^{+0.1}_{-0.0}$ & 0.2$^{+0.0}_{-0.1}$\\[1mm]
Keck\#2    & --18.38 & \ldots    & \ldots                     & --0.29 &--0.29                   & 7.95$^{+0.18}_{-0.19}$ & 8.29$^{+0.05}_{-0.14}$ & 7.30$^{+0.67}_{-0.30}$ & 8.30$^{+0.21}_{-0.60}$ & 0.9$^{+0.1}_{-0.3}$ & 0.4$^{+0.2}_{-0.4}$\\[1mm]
Keck\#3    & --21.12 & \ldots    & \ldots                     &  +0.67 & +1.81$^{+0.08}_{-0.07}$ & 9.42$^{+0.03}_{-0.06}$ & 9.46$^{+0.01}_{-0.05}$ & 7.50$^{+0.41}_{-0.17}$ & 7.70$^{+0.12}_{-0.06}$ & 1.0$^{+0.3}_{-0.8}$ & 0.6$^{+0.0}_{-0.3}$\\[1mm]
Keck\#4    & --19.45 & \ldots    & \ldots                     &  +0.55 & +0.55$^{+0.12}_{-0.12}$ & 8.60$^{+0.04}_{-0.15}$ & 8.47$^{+0.05}_{-0.09}$ & 7.30$^{+0.28}_{-0.23}$ & 7.10$^{+0.21}_{-0.10}$ & 1.0$^{+0.1}_{-0.1}$ & 1.2$^{+0.1}_{-0.1}$\\[1mm]
Keck\#5    & --18.05 & \ldots    & \ldots                     & --0.35 & +0.09$^{+0.38}_{-0.38}$ & 8.23$^{+0.18}_{-0.14}$ & 8.47$^{+0.18}_{-0.12}$ & 8.70$^{+0.44}_{-0.64}$ & 9.10$^{+0.20}_{-0.38}$ & 0.2$^{+0.4}_{-0.2}$ & 0.0$^{+0.2}_{-0.0}$\\[1mm]
Keck\#6    & --17.16 & \ldots    & \ldots                     & --0.42 & +0.02$^{+0.38}_{-0.38}$ & 7.52$^{+0.11}_{-0.15}$ & 7.45$^{+0.06}_{-0.07}$ & 7.30$^{+0.18}_{-0.30}$ & 7.00$^{+0.10}_{-0.00}$ & 1.1$^{+0.2}_{-0.2}$ & 1.3$^{+0.1}_{-0.1}$\\[1mm]
  \vspace{-3mm}
  \enddata
  \label{tab:SFR_Mass}
  \tablecomments{All values assume a \citetalias{chabrier03} IMF, and where applicable, dust attenuation follows \citetalias{calzetti00} formalism.
    68\% confidence measurement uncertainties are reported.}
  \tablenotetext{1}{One-fifth solar metallicity results are given Cols. (7), (9), and (11) while those obtained with solar metallicity models are provided in Cols. (8), (10), and (12).}
  \tablenotetext{2}{This source is likely a LINER, so SFR estimates are not trustworthy.}
  \tablenotetext{3}{Since the \Ha\ line falls at the edges of the filters, these
      \Ha\ SFRs are affected by accurate tracing of the NB filter profiles.}
\end{deluxetable*}
\end{center}

\clearpage
\begin{landscape}
\begin{deluxetable*}{cccccccccccccccc}
  \tabletypesize{\scriptsize}
  \tablewidth{0pc}
  \tablecaption{Spectral Energy Distributions}
  \tablehead{
    \colhead{ID}&
    \colhead{\FUV}&
    \colhead{\NUV}&
    \colhead{$U$}&
    \colhead{$B$}&
    \colhead{$V$}&
    \colhead{IA598}&
    \colhead{\Rc}&
    \colhead{IA679}&
    \colhead{$i$\arcmin}&
    \colhead{$z_b$}&
    \colhead{$z$\arcmin}&
    \colhead{$z_r$}&
    \colhead{$J$}&
    \colhead{$H$}&
    \colhead{$K$}\\
    \colhead{(1)}&\colhead{(2)}&\colhead{(3)}&\colhead{(4)}&\colhead{(5)}&\colhead{(6)}&\colhead{(7)}&
    \colhead{(8)}&\colhead{(9)}&\colhead{(10)}&\colhead{(11)}&\colhead{(12)}&\colhead{(13)}&\colhead{(14)}&\colhead{(15)}&\colhead{(16)}}
  \startdata
  MMT\#01 &  0.26$\pm$0.04 &  0.93$\pm$0.03 &  0.97$\pm$0.03 &  0.89 &  1.00 &  1.37$\pm$0.03 &  1.41 &  1.24$\pm$0.02 &  2.36 &  1.14 &  1.16 &  1.06$\pm$0.09 &  1.00$\pm$0.47 &  2.75$\pm$1.89 &  1.44$\pm$0.26\\
  \ldots  & \ldots & \ldots & \ldots & \ldots & \ldots &  -0.43 &  -0.28 &  -0.11 &  -1.09 & \ldots &  -0.04 & \ldots & \ldots & \ldots & \ldots\\[1mm]
  MMT\#02 &  0.62$\pm$0.05 &  1.46$\pm$0.04 &  1.86$\pm$0.03 &  2.01 &  3.57 &  4.51$\pm$0.03 &  4.97 &  5.01$\pm$0.02 &  5.61 &  5.93 &  6.59 &  7.17$\pm$0.10 &  9.46$\pm$0.66 & 10.13$\pm$1.63 & \ldots\\
  \ldots  & \ldots & \ldots & \ldots & \ldots &  -0.22 & \ldots &  -0.18 &  -0.09 &  -0.37 & \ldots &  -0.24 &  -0.08 & \ldots & \ldots & \ldots\\[1mm]
  MMT\#03 &  0.70$\pm$0.05 &  1.96$\pm$0.04 &  2.88$\pm$0.03 &  3.07 &  5.24 &  7.44$\pm$0.03 &  8.31 &  8.78$\pm$0.02 &  9.51 & 10.86 & 11.35 & 15.18$\pm$0.10 & 17.15$\pm$0.72 & 22.32$\pm$1.84 & \ldots\\ 
  \ldots  & \ldots & \ldots & \ldots & \ldots &  -0.25 &  -0.05 &  -0.05 & \ldots &  -0.24 & \ldots &  -0.67 &  -3.16 & \ldots & \ldots & \ldots\\ 
  MMT\#04 &  1.73$\pm$0.07 &  2.86$\pm$0.05 &  3.28$\pm$0.03 &  3.04 &  4.76 &  4.98$\pm$0.03 &  7.03 &  8.38$\pm$0.02 &  5.97 &  7.53 &  7.00 &  5.82$\pm$0.10 &  9.16$\pm$0.59 & 10.11$\pm$1.23 &  7.24$\pm$0.28\\
  \ldots  & \ldots & \ldots & \ldots & \ldots &  -0.69 &  -0.43 &  -1.97 &  -3.41 &  -1.09 &  -2.39 &  -1.82 &  -0.01 & \ldots & \ldots & \ldots\\
  MMT\#06 &  0.64$\pm$0.05 &  1.40$\pm$0.06 &  1.64$\pm$0.03 &  1.74 &  3.15 &  3.69$\pm$0.03 &  4.28 &  4.52$\pm$0.02 &  4.51 &  5.57 &  5.60 &  5.41$\pm$0.10 &  7.45$\pm$0.71 & 10.57$\pm$1.82 &  7.19$\pm$0.28\\ 
  \ldots  & \ldots & \ldots & \ldots & \ldots &  -0.23 &  -0.17 &  -0.37 &  -0.51 &  -0.22 &  -0.66 &  -0.57 & \ldots & \ldots & \ldots & \ldots\\ 
  MMT\#07 &  0.63$\pm$0.04 &  1.11$\pm$0.04 &  1.26$\pm$0.03 &  1.19 &  1.70 &  1.69$\pm$0.03 &  3.62 &  5.69$\pm$0.02 &  2.44 &  3.28 &  2.78 &  1.64$\pm$0.09 &  3.17$\pm$0.75 &  4.52$\pm$1.17 &  1.94$\pm$0.27\\ 
  \ldots  & \ldots & \ldots & \ldots & \ldots &  -0.42 &  -0.48 &  -2.29 &  -4.86 &  -1.13 &  -1.79 &  -1.31 & \ldots & \ldots & \ldots & \ldots\\ 
  MMT\#08 &  0.17$\pm$0.03 &  1.11$\pm$0.04 &  1.71$\pm$0.03 &  1.87 &  2.46 &  3.81$\pm$0.03 &  4.34 &  4.66$\pm$0.02 &  5.57 &  5.43 &  5.40 &  6.01$\pm$0.10 &  6.99$\pm$0.69 & 10.37$\pm$1.69 & \ldots\\
  \ldots & \ldots & \ldots & \ldots & \ldots &  -0.01 &  -1.07 &  -0.41 &  -0.08 &  -0.80 & \ldots &  -0.02 & \ldots & \ldots & \ldots & \ldots\\ 
  MMT\#09 &  0.85$\pm$0.05 &  2.12$\pm$0.06 &  2.23$\pm$0.03 &  2.03 &  3.29 &  4.43$\pm$0.03 &  4.73 &  4.93$\pm$0.02 &  5.50 &  5.47 &  5.76 &  7.49$\pm$0.10 &  8.99$\pm$0.67 & 10.56$\pm$1.69 & \ldots\\ 
  \ldots  & \ldots & \ldots & \ldots & \ldots &  -0.22 &  -0.05 &  -0.14 & \ldots &  -0.62 & \ldots &  -0.36 &  -1.71 & \ldots & \ldots & \ldots\\ 
  MMT\#10 &  0.82$\pm$0.06 &  1.15$\pm$0.03 &  1.89$\pm$0.03 &  2.68 &  4.98 &  3.57$\pm$0.03 &  4.75 &  4.22$\pm$0.02 &  4.84 &  4.56 &  4.66 &  5.33$\pm$0.10 &  5.64$\pm$0.69 &  9.85$\pm$1.49 &  3.38$\pm$0.28\\
  \ldots  & \ldots & \ldots &  -0.03 &  -0.27 &  -1.43 & \ldots &  -0.65 &  -0.18 &  -0.66 & \ldots & \ldots & \ldots & \ldots & \ldots & \ldots\\
  MMT\#11 &  1.93$\pm$0.09 &  2.77$\pm$0.09 &  3.37$\pm$0.03 &  4.05 &  6.99 &  9.59$\pm$0.03 &  6.18 &  6.16$\pm$0.02 &  7.41 &  6.95 &  6.85 &  6.90$\pm$0.10 & 10.02$\pm$0.70 &  9.74$\pm$1.61 &  7.46$\pm$0.28\\ 
  \ldots  & \ldots & \ldots & \ldots &  -0.44 &  -1.73 &  -3.73 &  -0.29 & \ldots &  -1.11 & \ldots & \ldots & \ldots & \ldots & \ldots & \ldots\\ 
  MMT\#12 &  0.34$\pm$0.04 &  1.82$\pm$0.08 &  2.01$\pm$0.03 &  1.91 &  2.26 &  3.15$\pm$0.03 &  3.60 &  3.80$\pm$0.02 &  4.41 &  4.10 &  4.02 &  4.20$\pm$0.10 &  4.16$\pm$0.74 &  4.22$\pm$1.63 & \ldots\\
  \ldots  & \ldots & \ldots & \ldots & \ldots & \ldots &  -0.90 &  -0.34 &  -0.04 &  -0.62 & \ldots &  -0.02 & \ldots & \ldots & \ldots & \ldots\\ 
  MMT\#13 &  0.51$\pm$0.04 &  0.84$\pm$0.06 &  0.66$\pm$0.03 &  0.66 &  0.85 &  0.89$\pm$0.03 &  0.88 &  0.77$\pm$0.02 &  1.26 &  0.60 &  0.94 &  2.06$\pm$0.10 &  0.79$\pm$0.66 &     $\leq$1.62 &  1.21$\pm$0.26\\
  \ldots  & \ldots & \ldots & \ldots & \ldots &  -0.09 &  -0.06 &  -0.09 & \ldots &  -0.62 & \ldots &  -0.33 &  -1.50 & \ldots & \ldots & \ldots\\ 
  MMT\#14 &  3.81$\pm$0.11 &  7.68$\pm$0.15 &  9.16$\pm$0.03 &  8.64 & 11.57 & 10.59$\pm$0.03 & 10.76 &  9.69$\pm$0.02 & 15.73 & 10.03 & 13.16 & 27.56$\pm$0.10 & 12.44$\pm$0.76 & 12.66$\pm$1.82 & 13.34$\pm$0.29\\
  \ldots  & \ldots & \ldots & \ldots & \ldots &  -1.54 &  -0.58 &  -0.82 & \ldots & \ldots & \ldots &  -4.84 & -20.59 & \ldots & \ldots &  -0.02\\ 
  Keck\#1 &     $\leq$0.03 &  0.93$\pm$0.05 &  0.55$\pm$0.03 &  0.51 &  0.54 &  0.55$\pm$0.03 &  0.65 &  0.79$\pm$0.02 &  0.66 &  1.85 &  1.65 &  0.31$\pm$0.09 &  1.17$\pm$0.74 &     $\leq$1.62 & \ldots\\
  \ldots  & \ldots & \ldots & \ldots & \ldots & \ldots & \ldots &  -0.07 &  -0.21 &  -0.14 &  -1.60 &  -1.39 &  -0.02 &  -0.56 & \ldots & \ldots\\ 
  Keck\#2 &  0.15$\pm$0.03 &  0.62$\pm$0.03 &  0.59$\pm$0.03 &  0.61 &  0.69 &  0.91$\pm$0.03 &  0.99 &  0.95$\pm$0.02 &  1.27 &  0.94 &  0.96 &  0.88$\pm$0.09 &  0.84$\pm$0.66 &  2.19$\pm$1.48 &  1.21$\pm$0.26\\
  \ldots  & \ldots & \ldots & \ldots & \ldots & \ldots &  -0.14 &  -0.07 &  -0.03 &  -0.26 & \ldots & \ldots & \ldots & \ldots & \ldots & \ldots\\ 
  Keck\#3 &     $\leq$0.03 &  3.70$\pm$0.13 &  4.20$\pm$0.03 &  3.81 &  4.46 &  4.77$\pm$0.03 &  5.91 &  7.65$\pm$0.02 &  7.54 & 10.23 &  9.36 &  8.38$\pm$0.10 & 13.20$\pm$0.71 & 15.65$\pm$1.83 & \ldots\\
  \ldots  & \ldots & \ldots & \ldots & \ldots & \ldots & \ldots &  -0.40 &  -1.38 &  -0.13 &  -2.55 &  -1.76 & \ldots &  -1.09 & \ldots & \ldots\\ 
  Keck\#4 &     $\leq$0.03 &  0.55$\pm$0.03 &  0.78$\pm$0.03 &  0.80 &  0.98 &  1.01$\pm$0.03 &  1.25 &  1.14$\pm$0.02 &  1.51 &  1.56 &  2.25 &  1.74$\pm$0.09 &  2.57$\pm$0.62 &  1.84$\pm$0.95 &  2.12$\pm$0.27\\
  \ldots  & \ldots & \ldots & \ldots & \ldots & \ldots & \ldots &  -0.14 &  -0.03 &  -0.33 &  -0.22 &  -0.62 &  -0.20 &  -0.63 & \ldots & \ldots\\ 
  Keck\#5 &     $\leq$0.03 &  0.28$\pm$0.02 &  0.20$\pm$0.03 &  0.21 &  0.24 &  0.18$\pm$0.03 &  0.30 &  0.36$\pm$0.02 &  0.40 &  0.68 &  0.63 &  0.25$\pm$0.08 &  1.05$\pm$0.58 &     $\leq$1.62 &  0.54$\pm$0.24\\
  \ldots  & \ldots & \ldots & \ldots & \ldots & \ldots & \ldots &  -0.03 &  -0.08 &  -0.05 &  -0.23 &  -0.19 & \ldots &  -0.08 & \ldots & \ldots\\ 
  Keck\#6 &     $\leq$0.03 &     $\leq$0.03 &  0.03$\pm$0.07 &  0.10 &  0.11 &  0.08$\pm$0.02 &  0.15 &  0.13$\pm$0.01 &  0.13 &  0.50 &  0.37 &     $\leq$0.10 &     $\leq$0.61 &     $\leq$1.62 & \ldots\\ 
  \ldots  & \ldots & \ldots & \ldots & \ldots & \ldots & \ldots & \ldots & \ldots &  -0.03 &  -0.19 &  -0.14 & \ldots &  -0.07 & \ldots & \ldots\\ 
  \vspace{-3mm}
  \enddata
  \label{tab:SED}
  \tablecomments{All fluxes are provided in units of $\mu$Jy. Flux uncertainties for $B$, $V$, \Rc,
    and $i$\arcmin\ are 0.01 $\mu$Jy, and 0.04 and 0.03 for $z_b$ and $z$\arcmin, respectively.
    All uncertainties are reported at the 68\% C.L.}
\end{deluxetable*}
\clearpage
\end{landscape}

\begin{appendix}

\section{Accurate Flux Calibration of MMT/Hectospec and Keck/DEIMOS Spectra}\label{sec:f_calib}

Since the spectroscopic data were obtained from two different
instruments, MMT/Hectospec and Keck/DEIMOS, using different
observational configurations (fibers versus slits) and in various
observational conditions (seeing: 0\farcs5--1\farcs5), the flux
calibration of our data is problematic. To better unify the
spectroscopic data, we use the following approach.

First, spectro-photometric standard stars were observed. These
observations were generally obtained on the same night; however,
in some cases, these calibration data were taken a few days apart.
Reducing the calibration data in identical or similar ways to
our science spectra, we determine the sensitivity function(s) using
standard \textsc{iraf} processing techniques. We then apply these
sensitivity functions to our data to yield ``first-pass''
flux-calibrated spectroscopic products.

Second, we examine the reliability of our flux calibration for
each slit mask or fiber configuration by comparing them to
broad-band photometric data. Here we only consider the brightest
galaxies such that the continuum is well-detected. For the MMT and
DEIMOS data, we restrict the calibration sample to $R_{\rm C}\leq22.0$
mag and $i\arcmin\leq23.0$ mag, respectively. This typically has 10
galaxies per set-up. To avoid the effects of OH skylines, we generate
smoothed spectra where a boxcar median is used. From there we
convolved our smoothed spectra against the filter responses and
determine fluxes. These fluxes are compared against the photometric
data to examine slit/fiber losses and if any second-order
wavelength-dependent corrections are needed. In most cases, the
comparison against photometric data only showed a systematic offset
due to slit/fiber losses.
This comparison showed that the rms is typically between 10\%--20\%.

\begin{figure}
  \epsscale{1.1}
  \plottwo{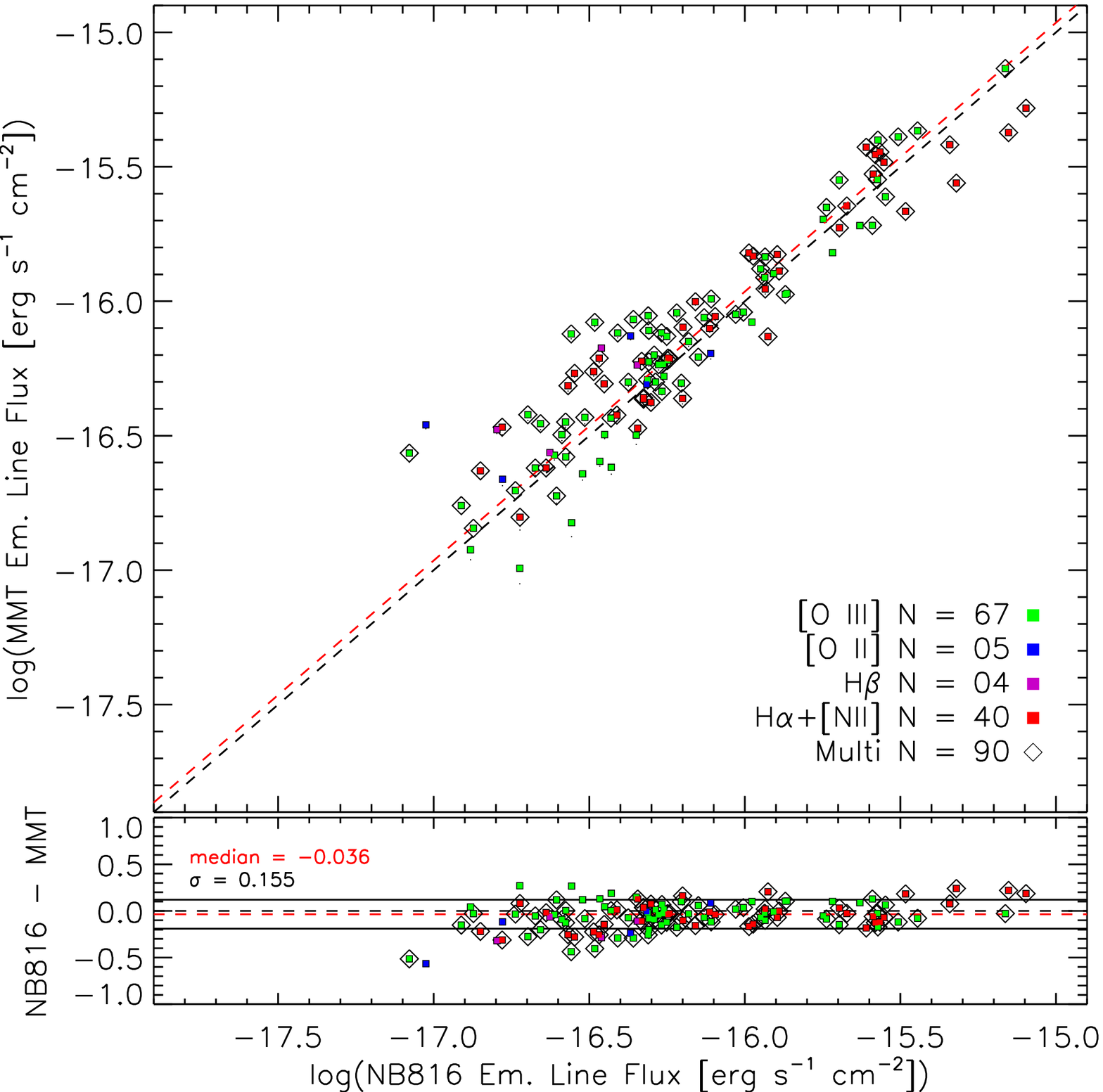}{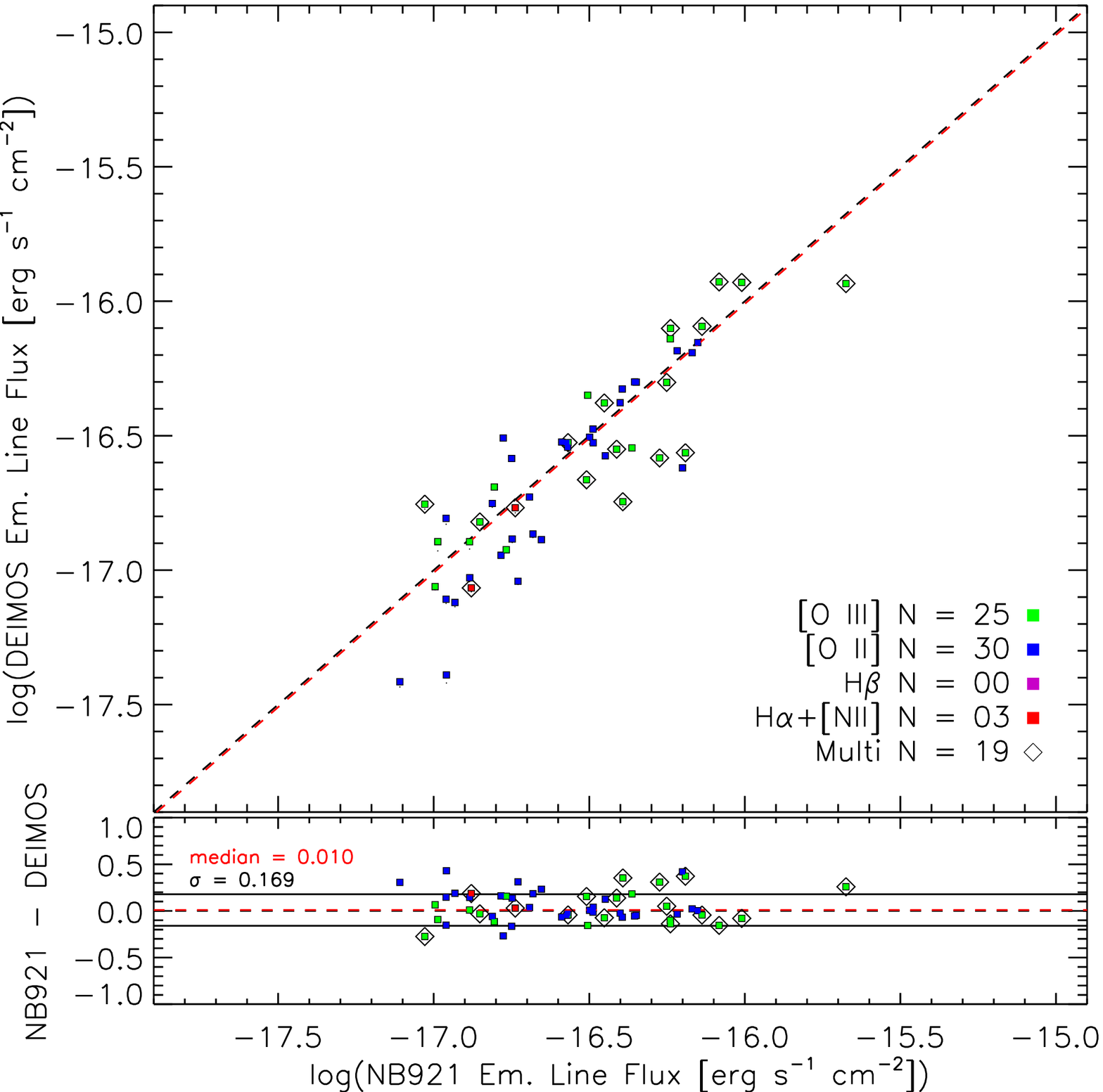}
  \caption{Comparison between spectroscopic flux measurements against
    NB fluxes. The left (right) panel shows the MMT (DEIMOS) flux
    measurements for NB816 (NB921) emitters. Measurements for 
    \Ha+\NII, \OIII, \Hb, and \OII\ are shown as red, green,
    purple, and blue, respectively. Diamonds denote those measurements
    where more than one emission line enters the NB filter (e.g.,
    \Ha\ and \NII, or \OIII\,$\lambda$4959,5007). These figures
    demonstrate that the flux calibration is precise and accurate
    for both MMT and DEIMOS. Other NB emitters have been inspected
    and similar results have been found.}
  \label{fig:calib1}
\end{figure}

Next, to ensure that the flux calibration is reliable, we use our
measurements of nebular emission lines from NB imaging data to
compare against the spectroscopic measurements. The NB fluxes
are derived from Equation~(\ref{eqn:Ha_flux}) but differs for
each NB filter. We use the spectroscopic redshift to correct
the NB flux for when the emission line falls in the filter's wing.
This comparison is illustrated in Figure~\ref{fig:calib1}.
For 116 NB816 emitters with MMT spectroscopy, we find excellent
agreement with a small offset (--0.036 dex) and an rms of 0.155
dex. Similar results are also seen for 108 NB704 emitters.
We note that these calibration results are consistent with
those obtained by the Hectospec instrument team \citep{fabricant08}.
In addition, the DEIMOS fluxes also show great agreement for 58
NB921 emitters.
Similar results are also seen for 9 NB704, 39 NB816, and 8 NB973
emitters. This consistent agreement at multiple wavelengths
supports our wavelength-dependent flux calibration.
Note that in our second step we {\it only} use broad-band data,
so this comparison against NB data is an independent test.

\begin{figure}
  \epsscale{1.1}
  \plottwo{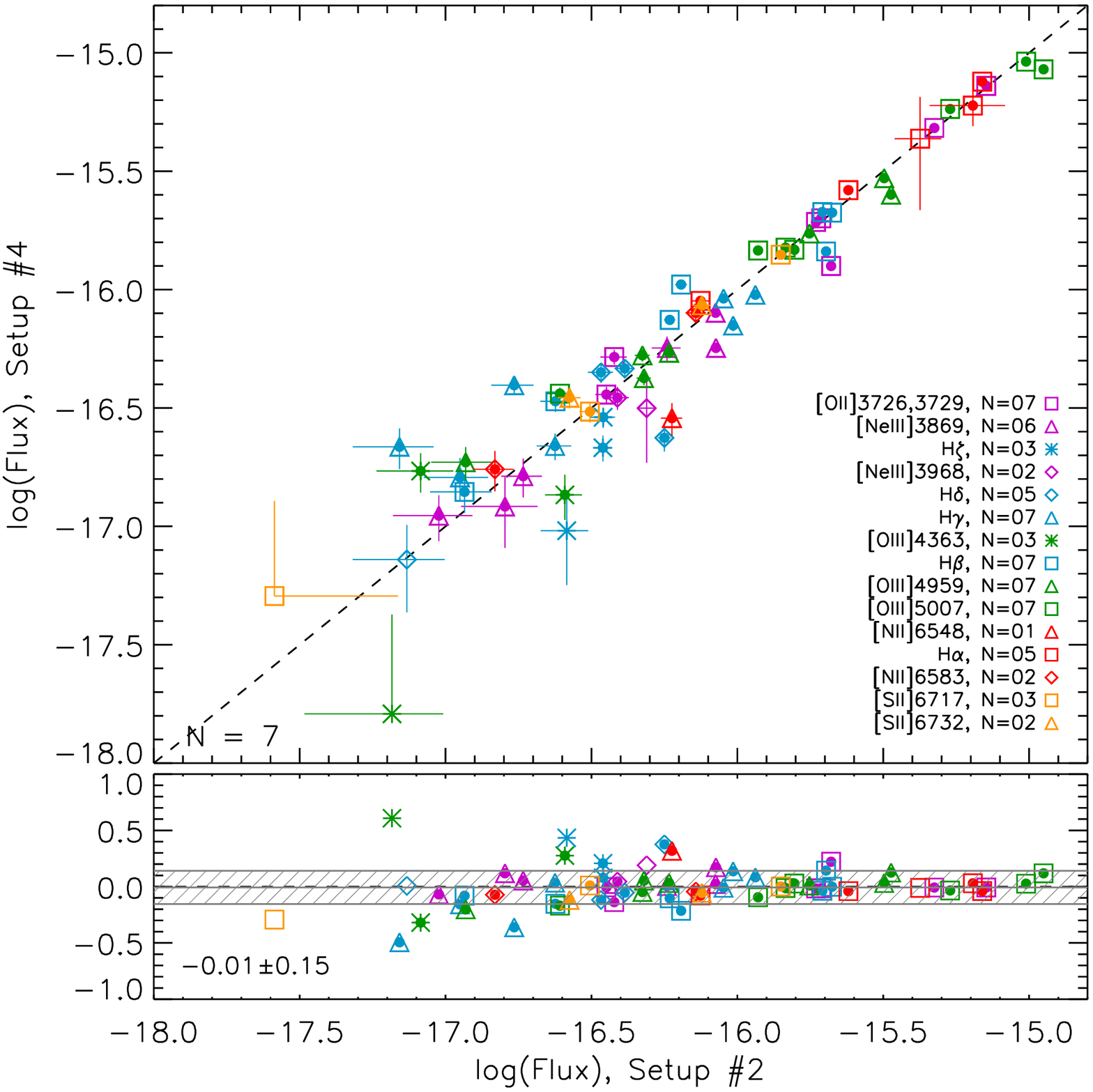}{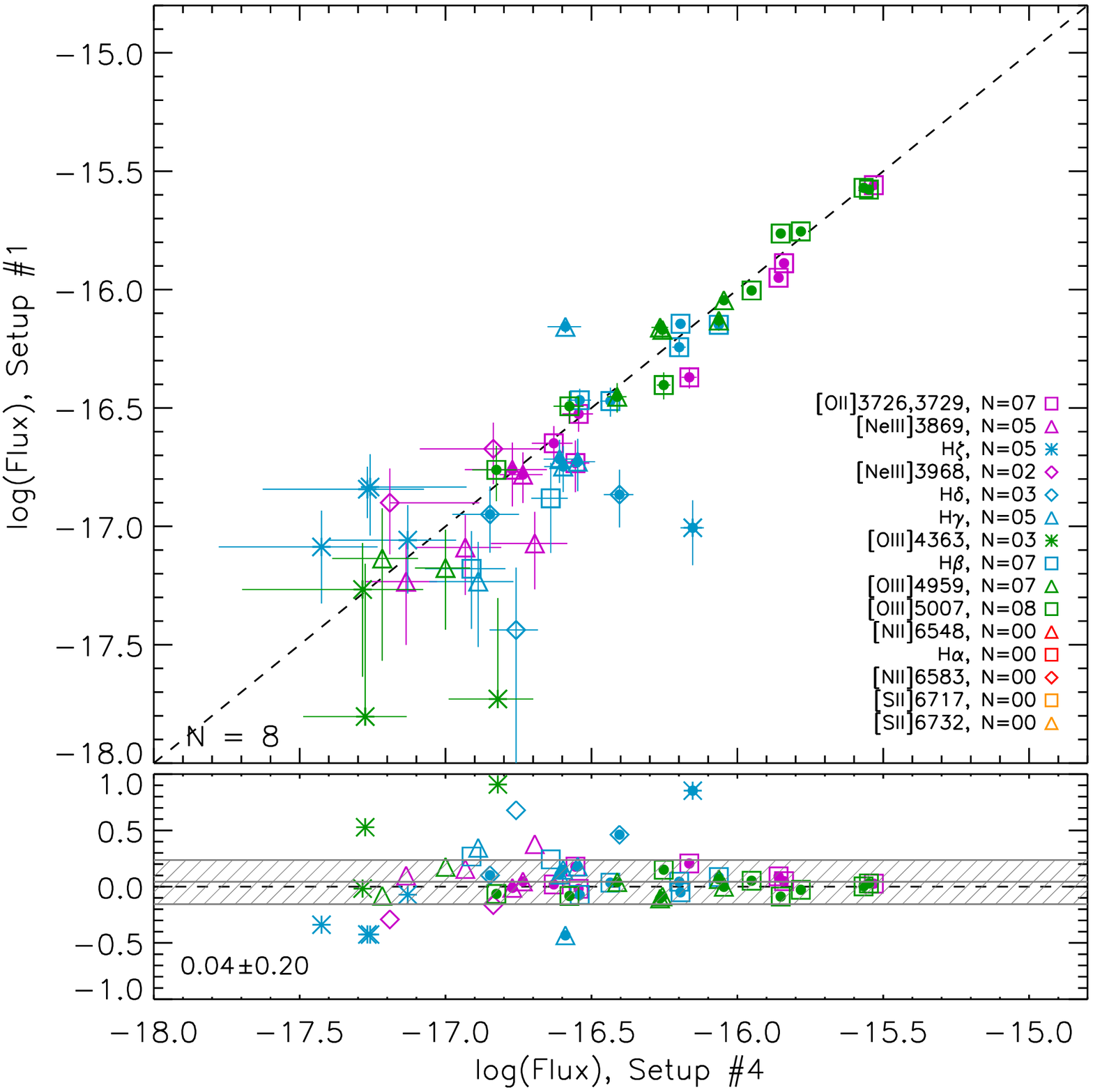}
  \caption{Comparison between emission-line fluxes for various MMT setups.
    The top panels show the flux comparisons while the bottom panels show the
    differences against the flux. The different colors and symbols correspond
    to a given emission line (e.g., purple squares for \OII\,$\lambda\lambda$3726,3729).
    Measurements that are reliable at $\geq$3$\sigma$ in both spectra
    are marked with filled circles. The shaded grey regions in the bottom
    panels show the median and $\pm1\sigma$. The median is computed using
    only $\geq$3$\sigma$ measurements.}
  \label{fig:calib2}
\end{figure}

Finally, a second test that we conducted is to examine the
emission-line flux determinations for the same galaxies in 
different observing conditions. This was only feasible with
MMT because of the wide field-of-view to target multiple
galaxies simultaneously. In total, we have 28 galaxies with
two spectra. We illustrate in Figure~\ref{fig:calib2} the
determination of fluxes for various emission lines that
span rest-frame wavelengths of 3700\AA--6730\AA\ and
observed wavelengths of 4000\AA--8300\AA.
We find that between different MMT setups, the same emission-line
fluxes are obtained with an rms of 0.15--0.20 dex across a wide
spectral range, which further supports our approach for accurate
flux calibration.

We note that the effects of differential refraction are
mitigated for the MMT spectra, as an atmospheric dispersion
compensator (ADC) is used and is known to eliminate issues
affecting the flux calibration of the spectra \citep{fabricant08}.
DEIMOS, however, does not have an ADC. The wavelength-dependent
effects of differential refraction are seen in some of our DEIMOS
observations when comparisons are made against broad-band data.
Since our data are calibrated against broad-band photometry,
as discussed above, these issues are resolved (to first order).

\section{Rest-frame Optical Spectra for  \OIIIa\ Galaxies}
\label{sec:other_sources}

\begin{figure*}
  \epsscale{1.1}
  \plotone{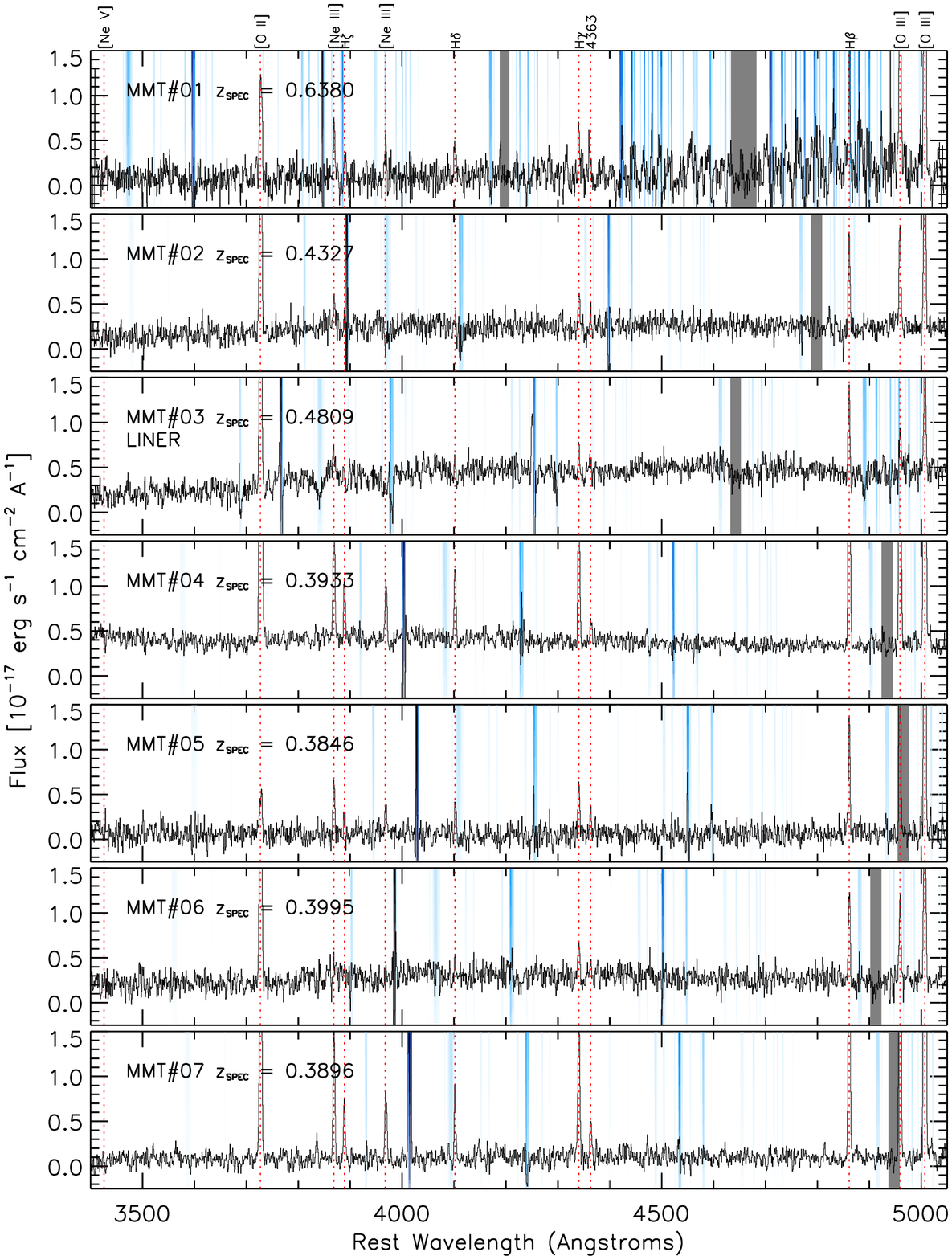}
  \caption{MMT/Hectospec spectra for seven galaxies with full spectral
    coverage and detection of \OIIIa\ at $\geq$3$\sigma$. The most common
    nebular emission lines between 3400\AA\ and 5010\AA\ are denoted by
    the red dashed lines. OH sky-lines are indicated by blue vertical
    bands where the strength of the sky-lines is denoted by their
    shade of blue (darker is stronger). Grey shaded regions are
    those affected by atmospheric absorption (A- and B-bands).}
  \label{fig:MMT_spec1}
\end{figure*}

\begin{figure*}
  \epsscale{1.1}
  \plotone{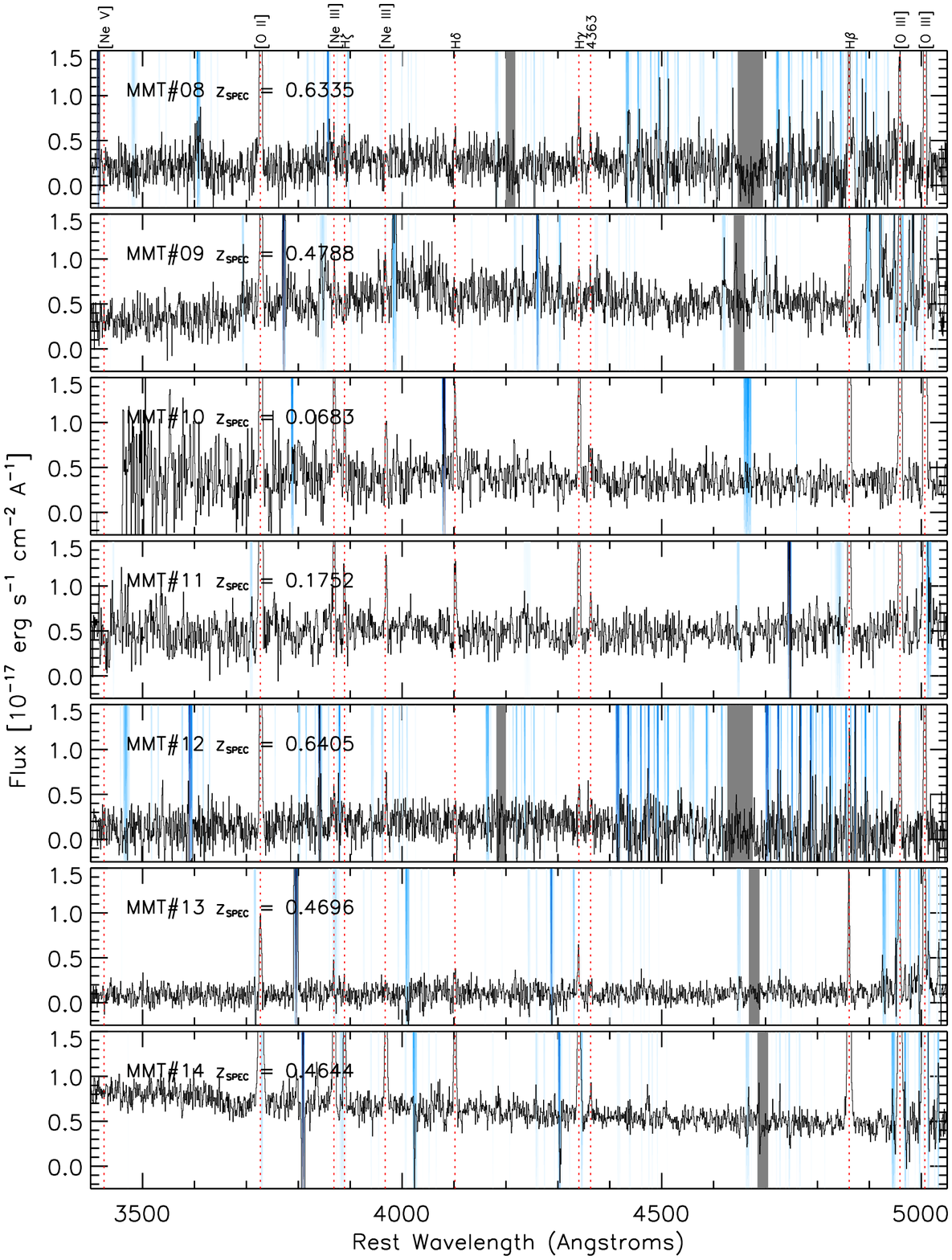}
  \caption{Same as Figure~\ref{fig:MMT_spec1} but for another seven
    galaxies observed with MMT/Hectospec.}
  \label{fig:MMT_spec2}
\end{figure*}

\begin{figure*}
  \epsscale{1.1}
  \plotone{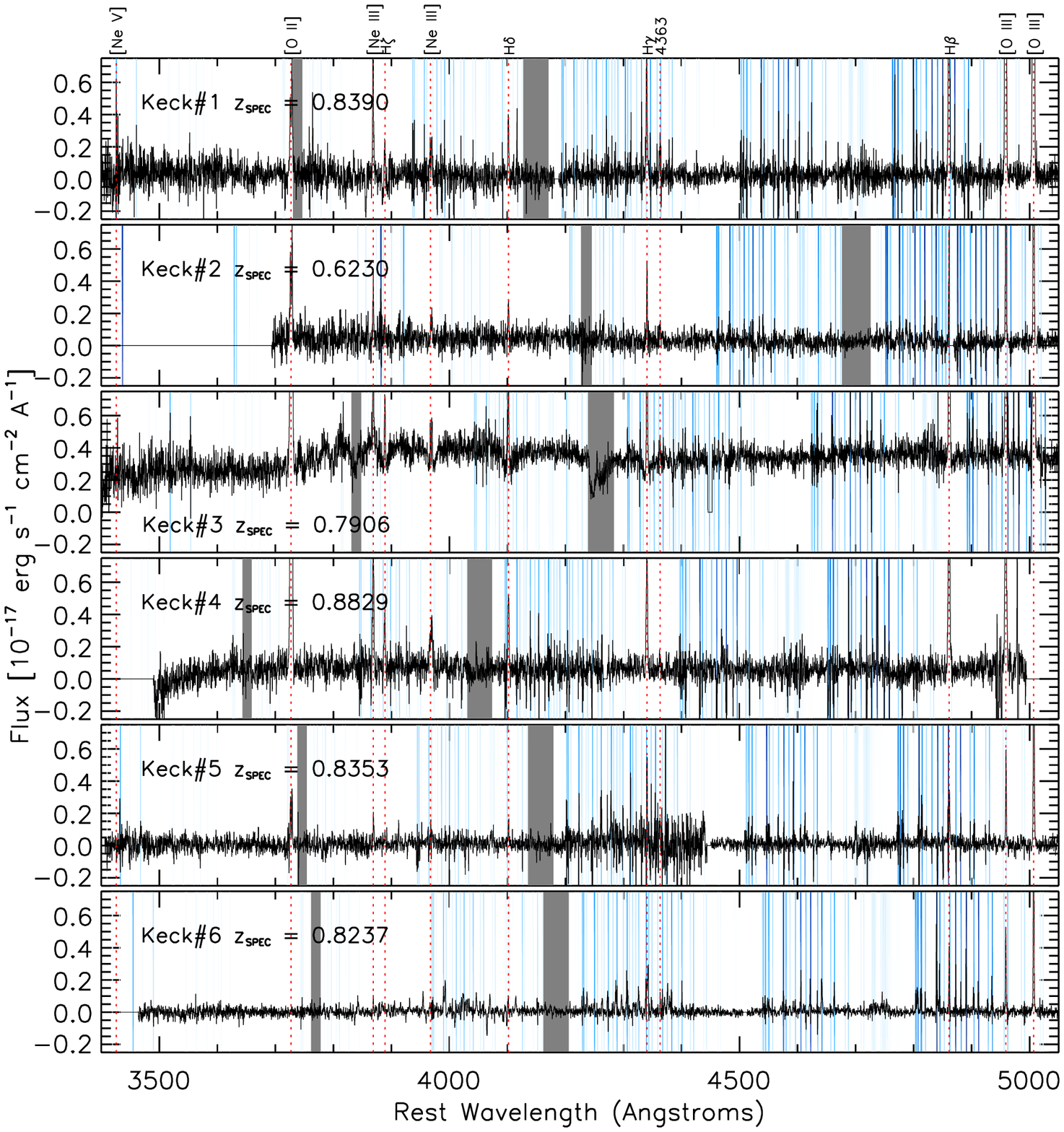}
  \caption{Keck/DEIMOS spectra for six galaxies with full spectra
    coverage and detection of \OIIIa\ at $\geq$3$\sigma$. The most
    common nebular emission lines between 3400\AA\ and 5010\AA\ are
    denoted by the red dashed lines. OH sky-lines are indicated by
    blue vertical bands where the strength of the sky-lines is
    denoted by their shade of blue (darker is stronger).
    Grey shaded regions are those affected by atmospheric
    absorption (A- and B-bands).}
  \label{fig:Keck_spec}
\end{figure*}

Here, we provide the rest-frame optical spectra for our sample of
\Ndet\ galaxies in Figures~\ref{fig:MMT_spec1}--\ref{fig:Keck_spec}.
These plots are limited to a rest-frame wavelengths of 3400\AA--5050\AA.
Our figures are limited toward lower fluxes to show the weak emission
lines. Since the strong lines are saturated in these figures, we refer
readers to Table~\ref{tab:em_lines}, which provides the emission-line
strengths.
Overlaid in blue on these spectra are the locations of OH sky-lines
with the OH line strengths indicated by the darkness of the blue shades.
We also overlay strong and weak nebular emission lines in red.
We previously discussed in Section~\ref{sec:spectra} the
acquisition and reduction of these spectra, and will discuss
the absolute flux calibration of our spectra in
Appendix~\ref{sec:f_calib}.
  

\end{appendix}

\end{document}